\documentclass[aps,11pt,nofootinbib]{revtex4-1}
\pdfoutput=1

\usepackage{graphicx}
\usepackage[margin=1in]{geometry}
\usepackage{amsmath}
\usepackage{amssymb}
\usepackage{xcolor}
\usepackage[utf8]{inputenc}

\usepackage{tensor}
\usepackage{braket}
\usepackage{float}
\usepackage[linktocpage=true,colorlinks=true,linkcolor=blue,citecolor=blue,urlcolor=blue]{hyperref}
\hypersetup{pdftitle={Primordial black hole merger rates: distributions for multiple LIGO observables},pdfauthor={Andrew D. Gow, Christian T. Byrnes, Alex Hall, John A. Peacock}}

\newcommand{\PBH}{\mathrm{PBH}}
\newcommand{\M}{\mathrm{M}}
\newcommand{\diffd}{\mathrm{d}} 
\newcommand{\ddv}[2]{\frac{\diffd #1}{\diffd #2}} 
\newcommand{\Msun}{\mathrm{M}_\odot}

\begin{document}

\author{Andrew D.~Gow$^1$}
\email{A.D.Gow@sussex.ac.uk}
\author{Christian T.~Byrnes$^1$}
\email{C.Byrnes@sussex.ac.uk}
\author{Alex Hall$^2$}
\email{ahall@roe.ac.uk}
\author{John A.~Peacock$^2$}
\email{jap@roe.ac.uk}

\affiliation{\\1) Department of Physics and Astronomy, University of Sussex,\\Brighton BN1 9QH, United Kingdom\\} 

\affiliation{\\2) \mbox{Institute for Astronomy, University of Edinburgh,} Royal Observatory, Blackford Hill,\\Edinburgh, EH9 3HJ, United Kingdom\\}

\date{28/11/2019}


\title{Primordial black hole merger rates: distributions for multiple LIGO observables}

\begin{abstract}
We have calculated the detectable merger rate of primordial black holes, as a function of the redshift, as well as the binary's mass ratio, total mass and chirp mass (observables that have not previously been explored in great detail for PBHs). We consider both the current and design sensitivity of LIGO and five different primordial black hole mass functions, as well as showing a comparison to a predicted astrophysical black hole merger rate. We show that the empirical preference for nearly equal-mass binaries in current LIGO/Virgo data can be consistent with a PBH hypothesis once observational selection effects are taken into account. However, current data do exclude some PBH mass distributions, and future data may be able to rule out the possibility that all observed BH mergers had a primordial origin.
\end{abstract}

\maketitle

\tableofcontents

\section{Introduction}
Primordial Black Holes (PBHs) were first considered by Zel'dovich and Novikov \cite{Zel'dovich_1967}, and were heavily studied by Hawking and Carr \cite{Hawking_1971,Carr_1974}. Since then, the field has generated an extensive literature; for a review of the current state of research, see \cite{Green_2015} and \cite{Sasaki_2018}. As they interact only via gravity, PBHs are a natural dark matter (DM) candidate without requiring physics beyond the standard model. The fraction of DM that can be composed of PBHs, $f_\PBH$, has been well constrained by a number of methods \cite{Sasaki_2018}. One of these methods is provided by the detection of gravitational wave signals by the Laser Interferometer Gravitational-wave Observatory (LIGO). During the O1 and O2 runs, LIGO detected 10 binary black hole (BBH) mergers \cite{LIGO_2019_GWTC-1}. The detector has recently finished its O3a run (the first half of the O3 sensitivity run) lasting from 1 April to 1 October 2019, and has detected 21 mergers with $>90\%$ probability of being BBHs. Some unexpected properties of the detected mergers in the O1O2 dataset, such as the high mass and low effective spin, led to the suggestion that the mergers may be primordial in origin \cite{Bird_2016,Clesse_2016,Sasaki_2016}, and that they could explain an excess of power in the cosmic infrared background, although this requires $f_\PBH\sim1$ \cite{Kashlinsky_2016}. The aim of our current paper is to investigate this possibility in more detail, attempting to model properties such as the BBH mass ratios, allowing for observational selection effects that bias the rates with which different binaries are detected.

A major goal of any PBH analysis is to place constraints on $f_\PBH$. There are a number of methods for placing limits on this parameter, over a large range of mass scales. In the LIGO range of $\sim$1-100 $\Msun$, the relevant constraining techniques are microlensing events \cite{Tisserand_2007,Wyrzykowski_2010,Oguri_2018,Zumalacarregui_2017} and CMB accretion effects \cite{AliHaimoud_2017a,Poulin_2017}. A summary plot of these limits (and other limits on different mass scales) can be seen in fig.~10 of \cite{Sasaki_2018} (published version). In this range, the tightest constraint on $f_\PBH$ comes from either CMB accretion limits or from the LIGO events themselves (fig.~17 of \cite{Sasaki_2018}, published version), depending on the sound speed of baryonic matter compared to the relative baryon to dark matter velocity \cite{Poulin_2017}. This means that, even with $f_\PBH<1$, all of the LIGO events could be of primordial origin. However, the constraints discussed above have all been determined for a monochromatic PBH mass distribution. This case is unrealistic based on the typical formation mechanisms, and so some effort has been made into determining equivalent limits for extended mass functions \cite{Carr_2017,Bellomo_2018}.

If PBHs exist, it is of extreme importance that their mass distribution be accurately characterised. We have made the first study of several parameters related to the masses in PBH scenarios, such as the total mass $M$, the chirp mass $\mathcal{M}$, and the mass ratio $q=m_2/m_1$. The LIGO convention is that $m_2$ is the smaller mass, and hence $0< q\leqslant1$. The PBH mass distribution is already observationally constrained, and with future data it will be possible to determine the distribution and its parameters to a high degree of accuracy, or perhaps even rule out any possible PBH mass distribution as the origin of all the detected LIGO events.

A particular motivation for considering the mass ratio $q$ is that the LIGO data have central $q$ values that are all statistically consistent with equal mass mergers. One may wonder whether such a strong correlation in mass is plausible for PBHs with a broad mass function, and much of our paper is devoted to considering this question. Of course, the same issue of principle arises if the BHs are of astrophysical origin, but it seems that $q$ could naturally be close to unity in this case \cite{Marchant_2016,Rodriguez_2016,O_Leary_2016,Kovetz_2017}, while PBH binaries, having a very different formation mechanism, would not necessarily have such a strong tendency. It is interesting to consider whether the current LIGO data favour a particular $q$ value, but the LIGO selection effects discussed below (which have a preference for equal mass mergers) must be taken into account. A recent analysis of the LIGO data showed that, for a mass distribution based on a power-law form, $q>0.6$ is favoured \cite{Fishbach_2019}.

Additional observables could also be used to distinguish between mergers of primordial and astrophysical origin. A recent paper by Gerosa et al \cite{Gerosa_2019} obtained merger rate distributions for astrophysical black holes against three observables: the total mass $M$, redshift $z$, and the mass ratio $q$. It is desirable to have the same distributions for PBHs so a comparison between the primordial and astrophysical cases can be drawn.

The intrinsic merger rate for PBHs is obtained by considering the number density of PBHs, and their interactions. A binary is formed when the gravitational attraction between two adjacent PBHs dominates over the Hubble flow. The surrounding PBHs, as well as other forms of DM, then generate a tidal force that determines the angular momentum of the binary, which in turn determines how long the binary takes to merge. The intrinsic merger rate as a function of time can then be obtained. This calculation has been carried out in a number of ways by various groups, for monochromatic \cite{Nakamura_1997,AliHaimoud_2017,Clesse_2017,Eroshenko_2018,Kavanagh_2018} and extended PBH mass distributions \cite{Raidal_2017,Chen_2018}, most recently by Raidal et al \cite{Raidal_2019}, whose method we use for the following calculations.

In section \ref{sec:IntrinsicMergerRate} the theoretical process for obtaining the intrinsic merger rate for PBHs is described, and is briefly shown in its numerical form. Section \ref{sec:Detectability} explains how to determine the rate of detections expected by LIGO for a given intrinsic merger rate. The resulting distributions for the detectable merger rate are shown in section \ref{sec:DetectableMergerRate}, and a comparison of different PBH mass distributions is considered in section \ref{sec:ComparisonOfMassDistributions}. Finally, the merger rate expected for the LIGO O1O2 sensitivity is compared to the detected merger events in section \ref{sec:CurrentLIGOData}.

\section{Intrinsic Merger Rate from PBHs}
\label{sec:IntrinsicMergerRate}
To determine the intrinsic merger rate of PBHs, a number of factors must be considered. First, there is the number density of PBHs of a given mass, which is related to the mass distribution $\psi(m)$. Then, the fraction of these that form binaries must be determined, and the angular momentum distribution must be taken into account to determine the number of PBH binaries that will result in mergers at time $t$. This procedure will give the merger rate assuming that the binaries are not disrupted between their formation and merger. Even a small alteration in the angular momentum $j$ of the binary will cause a significant change in the merger time $\tau$, due to the relation $\tau\propto j^7$ \cite{Peters_1964}. This assumption was considered by Ali-Haimoud et al \cite{AliHaimoud_2017}, who estimated that little disruption occurs, but Raidal et al \cite{Raidal_2019} carried out simulations and argued that significant disruption may occur for $f_\PBH\gtrsim10^{-1}$. A more recent work by Vaskonen and Veerm\"ae determined the lower bound on the merger rate including the impact of disruptions, for $f_\PBH\gtrsim0.1$, and found that it remained too large compared with the LIGO observed merger rate \cite{Vaskonen_2019}.

An additional consideration in finding the merger rate is the clustering of PBHs. If this is an important factor, then it could considerably alter the merger rate at a given time. This has been a topic of some debate, but it is now generally agreed that, for Gaussian initial conditions, the spatial distribution of PBHs is Poissonian. \cite{AliHaimoud_2018,Desjacques_2018,Ballesteros_2018,MoradinezhadDizgah_2019}. Primordial non-Gaussianity can strongly change the initial clustering of PBHs \cite{Tada_2015,Young_2015,Ding_2019,Suyama_2019,Matsubara_2019} and the subsequent merger rate \cite{Young_2019} (see also \cite{Bringmann_2018}).

The merger rate calculation performed by Raidal et al \cite{Raidal_2019} yielded the following equations, reproduced here for convenience:
\begin{align}
\diffd R = S \times \diffd R_0,
\end{align}
where $\diffd R$ is the differential merger rate, $S$ is a suppression factor (given by eq.~2.37 in \cite{Raidal_2019}, published version) that depends on the component masses $m_1$ and $m_2$, the fraction of dark matter in PBHs $f_\PBH$, and the rescaled deviation of matter density perturbations $\sigma_\M$, and
\begin{align}
\diffd R_0 = \frac{1.6\times10^6}{\mathrm{Gpc}^3\ \mathrm{yr}}\,f_\PBH^\frac{53}{37}\,\eta^{-\frac{34}{37}}\left(\frac{M}{\Msun}\right)^{-\frac{32}{37}}\left(\frac{t}{t_0}\right)^{-\frac{34}{37}}\psi(m_1)\,\psi(m_2)\ \diffd m_1\, \diffd m_2
\end{align}
is the unsuppressed differential merger rate, where $\eta$ is the symmetric mass ratio, $M$ is the total mass of the system, $t$ is the proper time, $t_0$ is the age of the universe, and $\psi(m)$ is the mass distribution of PBHs, normalised to unity. The suppression factor $S$ depends on the average number $\bar{N}(y)$ of PBHs in a spherical shell of radius $y$. Raidal et al determine a value of this to ensure minimal disruption of the binary for $f_\PBH<10^{-1}$, given by eq.~3.5 in \cite{Raidal_2019} (published version). We use this value for the following calculations.

\begin{figure}[H]
\centering
\includegraphics[width=0.6\textwidth]{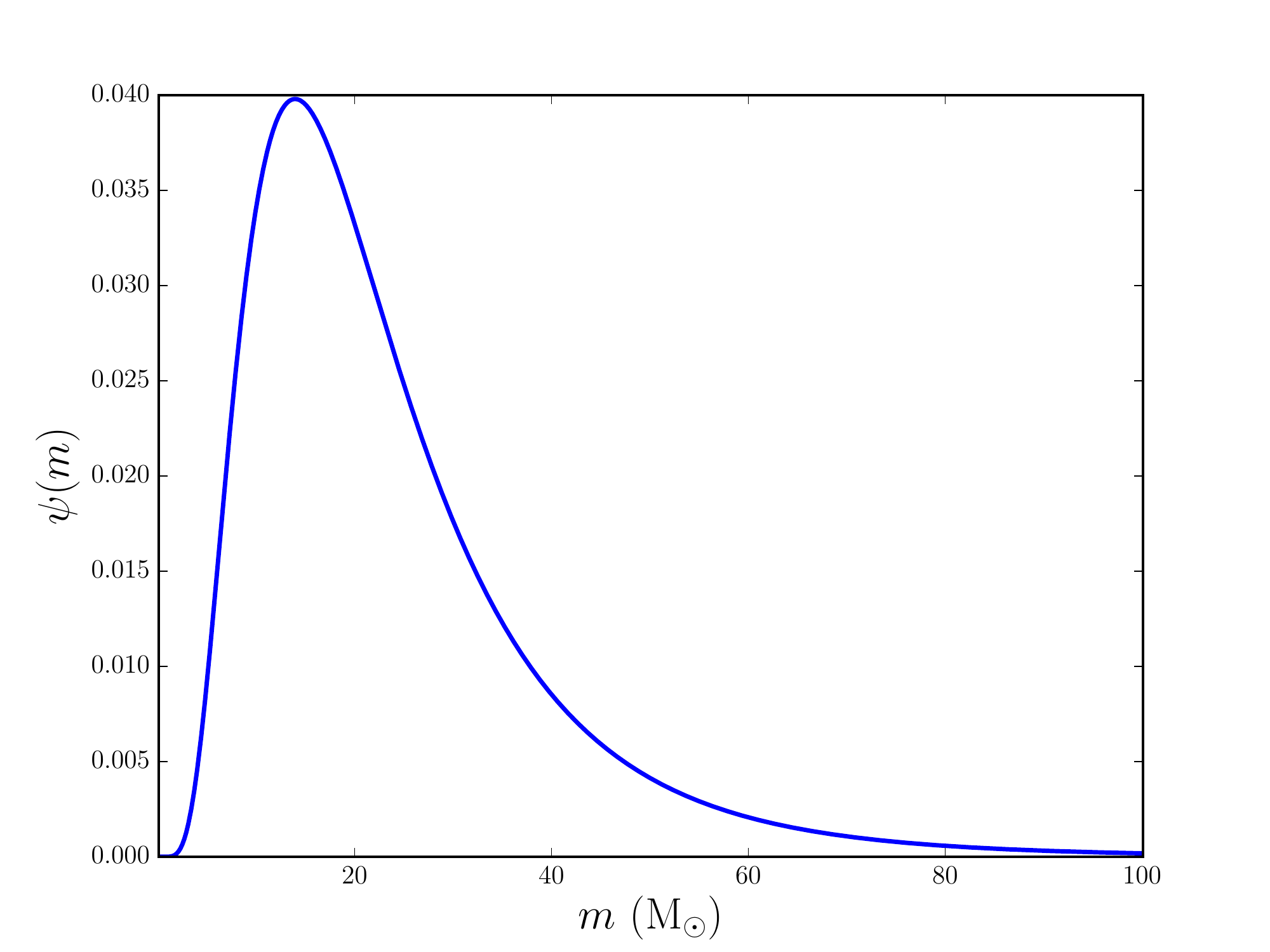}
\caption{Lognormal mass distribution with $m_c=20\ \Msun$ and $\sigma=0.6$.}
\label{fig:Mass-Function-Lognormal-s06}
\end{figure}

The mass distribution used in \cite{Raidal_2019} is a lognormal, given by the form
\begin{align}
\psi(m) &= \frac{1}{\sqrt{2\pi}\sigma m}\exp\left(-\frac{\ln^2(m/m_c)}{2\sigma^2}\right),
\label{eq:PsiLognormal}
\end{align}
where $m_c$ is the median of the distribution (also the mean of $m\psi(m)$) and $\sigma$ describes the width. This is a common choice for the PBH mass distribution, as it well approximates the class of distributions for PBHs formed from peaks in the power spectrum \cite{Carr_2017}. Raidal et al \cite{Raidal_2019} carried out a fit to the LIGO data for this mass distribution, although they did not incorporate the full detectability procedure described in section \ref{sec:Detectability}, instead using a step function in the signal to noise ratio. Their best fit parameters are $m_c=20 \Msun$ and $\sigma=0.6$, and we will begin by considering these values for the mass distribution. The mass distribution with these parameters is shown in fig.~\ref{fig:Mass-Function-Lognormal-s06}.

\vspace{-1em}
\vspace{-1em}

A common alternative form for the mass distribution is $f(m)$, which satisfies
\begin{align}
\int \diffd\ln(m)\ f(m) &= f_\PBH.
\end{align}
This is related to the above mass distribution by the relation
\begin{align}
\psi(m) &= \frac{1}{f_\PBH}\frac{f(m)}{m}.
\end{align}
The total intrinsic (source-frame) merger rate in Gpc$^{-3}$ yr$^{-1}$ can be obtained from the differential merger rate by applying the integration
\begin{align}
R &= \int \diffd m_1 \int \diffd m_2\ \frac{\diffd R}{\diffd m_1\diffd m_2} \label{eq:TotalIntrinsicMergerRate}
\end{align}
for source-frame masses $m_1$ and $m_2$, and with a fixed value of proper time $t$. The result of this (or a similar) equation with $t=t_0$ is often compared with the LIGO estimate for the intrinsic merger rate \cite{Sasaki_2016,AliHaimoud_2017,Clesse_2017,Eroshenko_2018,Kavanagh_2018,Chen_2018}. However, in obtaining their estimate of the intrinsic merger rate, the LIGO collaboration assumes a mass distribution, so this estimate could differ significantly compared to the PBH calculation if a very different mass distribution is used. For further details on this estimation method, see section VII of \cite{LIGO_2019_GWTC-1} (published version). To overcome this problem, the ground-based rate of detections in yr$^{-1}$ can be found, which can then be directly compared to the LIGO measurements, rather than their intrinsic rate estimate. Distributions of this type in the component masses can be seen in \cite{Chen_2018} with $f_\PBH$ chosen to fix the intrinsic merger rate to $R=100$ Gpc$^{-3}$ yr$^{-1}$, although these distributions do not take into account the dependence of the detectability on the component masses. The process for finding the ground-based detection rate is described in the following section.

\section{Detectability and Ground-based Detection Rate}
\label{sec:Detectability}
\begin{figure}[H]
\centering
\includegraphics[width=0.49\textwidth]{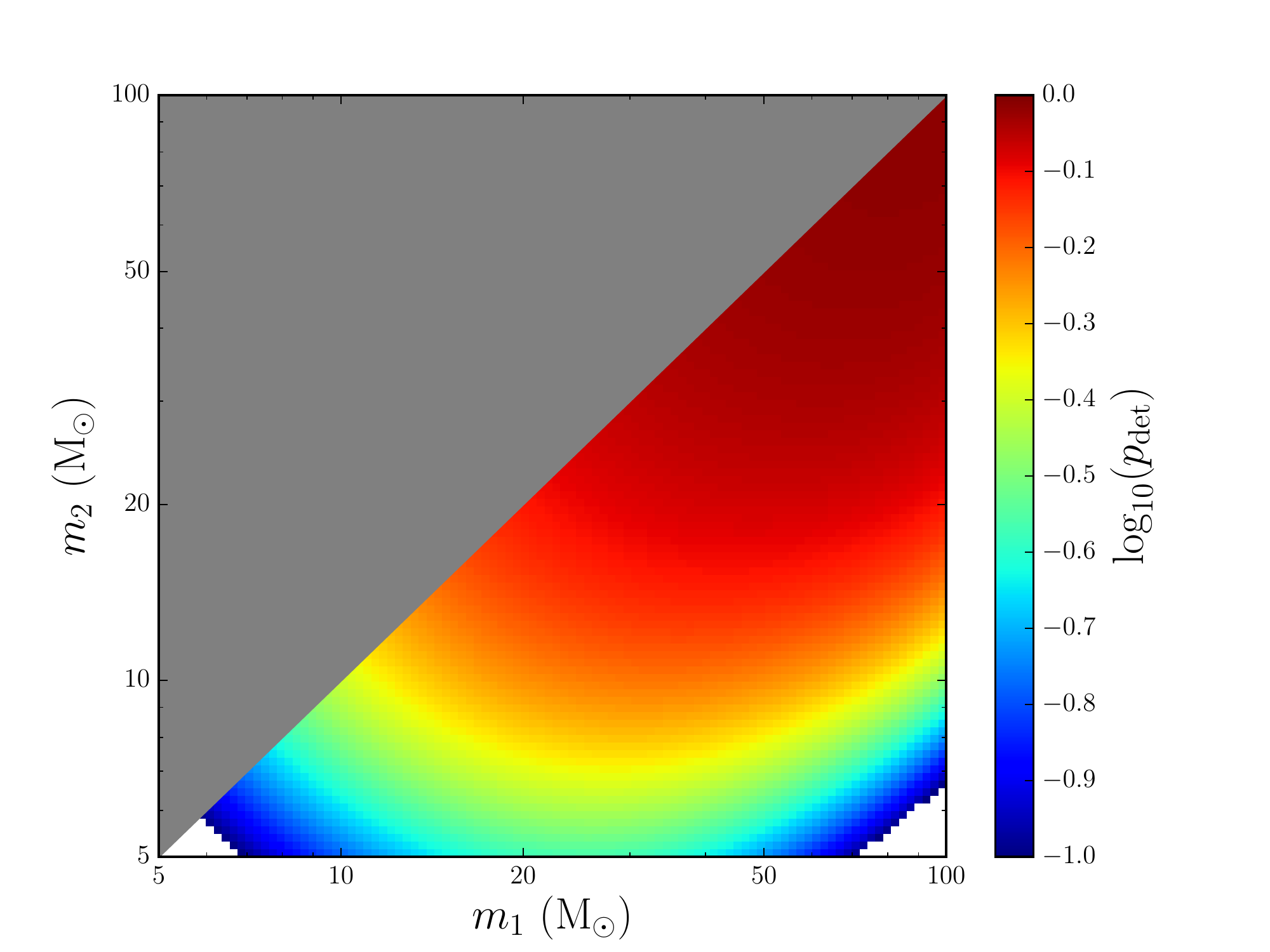}
\includegraphics[width=0.49\textwidth]{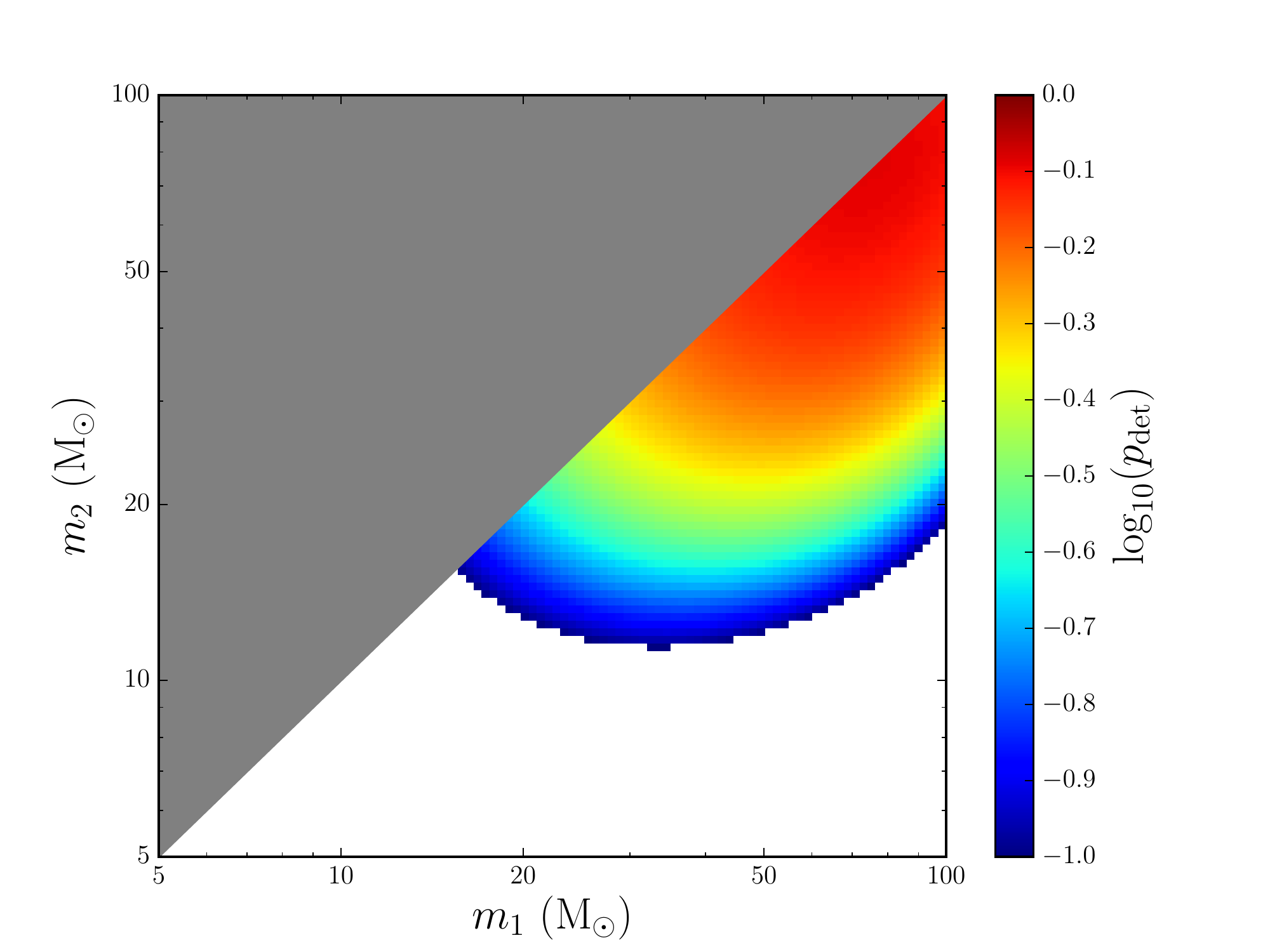}
\caption{Detection probability $p_\text{det}(m1,m2)$ at $z=0.2$ (left) and $z=0.5$ (right). Note that all three scales are in log-space. The white area indicates $p_\text{det}<0.1$, and the grey triangle indicates the case $m_2<m_1$, chosen by LIGO for their analysis.}
\label{fig:pdet}
\end{figure}

To obtain the ground-based detection rate, we must weight the different parts of the intrinsic merger rate with the ability of the LIGO instrument to detect the resulting waveform. This detectable merger rate is calculated using
\begin{align}
R_\text{det} = \int \diffd z \int \diffd m_1 \int \diffd m_2\ \frac{1}{1+z}\,\ddv{V_c}{z}\, p_\text{det}(z,m_1,m_2)\,\frac{\diffd R}{\diffd m_1\diffd m_2}(z), \label{eq:Rdet}
\end{align}
where $V_c$ is the comoving volume and $p_\text{det}$ is the detection probability \cite{Dominik_2015}. This detection probability is obtained by simulating merger waveforms and passing them through the LIGO detection pipeline to find the signal to noise ratio (SNR) in a single detector for a certain set of parameters. The angular dependence of the detection probability may be well approximated by the function $p(\omega)$ \cite{Gerosa_2019}, with the detection probability given by
\begin{align}
p_\text{det}(z,m_1,m_2) = \int_{\rho_{\textrm{thr}}/\rho_{\textrm{opt}}(z,m_1,m_2)}^{1} \diffd\omega\ p(\omega),
\end{align}
where $\rho_{\text{opt}}$ is the SNR for a merger happening face-on to the detector located directly above the detector and $\rho_{\text{thr}}$ is a threshold SNR above which it is assumed that the signal is detected, typically taken as $\rho_{\text{thr}}=8$ \cite{Gerosa_2019}. The noise curve used is the design sensitivity curve (\texttt{aLIGOZeroDetHighPower}). This process is carried out using the public code \texttt{gwdet} written by Davide Gerosa \cite{Gerosa_2017}. The resulting probability is plotted against the component masses in fig.~\ref{fig:pdet}, at $z=0.2$ and $z=0.5$.

The method described above does not take into account the spin of the component BHs. In principle the dependence of the SNR on these spins should be taken into account. Since PBH spins are expected to be very small at formation \cite{Chiba_2017,Mirbabayi_2019,He_2019}, we avoid this computationally expensive step by computing waveform approximants having zero spin. The difference in spin has also been considered as another observable that could be used for distinguishing between mergers of astrophysical and primordial origin \cite{Clesse_2017a,Fernandez_2019}. There is the possibility that, although the PBH spin is small at formation, they could spin up between formation and merger. However, this is likely to be a small effect \cite{Postnov_2019}. Assuming the detection probability varies little over the range around zero where PBH spins are expected to lie, then taking zero spin is a good approximation.

\section{Detectable Merger Rate for LIGO Observables}
\label{sec:DetectableMergerRate}
\begin{figure}[H]
\centering
\includegraphics[width=0.49\textwidth]{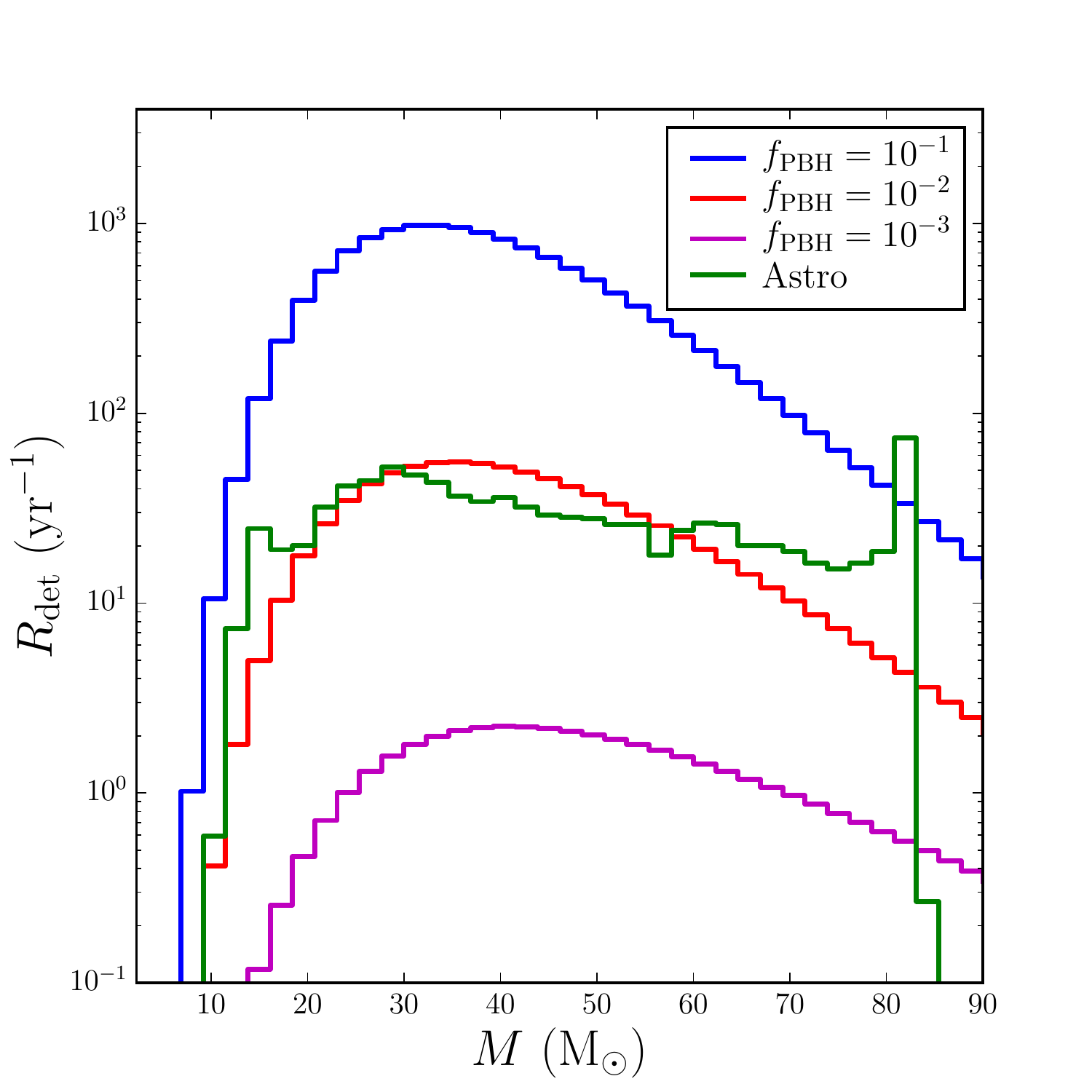}
\includegraphics[width=0.49\textwidth]{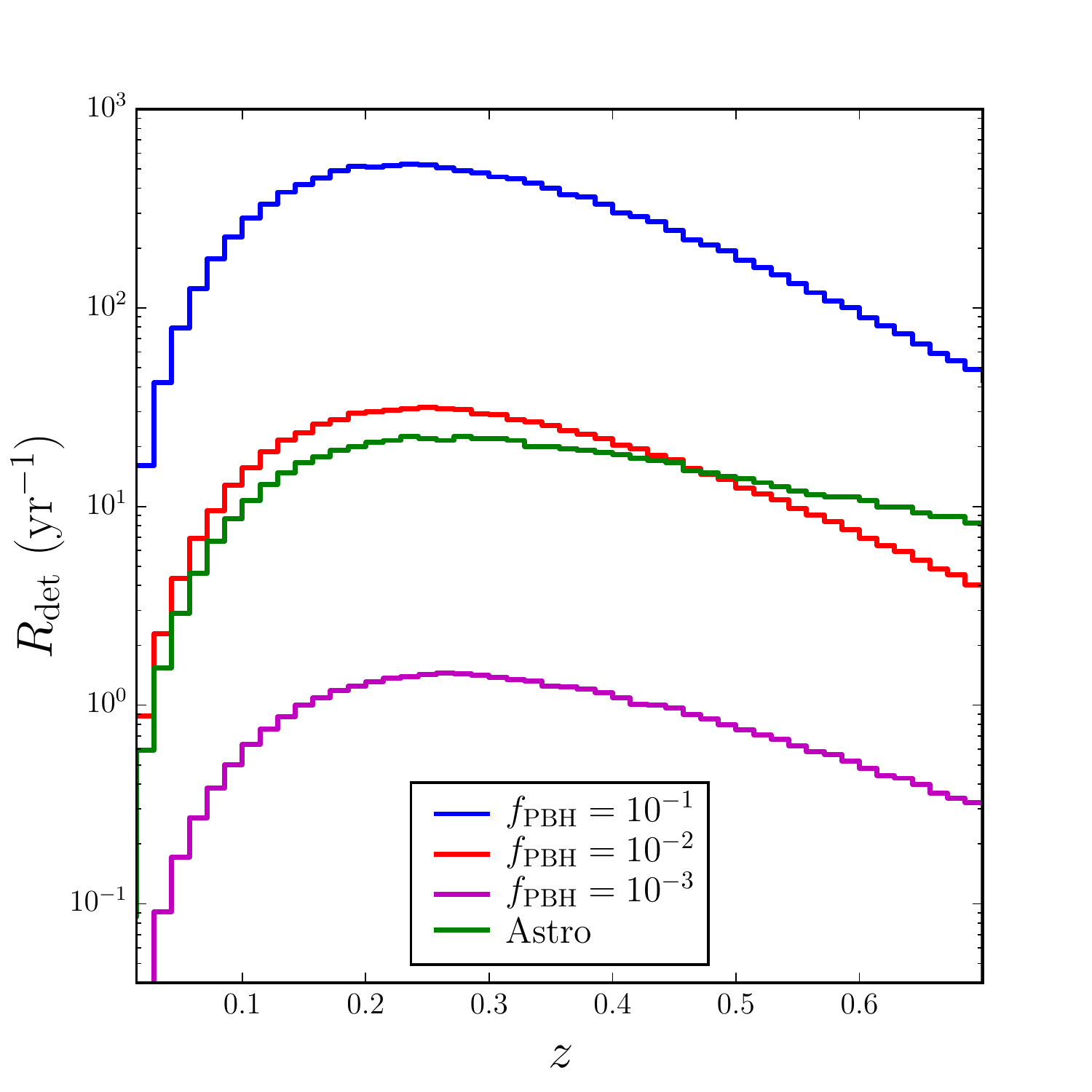}
\vspace*{-1em}
\includegraphics[width=0.49\textwidth]{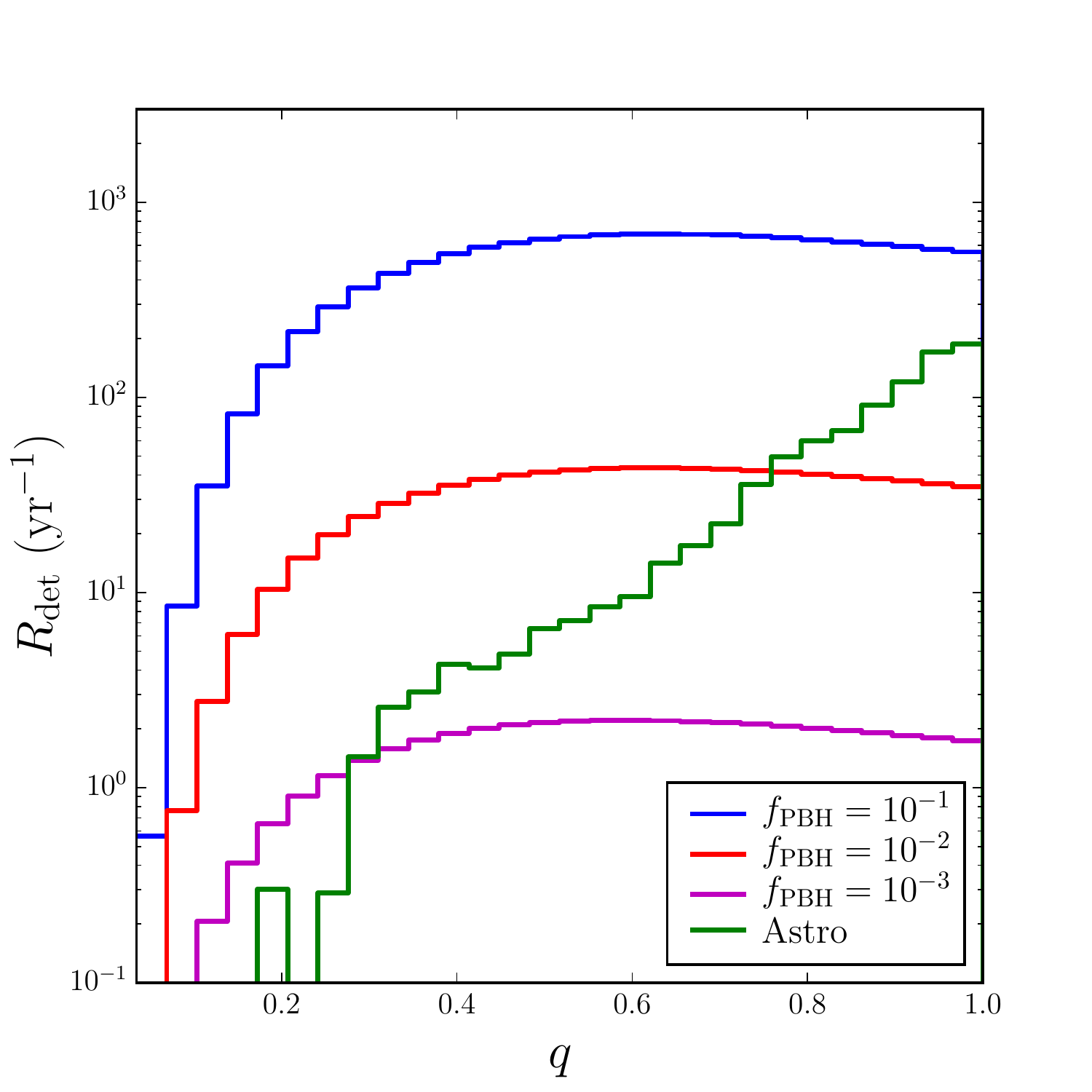}
\includegraphics[width=0.49\textwidth]{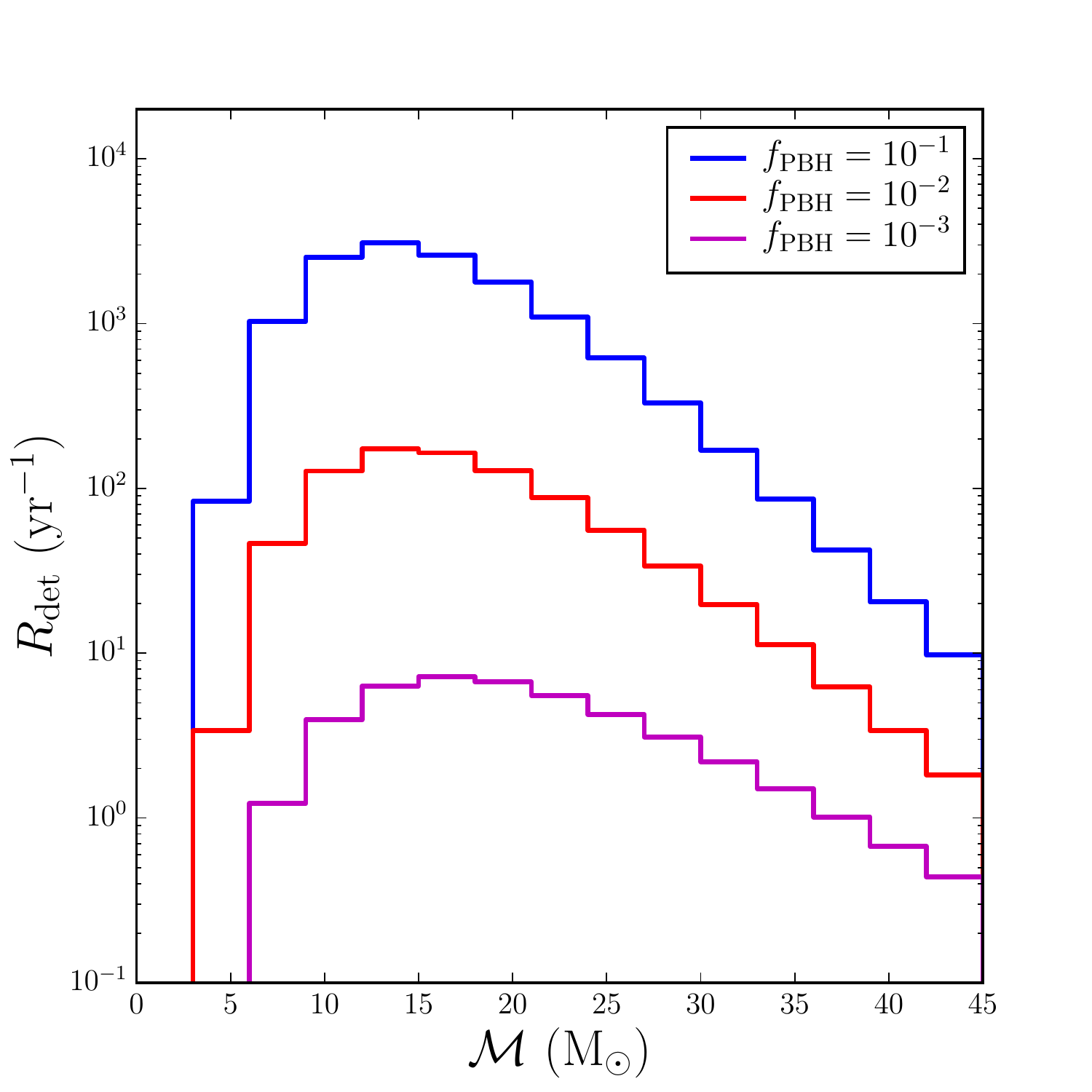}
\caption{Merger rate distributions in total mass $M$, redshift $z$, mass ratio $q$, and chirp mass $\mathcal{M}$ for a lognormal mass distribution. The distributions for astrophysical black holes from \cite{Gerosa_2019} are shown in green for the first three plots.}
\label{fig:Design-Lognormal-1D-s06}
\end{figure}

Distributions of the detectable merger rate against four observables have been generated by Monte Carlo integration of eq.~\eqref{eq:Rdet}. These distributions are shown in fig.~\ref{fig:Design-Lognormal-1D-s06}. The four observables are total mass $M$, redshift $z$, mass ratio $q$, and chirp mass $\mathcal{M}$. The first three of these observables are chosen for comparison with detectable merger rate distributions recently determined by Gerosa et al \cite{Gerosa_2019} for astrophysical black holes, using the \textsc{Startrack} code for stellar evolution and the \textsc{Precession} code to add spins, with the same detectability process described above \cite{Gerosa_2019}. The fourth observable, $\mathcal{M}$, is chosen because this is the observable best constrained by LIGO for lower mass mergers. There is no astrophysical curve publicly available at the time of writing for this observable, as it was not calculated in \cite{Gerosa_2019}. Three values of $f_\PBH$ are shown, with the largest being $10^{-1}$. Above this value, the merger rate calculation is unreliable due to the high probability of the binary being disrupted between formation and merger \cite{Raidal_2019}.

The distributions for the total mass $M$ and the chirp mass $\mathcal{M}$ follow the component mass distribution shown in fig.~\ref{fig:Mass-Function-Lognormal-s06} closely, with the peaks lying where one would expect by taking the peak of the individual lognormal mass distribution ($m_c=20\ \Msun$, $\sigma=0.6$) and calculating the resulting values of $M$ and $\mathcal{M}$. The distribution for the mass ratio $q$ seems to favour $q\approx0.6$, but is fairly flat from $q=1$ down to $q\approx0.4$. After this, there is a steep drop-off, due to a combination of the width and detectability factors. It can be seen that the major dependence on $f_\PBH$ is simply a global multiplier, scaling the curves up or down. However, there are other dependencies, such as the $f_\PBH=10^{-3}$ curve being flatter at high observable values than the curves with higher $f_\PBH$ values for the total mass $M$ and chirp mass $\mathcal{M}$.

\vspace{-1em}

As can be seen in fig.~\ref{fig:Design-Lognormal-1D-s06}, the distributions for total mass $M$ and redshift $z$ seem to match the astrophysical distribution quite closely at low redshift, but the astrophysical rate drops for redshifts above $z\approx1.5$ as it follows the stellar formation rate, while the primordial rate continues growing and becomes the dominant merger source, as can be seen in fig.~10 of \cite{Raidal_2019} (published version). All the rates tend to zero as the redshift tends to zero, due to the volume factor in eq.~\eqref{eq:Rdet}. In contrast to the above two cases, the distribution for the mass ratio $q$ shows a clear difference between the astrophysical distribution and any of the primordial curves. This could therefore be a useful observable for distinguishing between mergers of astrophysical and primordial origin. As can be seen, the astrophysical distribution tends to favour higher $q$ values, which is to be expected considering the formation mechanisms \cite{Marchant_2016,Rodriguez_2016,O_Leary_2016,Kovetz_2017}. The current LIGO data also favour high mass ratios \cite{Fishbach_2019}, and with future data, the allowed width and shape of the PBH mass distribution could be seriously constrained.

\begin{figure}[H]
\centering
\includegraphics[width=0.49\textwidth]{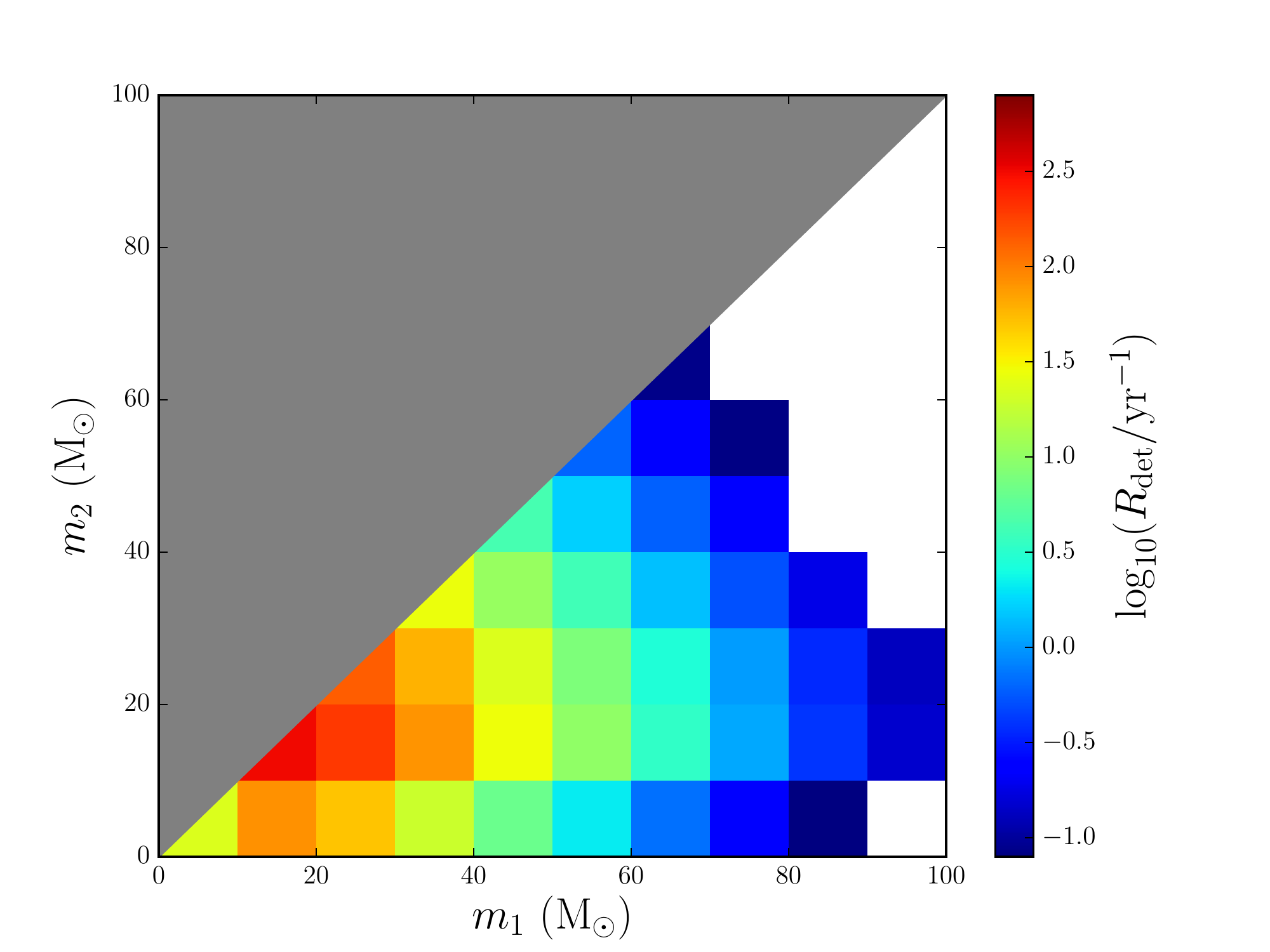}
\includegraphics[width=0.49\textwidth]{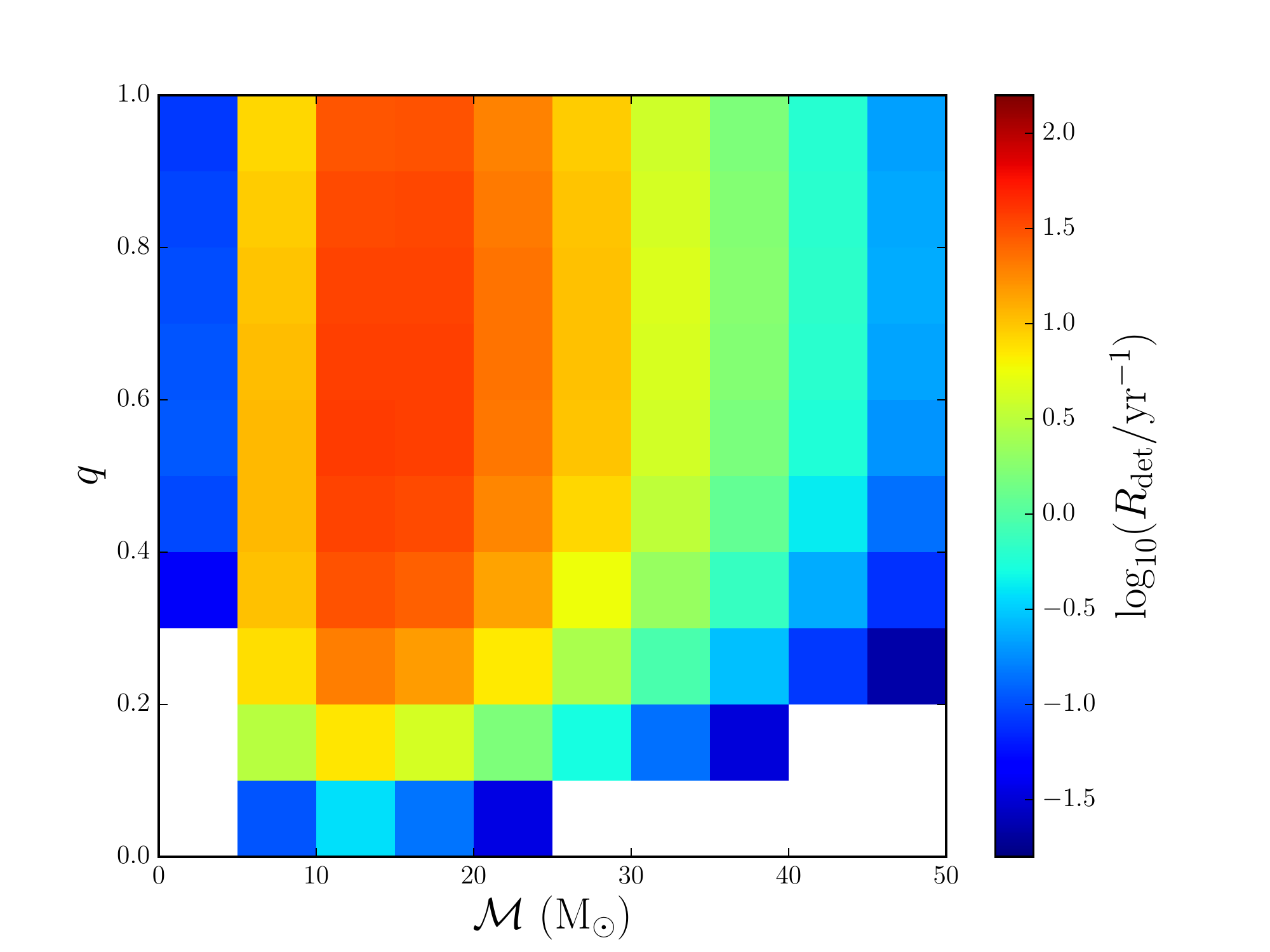}
\includegraphics[width=0.49\textwidth]{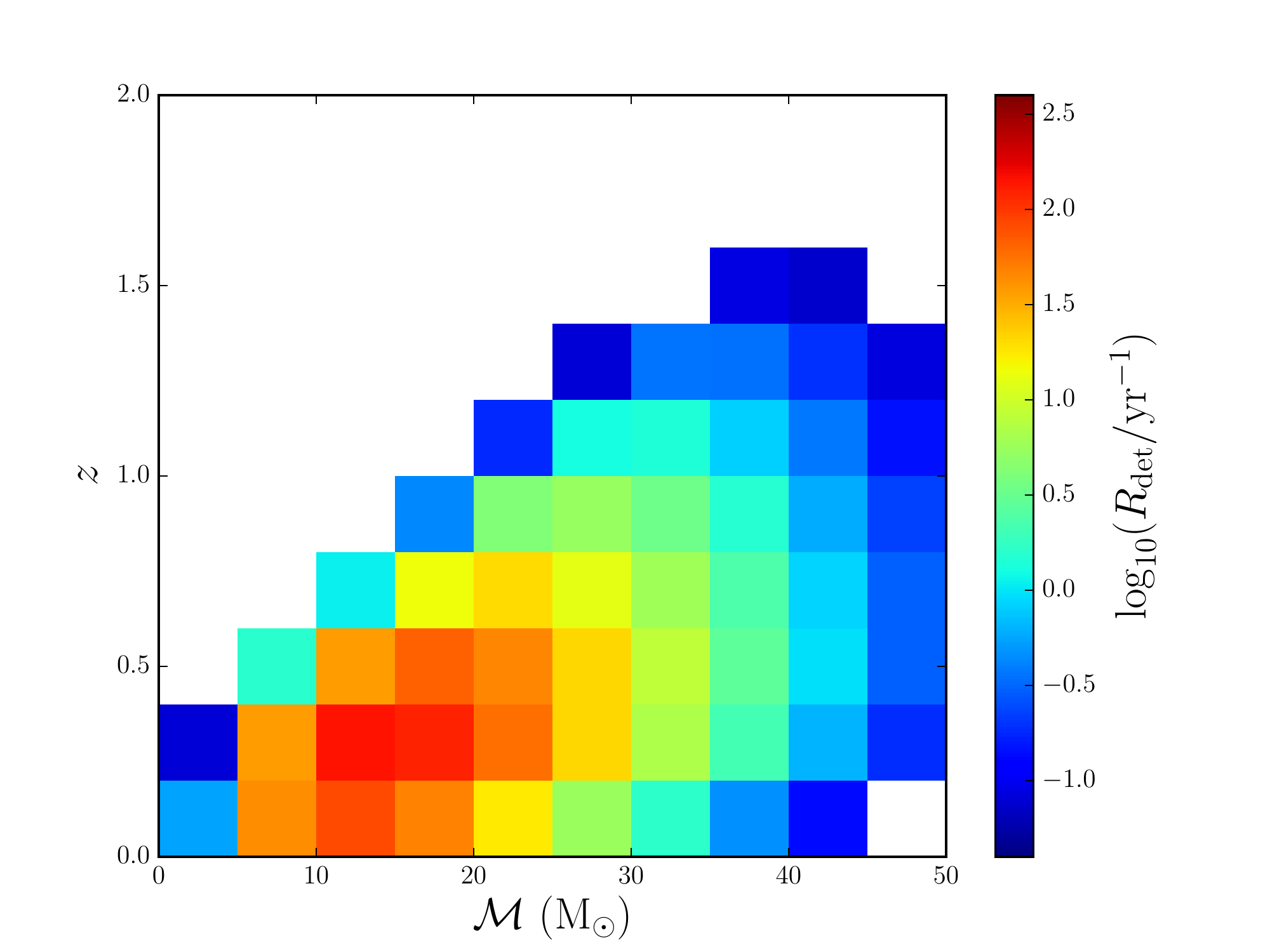}
\includegraphics[width=0.49\textwidth]{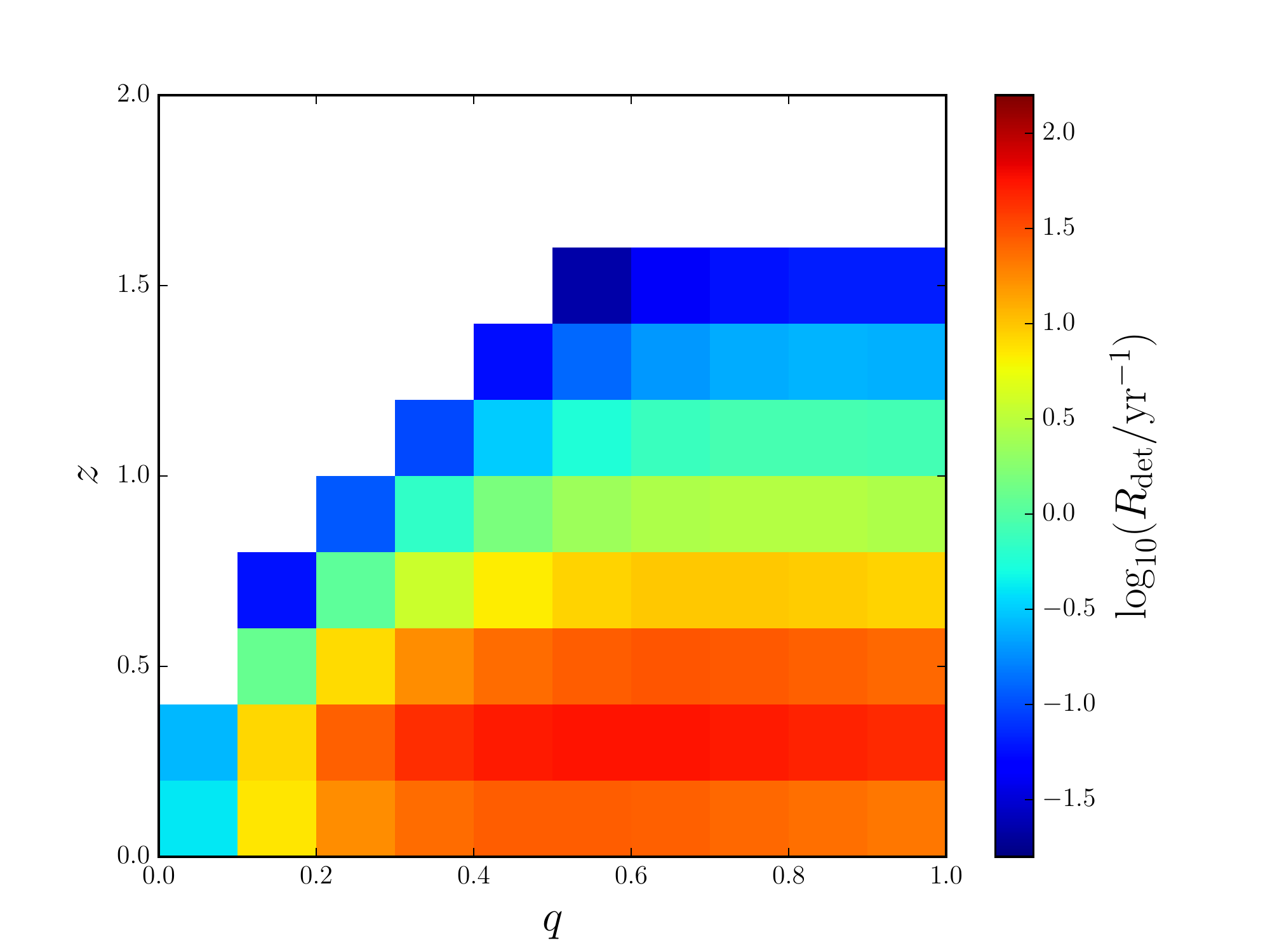}
\caption{2D Merger rate distributions in individual masses, mass ratio $q$, chirp mass $\mathcal{M}$ and redshift $z$ for a lognormal distribution. All plots have $f_\PBH=10^{-2}$. The white area corresponds to no significant merger rate, and the grey triangle indicates the case $m_2<m_1$.}
\label{fig:Design-Lognormal-2D-s06}
\end{figure}

It is also interesting to know how the merger rate is distributed across multiple observables. 2D plots of the merger rate against four sets of parameters were created. These are the component masses ($m_1$, $m_2$), and the three combinations of the redshift $z$, the chirp mass $\mathcal{M}$, and the mass ratio $q$. These distributions are shown in fig.~\ref{fig:Design-Lognormal-2D-s06}. The distribution in ($m_1$, $m_2$) exhibits the expected behaviour following the mass distribution, with a peak between $10\ \Msun$ and $20\ \Msun$, and then a drop off to higher masses. The distribution in ($q$, $\mathcal{M}$) shows that, away from the chirp mass peak, there is no detectable merger rate for low $q$ values.

The distributions involving redshift show that nothing can be detected past $z\approx1.6$, even at design sensitivity. The ($z$, $\mathcal{M}$) distribution peaks at the same chirp mass value as the other 1D and 2D distributions, but drops off very rapidly with redshift. Also, it can be seen that the larger chirp mass values can be detected out to much higher redshifts, due to the larger amplitude of the gravitational wave produced. The same is true of the higher $q$ values in the ($z$, $q$) distribution.

\begin{figure}[H]
\centering
\includegraphics[width=0.45\textwidth]{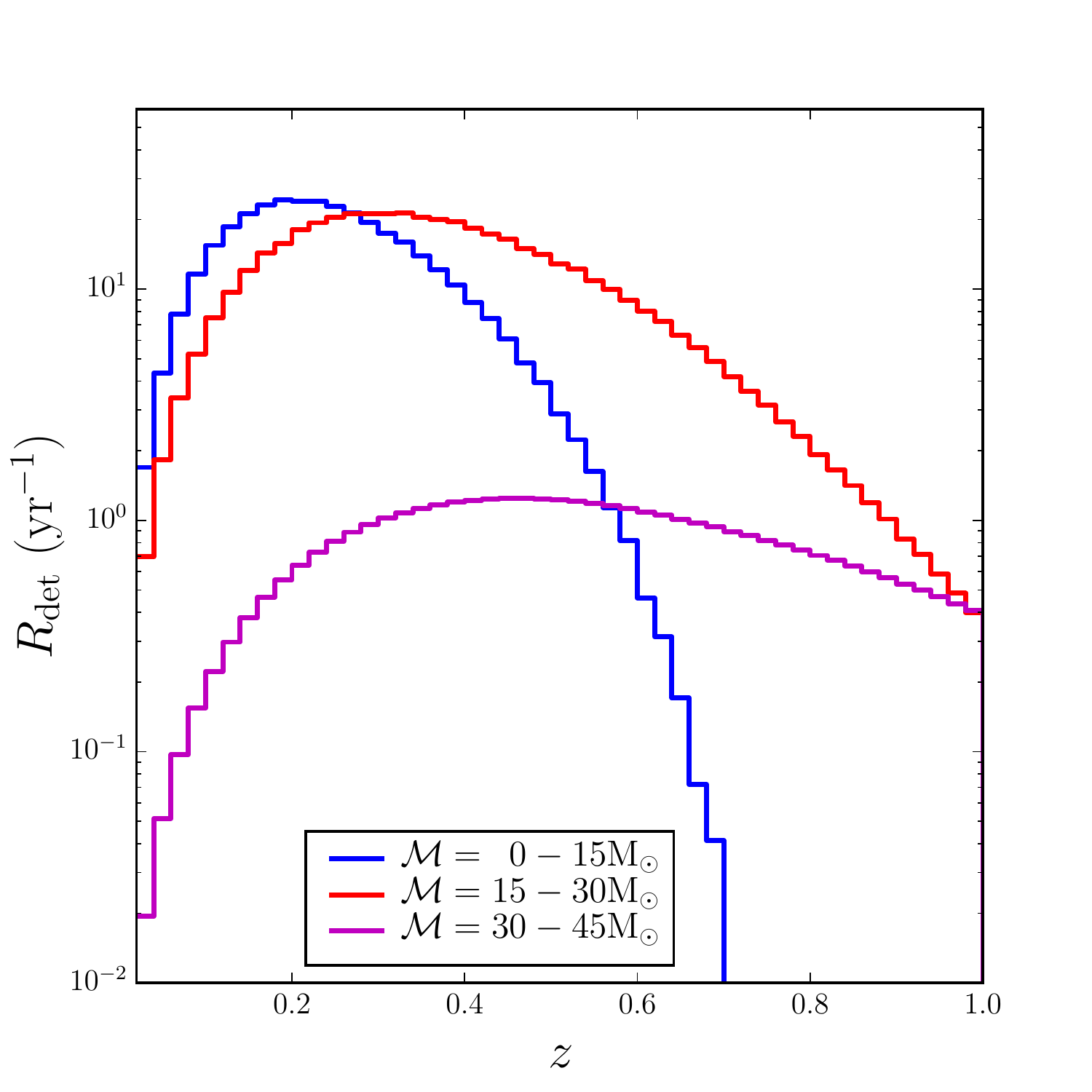}
\includegraphics[width=0.45\textwidth]{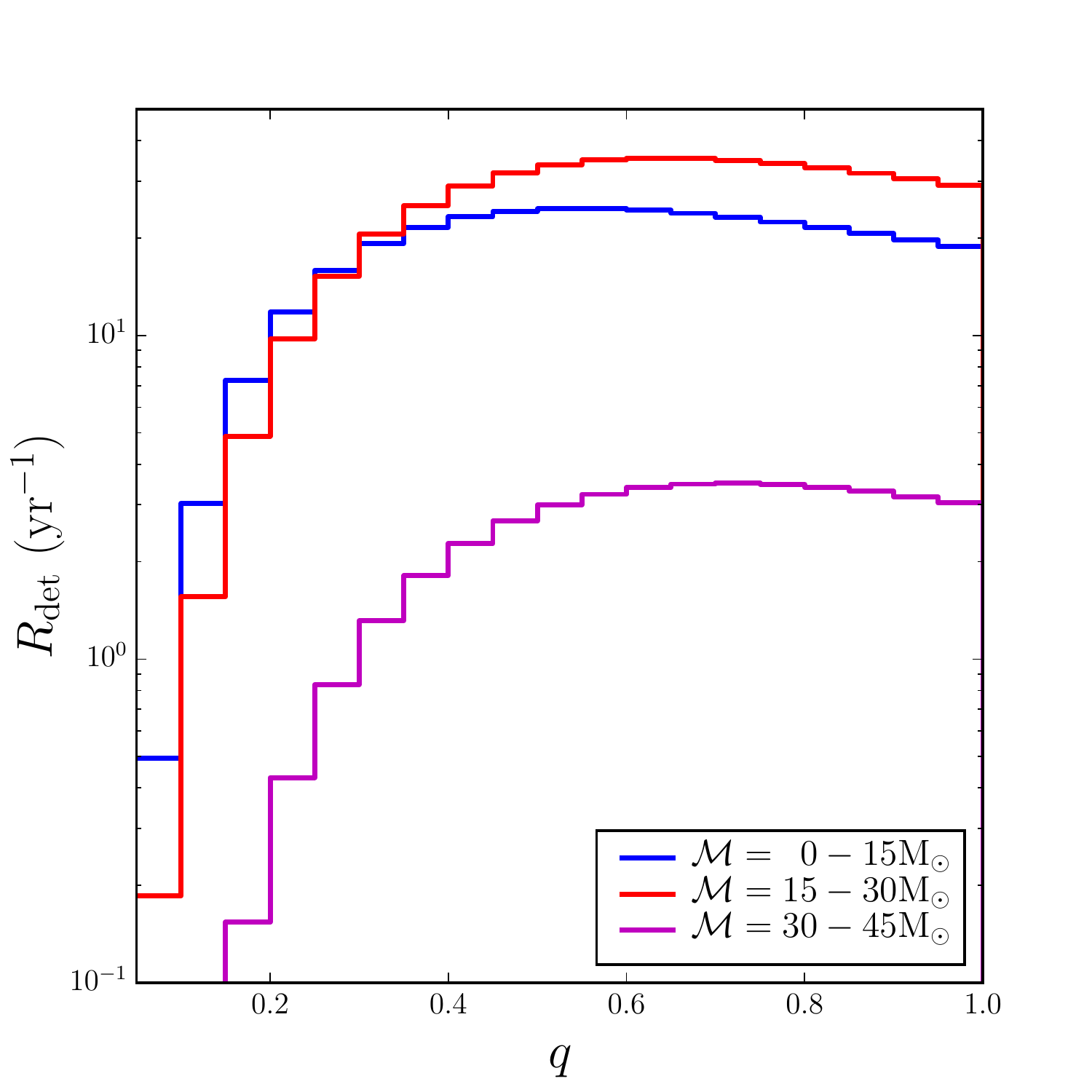}
\caption{Merger rate distributions in redshift $z$ and mass ratio $q$ for a lognormal distribution, binned by chirp mass $\mathcal{M}$. Both plots have $f_\PBH=10^{-2}$.}
\label{fig:Design-Lognormal-1D-s06-Mc-Binned}
\end{figure}

For the two distributions involving the chirp mass $\mathcal{M}$, we can take vertical slices and produce 1D distributions over a given chirp mass range, to better demonstrate the sensitivity of the merger rate to the chirp mass. These are shown in fig.~\ref{fig:Design-Lognormal-1D-s06-Mc-Binned}. It can be seen for both of the observables that the low chirp mass curve begins above the medium chirp mass curve, but then drops below as the value on the respective horizontal axes is increased. For the redshift distribution, this is because the low chirp mass binaries have a low detection probability, and so have a very limited $z$-range.

1D and 2D distributions of the detectable merger rate have been generated for the lognormal mass distribution in eq.~\eqref{eq:PsiLognormal}, with the parameters $m_c=20\ \Msun$ and $\sigma=0.6$ given in \cite{Raidal_2019}. However, while this is a plausible form for the mass function and its parameters, it is not the only option. Therefore, it is interesting to consider how these merger rate distributions change for other PBH mass distributions.

\section{Comparison of different PBH mass distributions}
\label{sec:ComparisonOfMassDistributions}
\subsection{Lognormal width parameter $\sigma$}
\begin{figure}[H]
\centering
\includegraphics[width=0.49\textwidth]{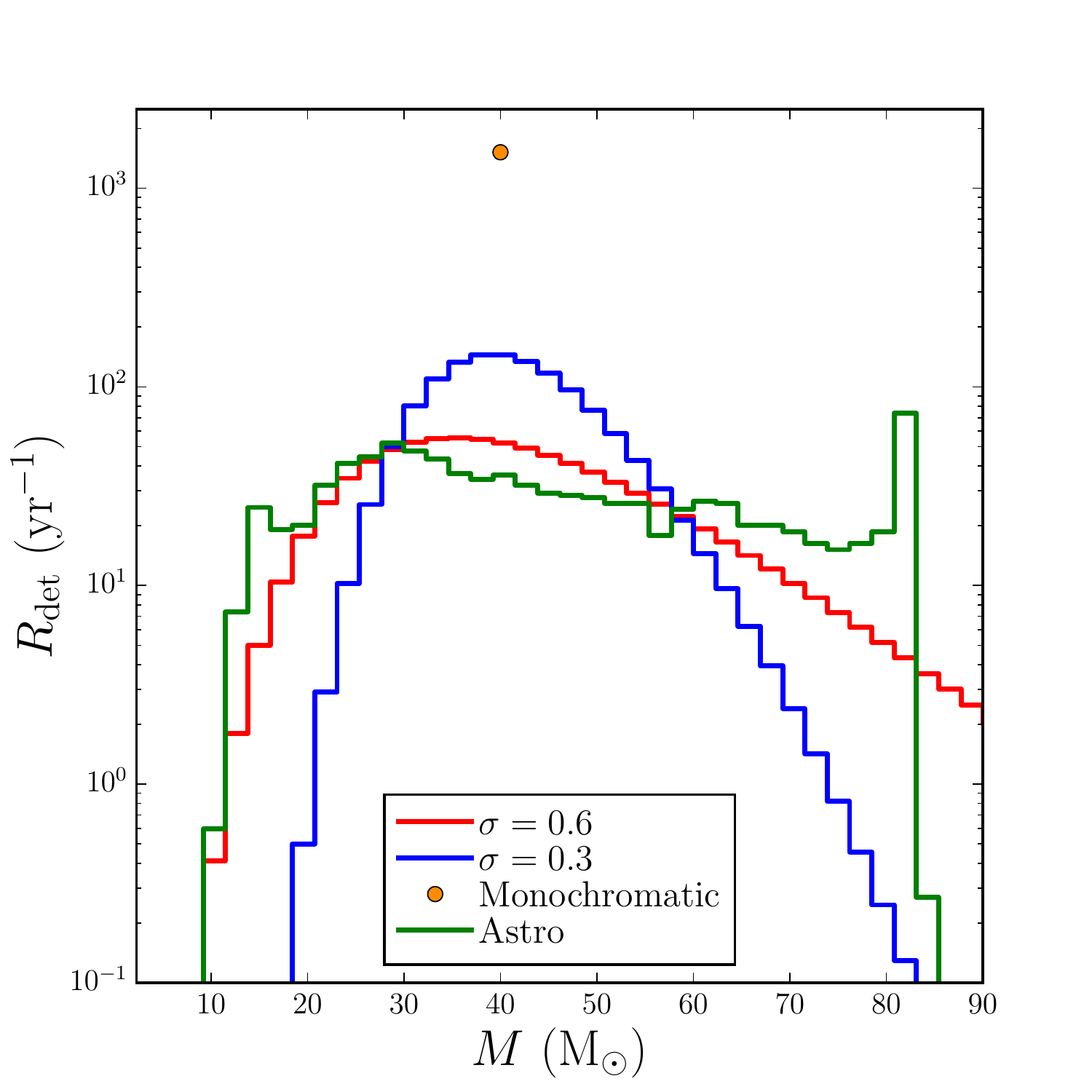}
\includegraphics[width=0.49\textwidth]{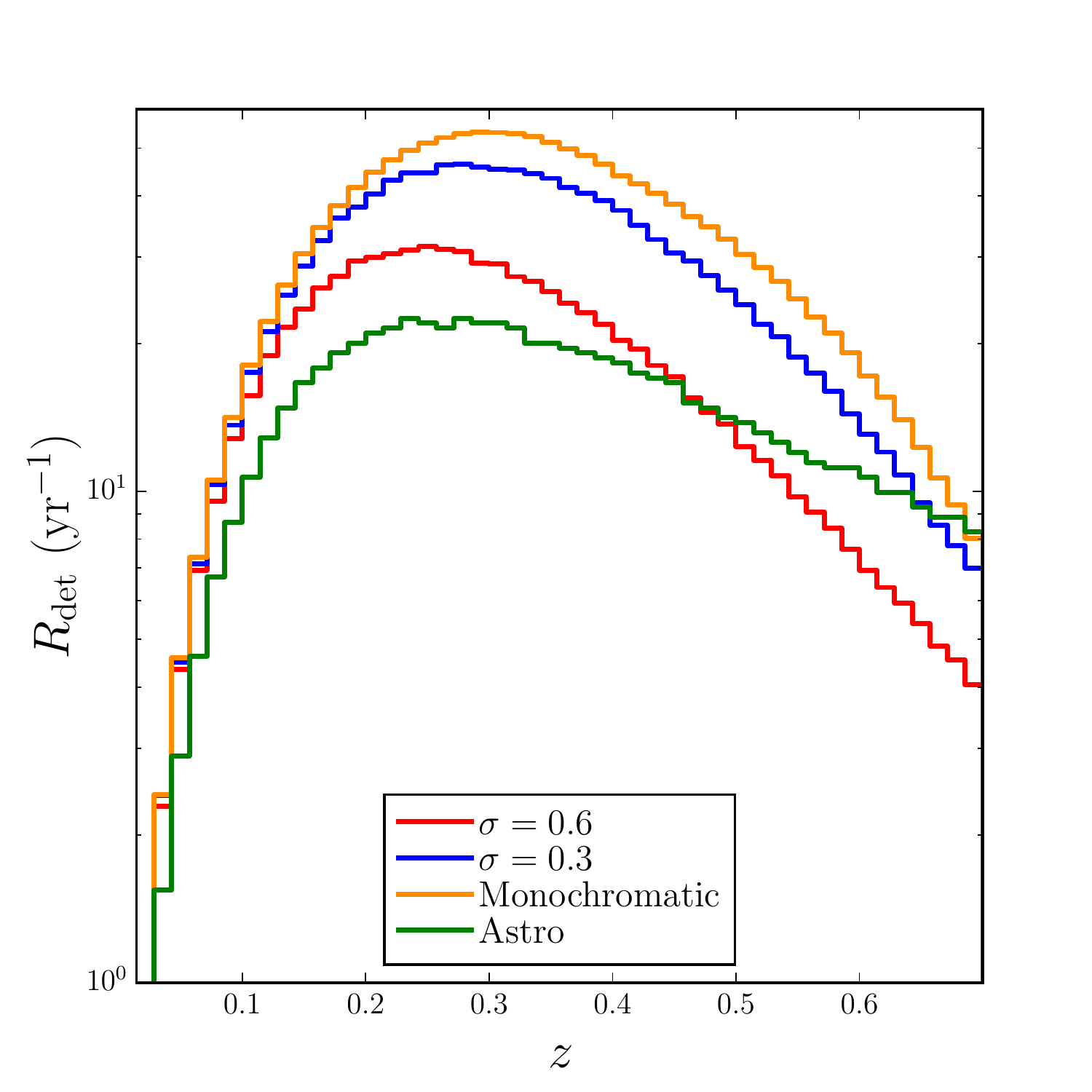}
\includegraphics[width=0.49\textwidth]{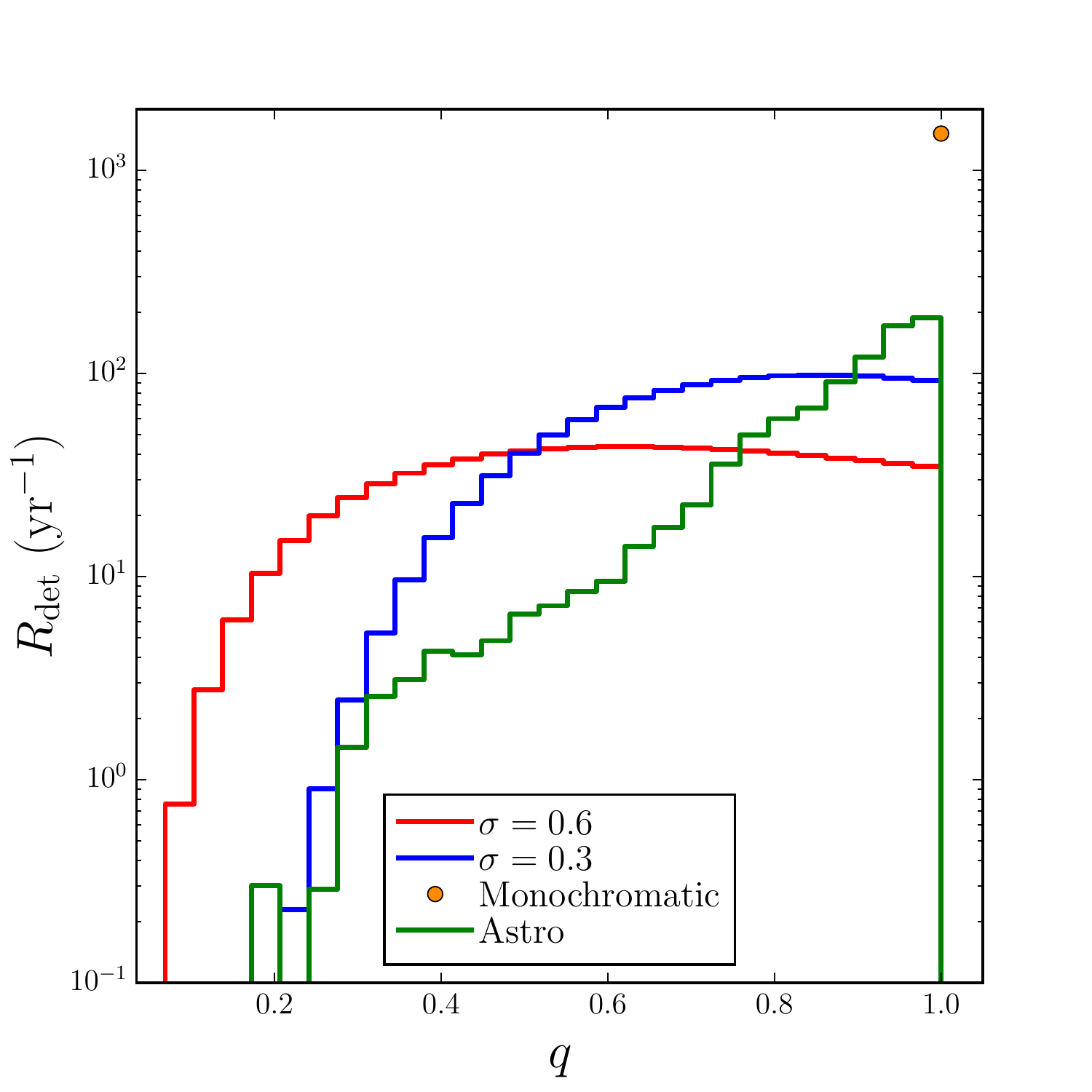}
\includegraphics[width=0.49\textwidth]{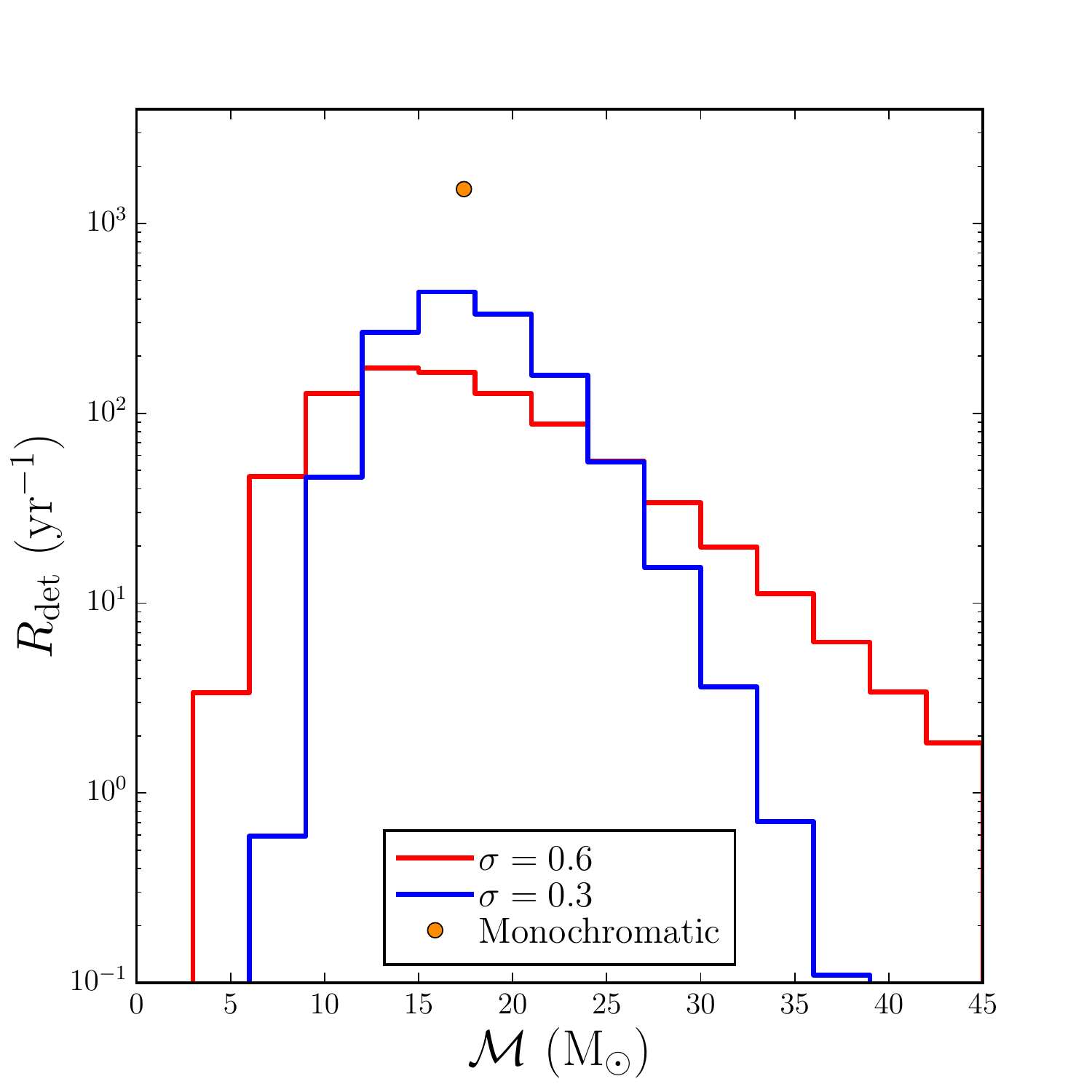}
\caption{Comparison of merger rate distributions in total mass $M$, redshift $z$, mass ratio $q$ and chirp mass $\mathcal{M}$ for different widths of mass distribution. All plots have $f_\PBH=10^{-2}$.}
\label{fig:Design-Lognormal-1D-width}
\end{figure}

While the best fit values in \cite{Raidal_2019} for the lognormal mass distribution are $m_c=20\ \Msun$ and $\sigma=0.6$, the full angular dependence for the detectability was not incorporated, and so it is of interest to consider the merger rate distributions for other widths. The method above was carried out for a second lognormal distribution with the same $m_c$ but with $\sigma=0.3$, and also for a monochromatic distribution $\delta(m-m_c)$ which is the limit of the lognormal distribution as $\sigma\to0$. Both of these additional distributions had the same $m_c=20\ \Msun$. For typical PBH formation scenarios, critical collapse imposes a minimum width on the mass distribution, and so a monochromatic distribution is not realistic. However, the monochromatic distribution is still useful for comparison and demonstration of the properties affecting the merger rate. A comparison of the distributions is shown in fig.~\ref{fig:Design-Lognormal-1D-width}.

For $M$, $q$, and $\mathcal{M}$, the monochromatic distribution is represented by a single point, since there is only one value of each observable it can take. In $M$ and $\mathcal{M}$, it can be seen that reducing the width of the mass distribution leads to a reduction of the width of the merger rate distribution, as expected. It also leads to an enhancement of the peak. For $q$, the width reduction suppresses low $q$ values, also to be expected, since a narrower distribution has a smaller difference between the highest and lowest probable masses. For $z$, the narrower mass distribution leads to an enhancement across the whole range because, for a given total mass, equal mass mergers are easier to detect. 2D distributions for the lognormal distribution with $\sigma=0.3$ are shown in fig.~\ref{fig:Design-Lognormal-2D-s03}.

For the monochromatic distribution, there are some further parameters that can be considered due to the simplicity of the function. These are $f_\PBH$, the mass of the monochromatic distribution $m_c$, and the rescaled deviation of matter density perturbations at the time of binary formation $\sigma_\M$, given just after eq.~2.24 in \cite{Raidal_2019} (published version). The value of $\sigma_\M$ is usually taken as 0.006 on scales relating to black hole masses of order 1-10$^3\ \Msun$, corresponding to the deviation of density perturbations $\sigma_\mathrm{eq}=0.005$ in \cite{AliHaimoud_2017} and \cite{Eroshenko_2018}. However, this value is found by extrapolating the power spectrum amplitude and spectral index measured from the CMB. Since PBH formation typically requires some type of peak in the power spectrum on relevant scales, this could quite dramatically change the value of $\sigma_\M$, or even give it a strong dependence on the black hole mass, and so this value is very uncertain \cite{AliHaimoud_2017,Raidal_2019}. A detailed study on the angular momentum sources excluding PBHs outside the binary, including the variability of $\sigma_\M$, was carried out by Garriga et al \cite{Garriga_2019}.

\begin{figure}[H]
\centering
\includegraphics[width=0.49\textwidth]{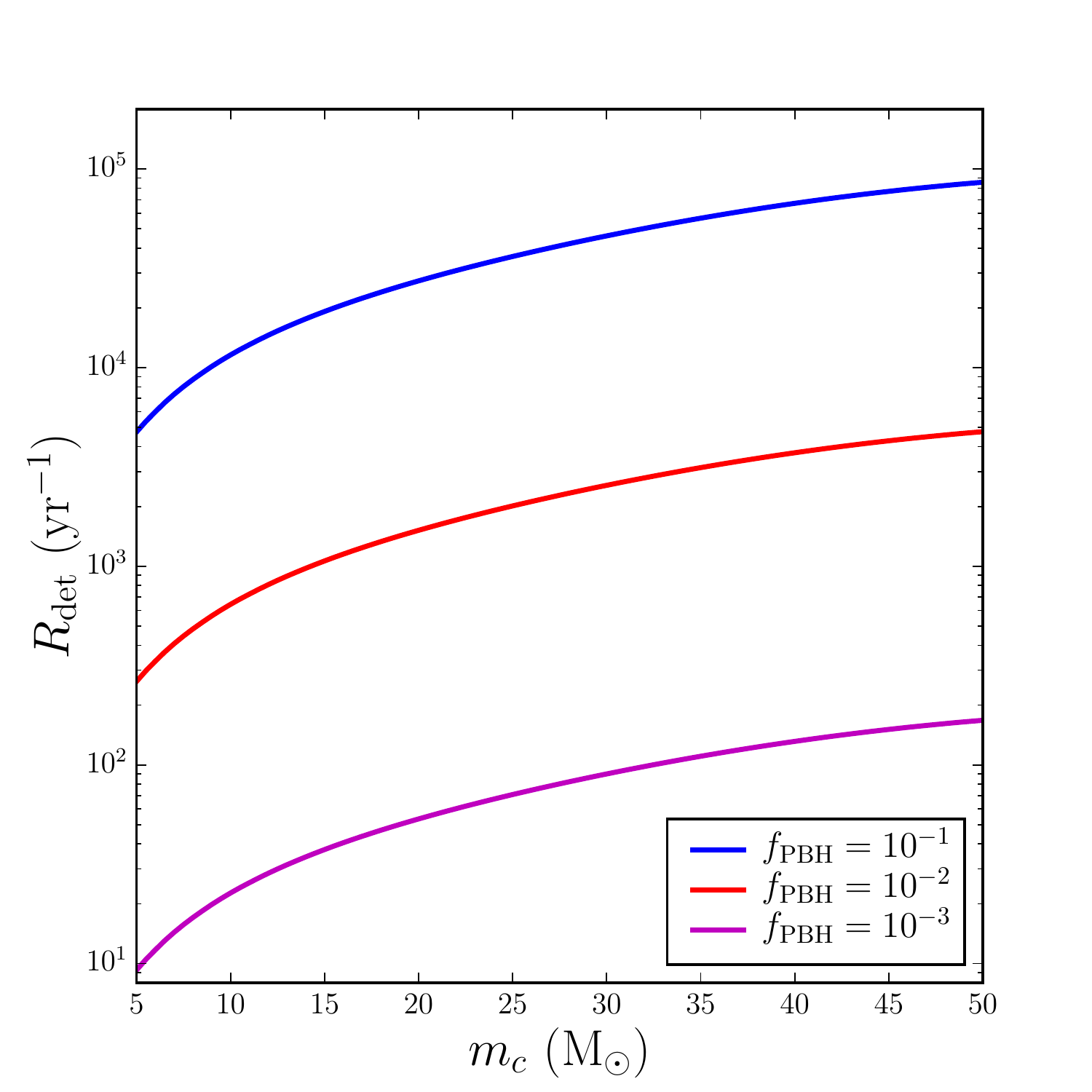}
\includegraphics[width=0.49\textwidth]{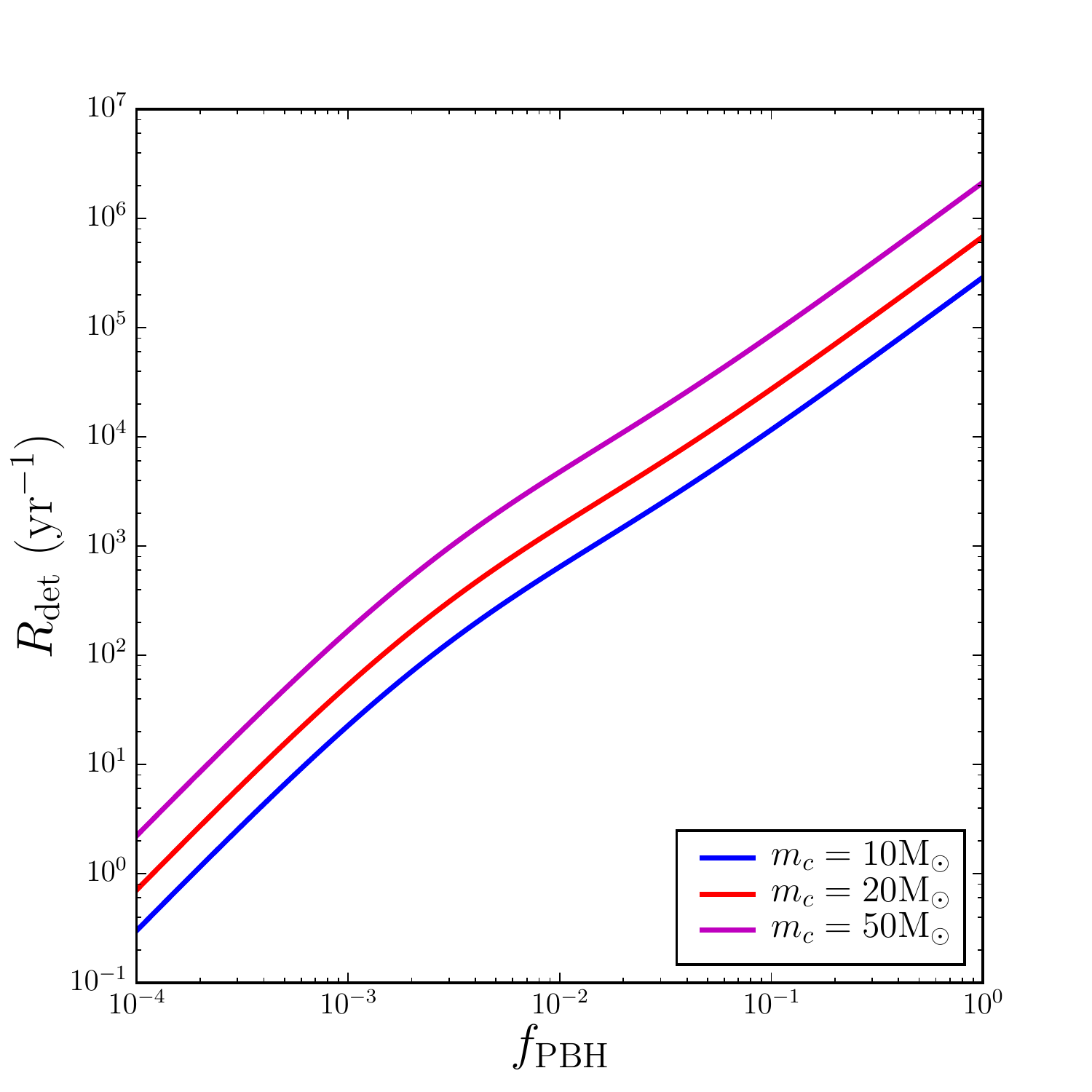}
\includegraphics[width=0.49\textwidth]{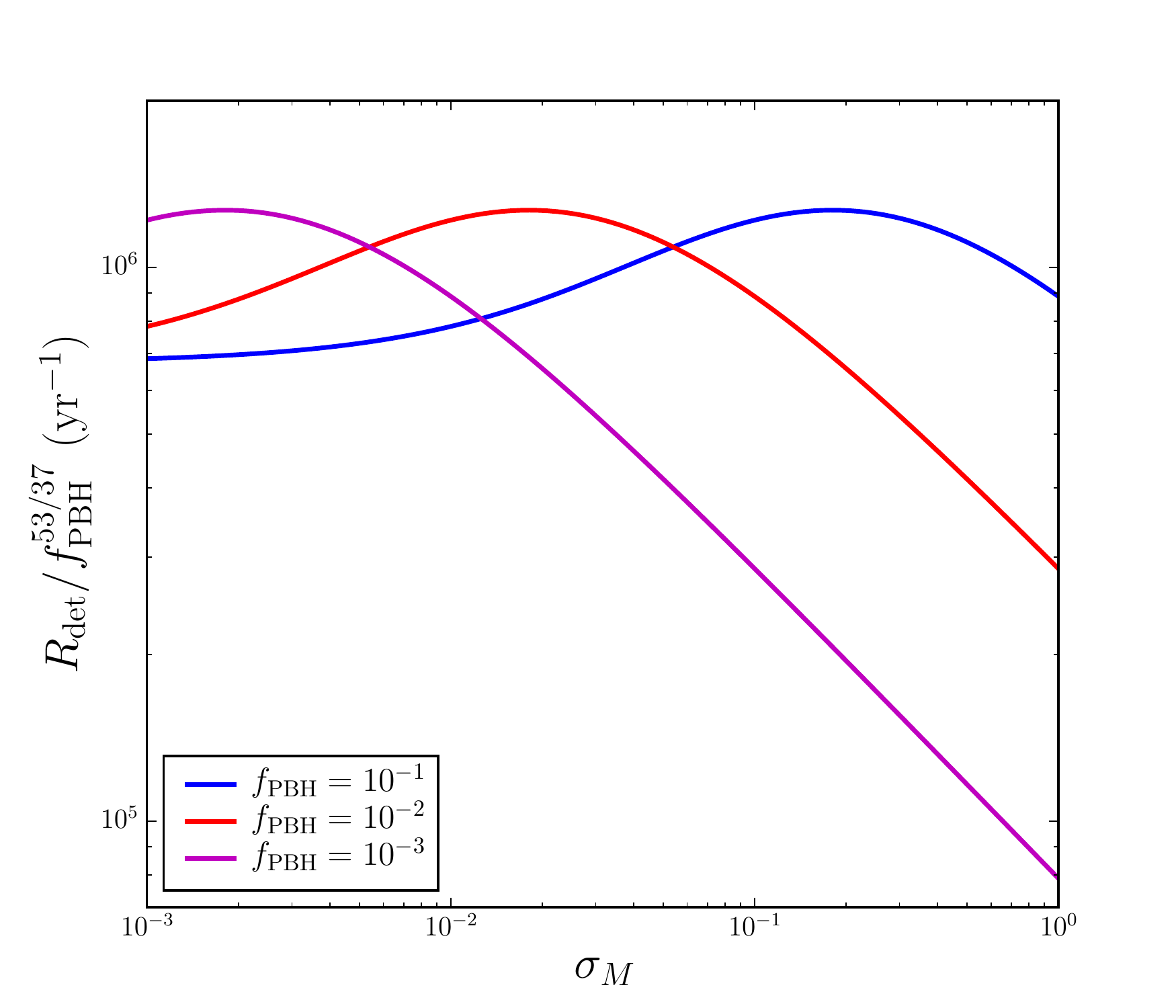}
\caption{Merger rate dependence on the BH mass $m_c$, the fraction of dark matter in PBHs $f_\PBH$ and the rescaled variance of matter density perturbations $\sigma_\M$ for a monochromatic mass distribution. The top two plots have $\sigma_\M=0.006$ and the bottom plot has $m_c=20\ \Msun$. The plots are generated using the method in \cite{Raidal_2019}, and would change with the additional effects considered in \cite{Garriga_2019}.}
\label{fig:Design-Monochromatic}
\end{figure}

Figure \ref{fig:Design-Monochromatic} shows the dependence of the detectable merger rate for a monochromatic distribution on the three parameters $m_c$, $f_\PBH$, and $\sigma_\M$. The dependence on $f_\PBH$ looks very similar to the plot of the intrinsic merger rate on this parameter in fig.~3 of \cite{Raidal_2019} (published version). This is expected, since the detectability and comoving volume factors do not introduce any additional dependence on $f_\PBH$, so the only difference is the monochromatic vs. lognormal mass distribution.

In the bottom panel of fig.~\ref{fig:Design-Monochromatic}, the detectable merger rate has been normalised by removing the global dependence on $f_\PBH$, leaving only the $f_\PBH$ dependence in the suppression factor $S$. This is to better highlight the relationship between the merger rate and the two parameters $\sigma_\M$ and $f_\PBH$, which have a degeneracy in $S$. The curves follow the same shape as each other, but with the peak in a different place that is determined by the relationship between $f_\PBH$ and $\sigma_\M$. To the right of the peak, the torque on the binary is dominated by the matter perturbations, leading to binaries that have not yet merged. To the left of the peak, there is insufficient torquing from matter perturbations, and the binaries merged at redshifts too high to be detected by LIGO. However, on the left of the peak, the merger rate tends to a constant value, determined by the torquing generated by other PBHs, which is fixed by the value of $f_\PBH$. It can also be seen that the normalised merger rate varies by up to an order of magnitude for different values of $\sigma_\M$. This is enough to shift the LIGO constraints yielding different optimised mass distribution parameters, and so further study of the degeneracy of observables with the currently unknown value of $\sigma_\M$ is required.

\subsection{Power-law distribution}
\label{ssec:Power-law}
Another commonly considered distribution for the masses of primordial (and astrophysical) black holes is a power-law ($\propto m^{-\alpha}$). The parameters for this model are the power to which the mass is raised and the lower/upper mass cutoffs if applicable. A scale-invariant primordial power spectrum generates $\alpha=3/2$, due to the enhancement of the PBH energy density relative to the background radiation energy density after they have formed \cite{Carr_2016,Carr_2017}. With both a lower and upper mass cutoff\footnote{In practice, unless the mass function is close to scale invariant, only one cutoff is important.}, the normalised mass distribution is
\begin{align}
\psi(m) = \left(\alpha - 1\right)\left[m_\text{min}^{-(\alpha-1)} - m_\text{max}^{-(\alpha-1)}\right]^{-1} m^{-\alpha}.
\end{align}
However, if the minimum mass is chosen to be too small, the suppression factor $S$ in the merger rate calculation described above heavily suppresses the result. This is probably because a power-law mass distribution heavily favours the lighter end of the mass spectrum, meaning there are far more of these than there are heavier black holes. Physically, if there is a large population of lighter black holes and a smaller population of heavier black holes, then it will still be the heavier black holes that will merge as the lighter ones will not contribute significantly to the gravitational force that causes a binary to form. However, the calculation described in \cite{Raidal_2019}, while being very thorough, does not capture this effect because it assumes that a PBH will form a binary with its nearest neighbour, rather than the neighbour contributing the largest gravitational force. In the equations, this manifests itself as a strong dependence on the average mass $\bar{m}$, and a suppression of the resulting merger rate.

To determine if this effect is important, and if so how problematic it is, the merger rates of three power-law mass distributions were calculated. Each mass distribution had the same values of $\alpha=3/2$ and $m_\text{max}=100\ \Msun$, but had different $m_\text{min}$ values ($5\ \Msun$, $1\ \Msun$, and $0.1\ \Msun$). For each distribution, the value of $f_\PBH$ was chosen such that the number density of PBHs with masses in the range $1$-$100\ \Msun$ was the same. Physical intuition would suggest that the total number of merger events with masses in this range would be similar for all three distributions. However, the intrinsic merger rate calculated using eq.~\eqref{eq:TotalIntrinsicMergerRate} varied by two orders of magnitude between the distributions with minimum masses of $5\ \Msun$ and $1\ \Msun$, and by 30 orders of magnitude between $5\ \Msun$ and $0.1\ \Msun$. This is clearly a very significant problem that prevents the study of very broad mass distributions.

While it remains unclear how broad the mass distribution can be before this effect starts to become important, it is still desirable to compare the lognormal distribution with a power-law distribution. Therefore, the analysis was rerun with two power-law distributions. Both had $\alpha=3/2$ and $m_\text{max}=100\ \Msun$. The minimum mass for the two distributions was $m_\text{min} = 5\ \Msun$ and $10\ \Msun$ respectively. However, due to the problem described above, it is not clear if these results are reliable. The 1D merger rate distributions for these mass distribution can be seen in fig.~\ref{fig:Design-Combined} in the appendix, and the 2D distributions are shown in figs.~\ref{fig:Design-PL-2D-mMin5}~and~\ref{fig:Design-PL-2D-mMin10}.

Another mass distributions of interest is that of PBHs generated at the QCD phase transition. For a scale invariant power spectrum, this formation gives a mass distribution like a power-law with $\alpha=3/2$, but with an excess at around $1\ \Msun$, caused by the reduction in pressure at the QCD transition \cite{Byrnes_2018}. However, the relevant range of this mass distribution is extremely large (four orders of magnitude), and so it is affected by the broadness problem discussed above. This mass distribution remains of interest due to its physical motivation, and should be studied once a reliable calculation of the merger rate for broad mass distributions is available.

\section{Current LIGO data and constraints}
\label{sec:CurrentLIGOData}

\begin{figure}
\centering
\includegraphics[width=0.49\textwidth]{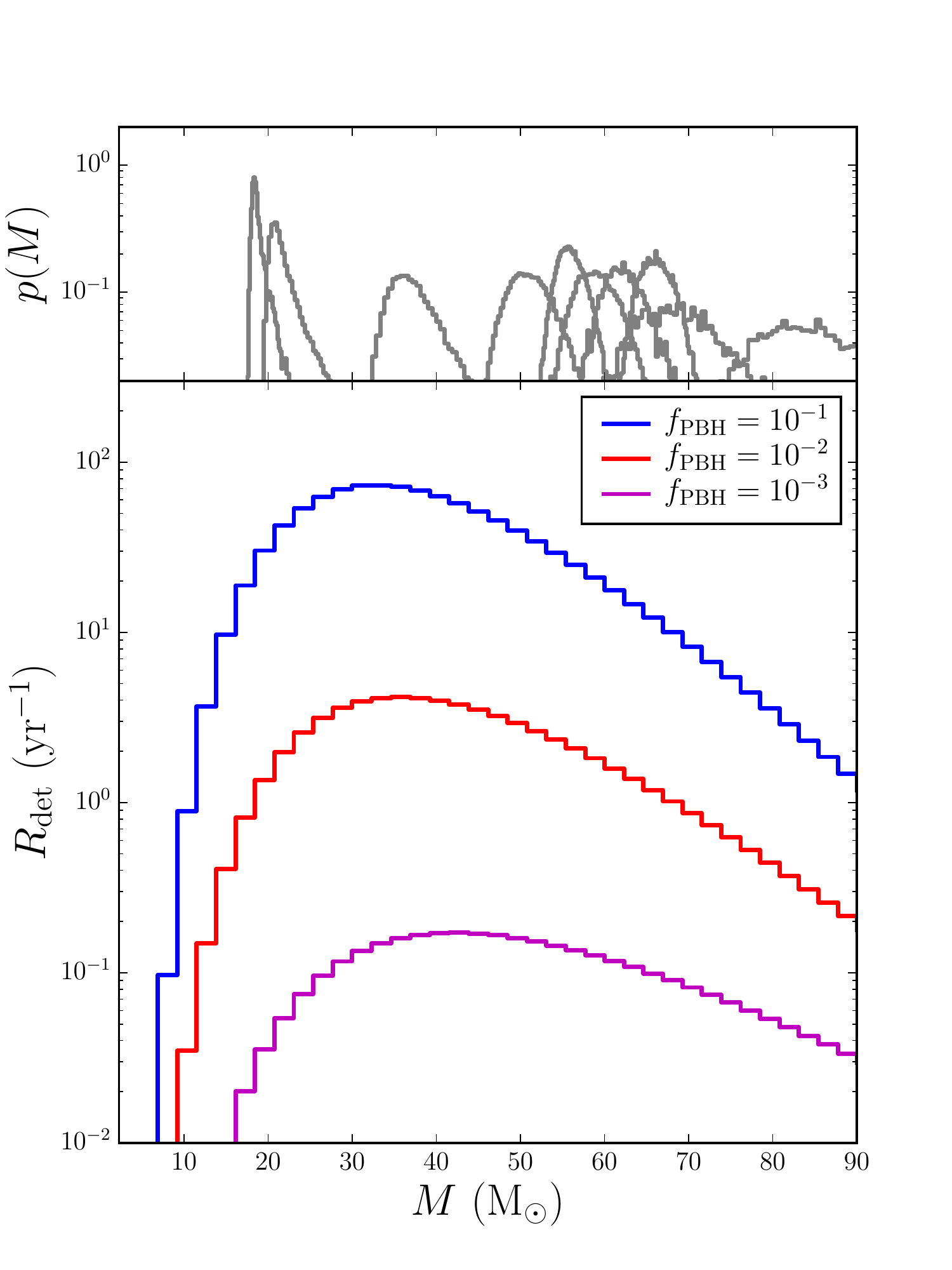}
\includegraphics[width=0.49\textwidth]{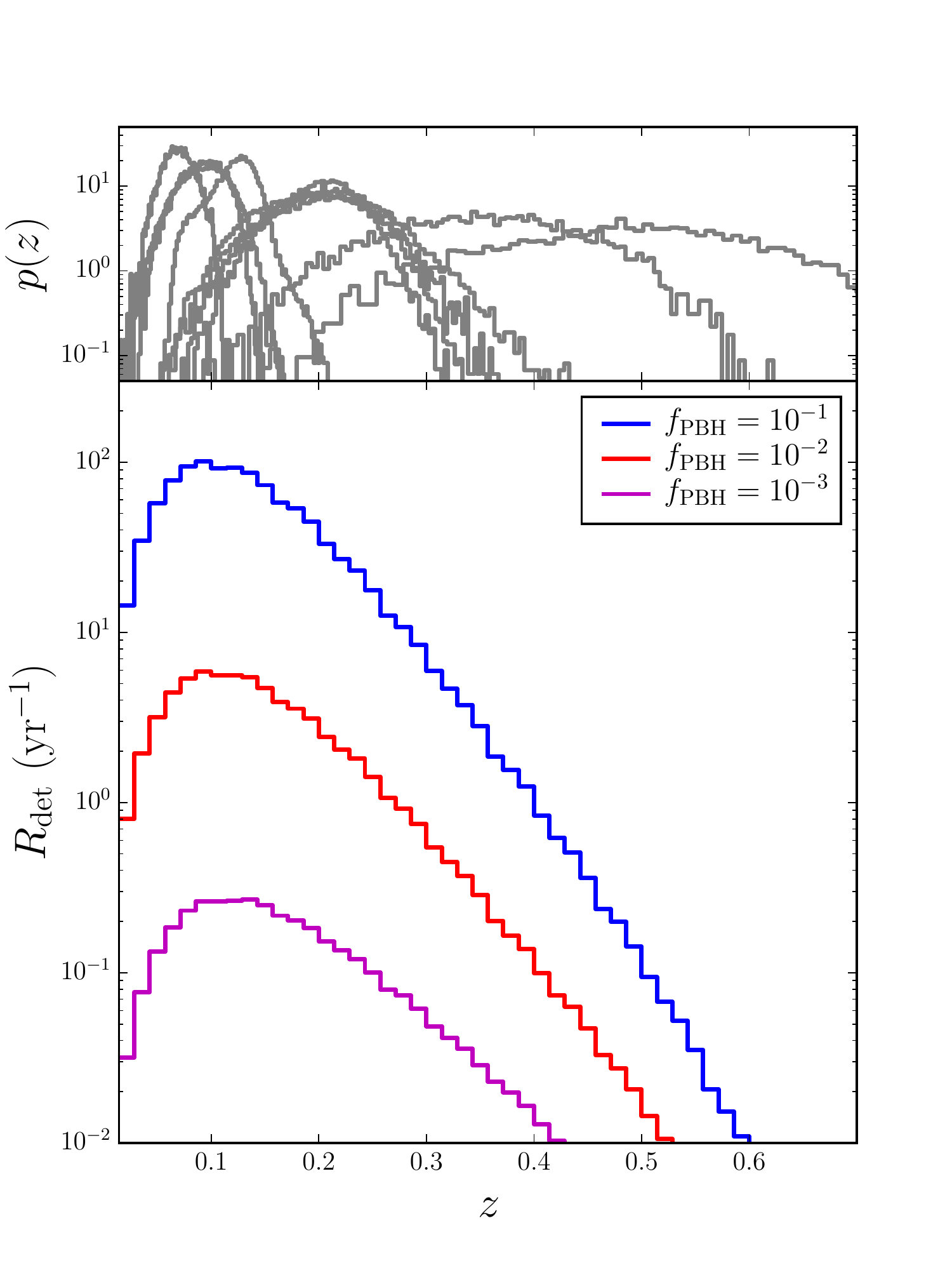}
\vspace*{-1em}
\includegraphics[width=0.49\textwidth]{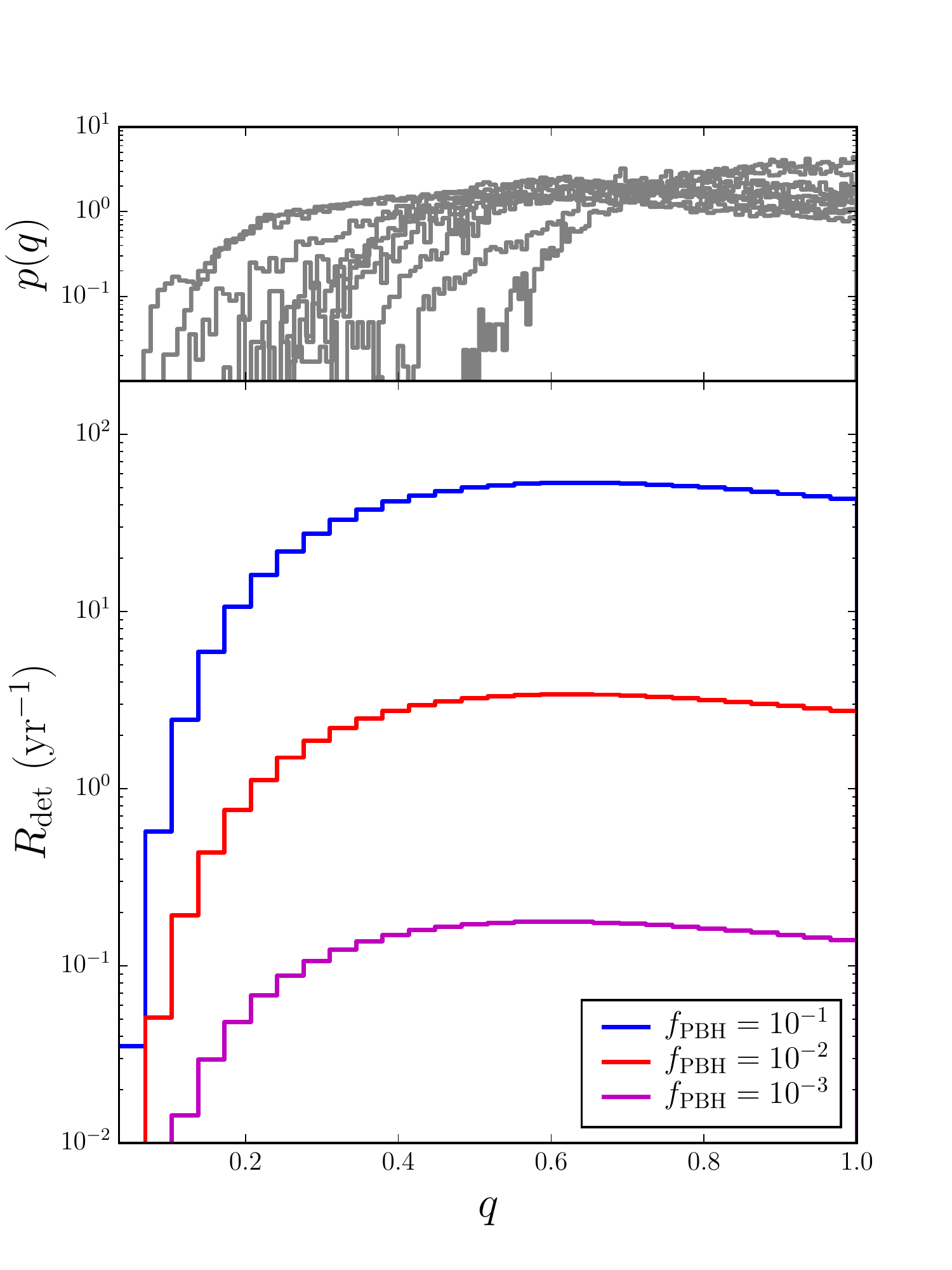}
\includegraphics[width=0.49\textwidth]{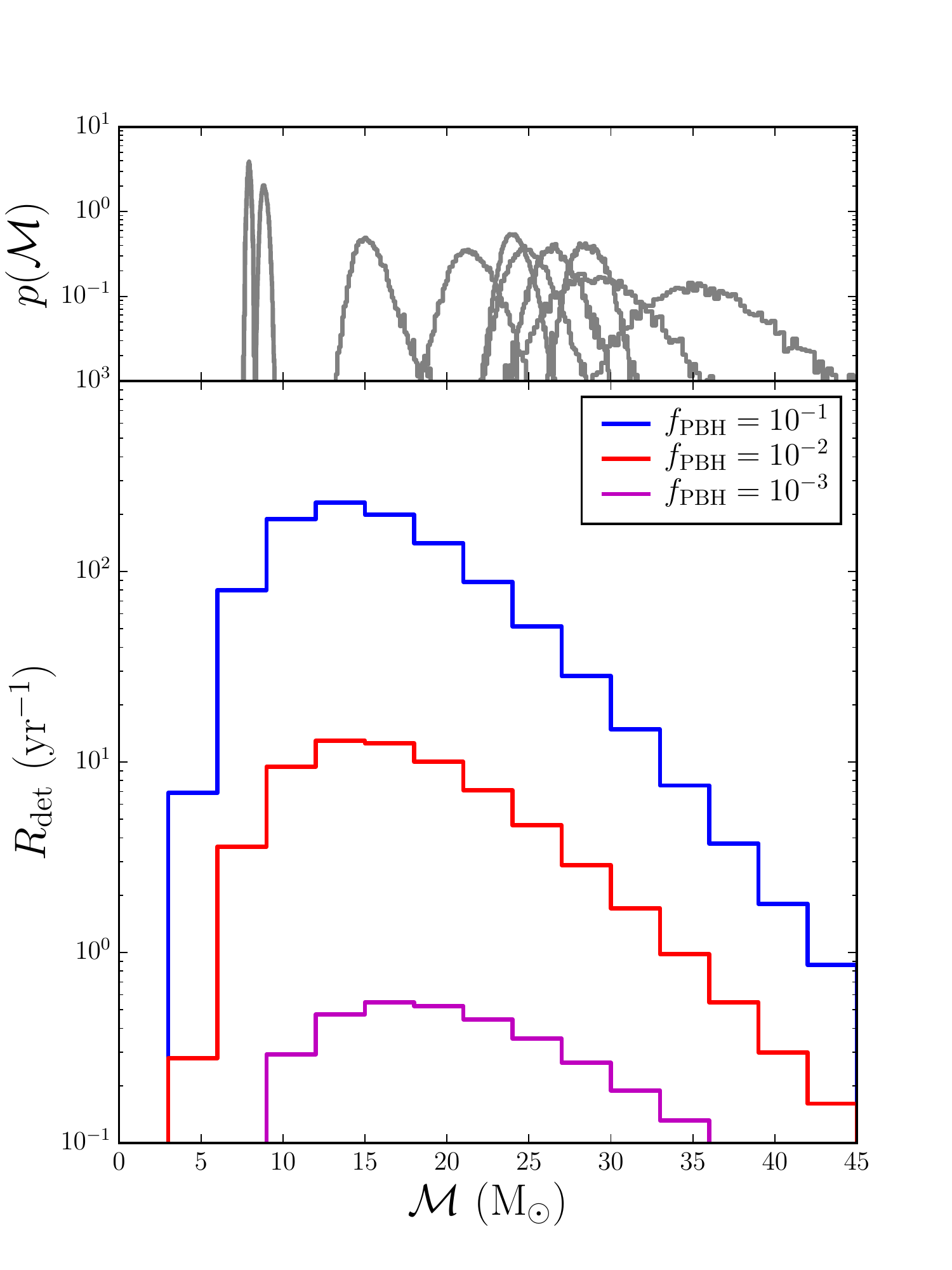}
\caption{Merger rate distributions in total mass $M$, redshift $z$, mass ratio $q$, and chirp mass $\mathcal{M}$ for the lognormal ($m_c=20\ \Msun$, $\sigma=0.6$) distribution with the O1O2 sensitivity. The top panel in each plot shows the LIGO posteriors for the 10 BBH events.}
\label{fig:O1O2-Lognormal-1D-s06}
\end{figure}

In the above sections, the detection probability $p_\text{det}$ was calculated using the default power spectral density (PSD) for the LIGO noise. This is the design sensitivity noise curve (\texttt{aLIGOZeroDetHighPower}). To compare to current LIGO data, a different PSD must be used. Therefore, the process above was carried out again with a detection probability generated using the \texttt{aLIGOEarlyHighSensitivityP1200087} PSD, which is a good approximation of the O1 and O2 sensitivities.

\begin{figure}[H]
\centering
\includegraphics[width=0.49\textwidth]{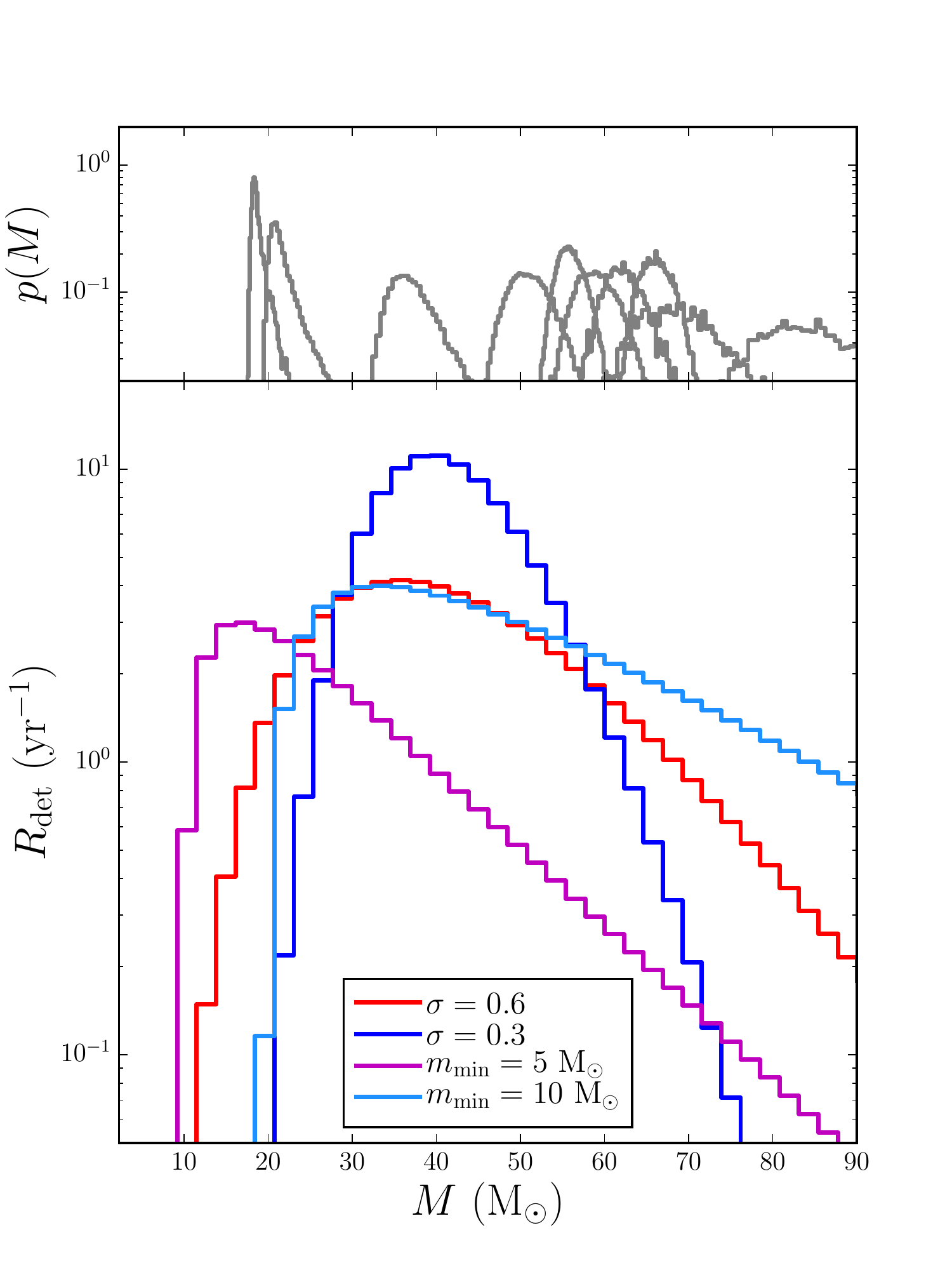}
\includegraphics[width=0.49\textwidth]{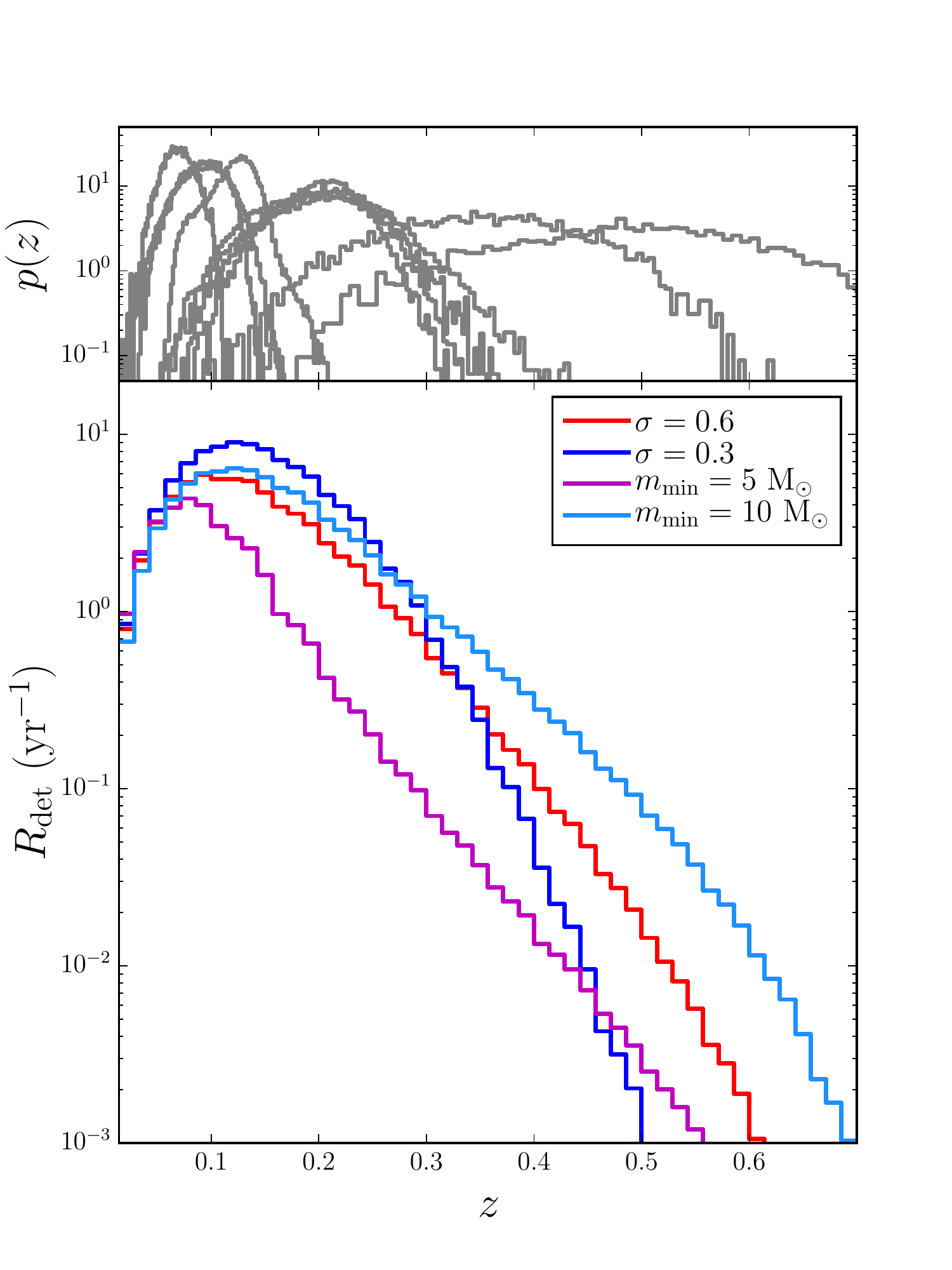}
\vspace*{-2em}
\includegraphics[width=0.49\textwidth]{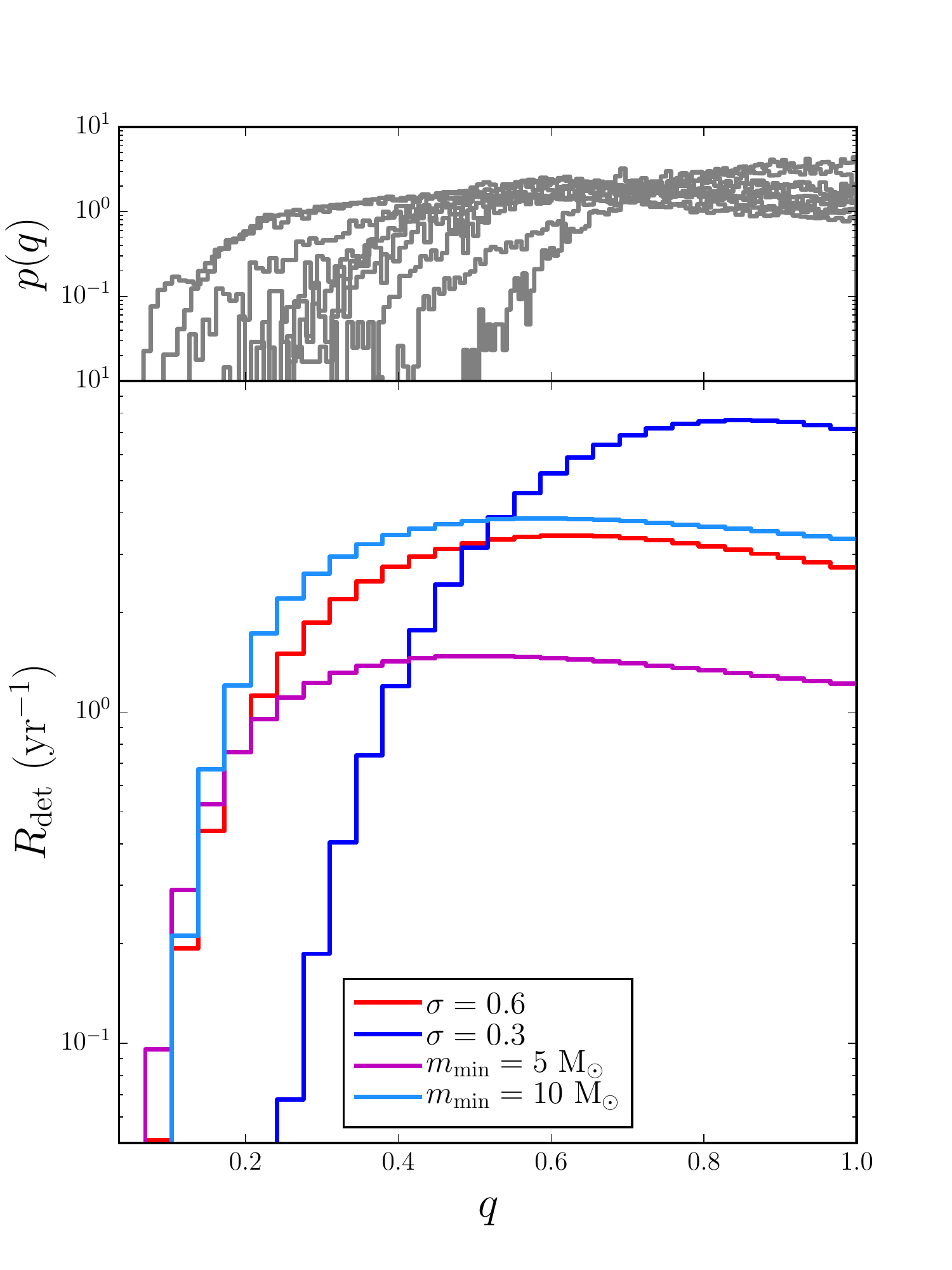}
\includegraphics[width=0.49\textwidth]{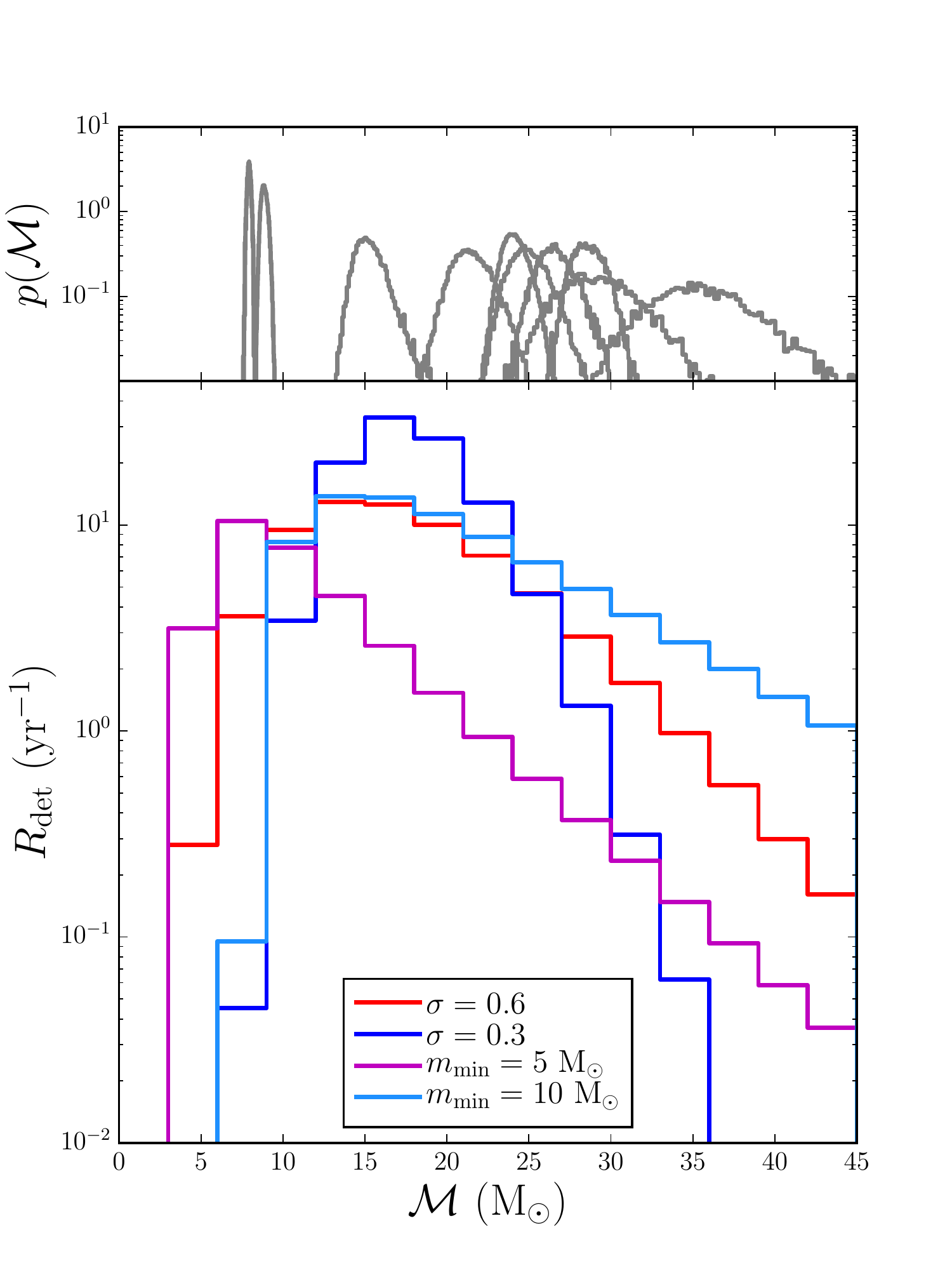}
\caption{Merger rate distributions in total mass $M$, redshift $z$, mass ratio $q$, and chirp mass $\mathcal{M}$ for a lognormal mass distribution with $\sigma=0.3$ and $\sigma=0.6$, and a power-law mass distribution with $m_\mathrm{min}=5\ \Msun$ and $m_\mathrm{min}=10\ \Msun$, with the O1O2 sensitivity. The top panel in each plot shows the LIGO posteriors for the 10 BBH events. All plots have $f_\PBH=10^{-2}$.}
\label{fig:O1O2-Combined}
\end{figure}

Figure~\ref{fig:O1O2-Lognormal-1D-s06} shows the same plots as in fig.~\ref{fig:Design-Lognormal-1D-s06}, but for the O1O2 detectability. The posterior probability distributions for the 10 LIGO binary black hole (BBH) events are shown in the top panel of each plot \cite{LIGO_2019_GWTC-1}. For all four observables, we can see that none of the posteriors has a distribution that drastically disagrees with the shapes of the merger rate curves. The expected number of events in a given observable range can be found by summing the merger rate curves over this range and multiplying by the total observing time of the O1O2 dataset, which is 0.46 yr. We can also consider the other mass distributions. Figure \ref{fig:O1O2-Combined} shows the O1O2 merger rate distributions for the lognormal distribution with the widths $\sigma=0.6$ and $0.3$, and the power-law distribution with $m_\mathrm{min}=5\ \Msun$ and $m_\mathrm{min}=10\ \Msun$.

\begin{figure}[H]
\centering
\includegraphics[width=0.49\textwidth]{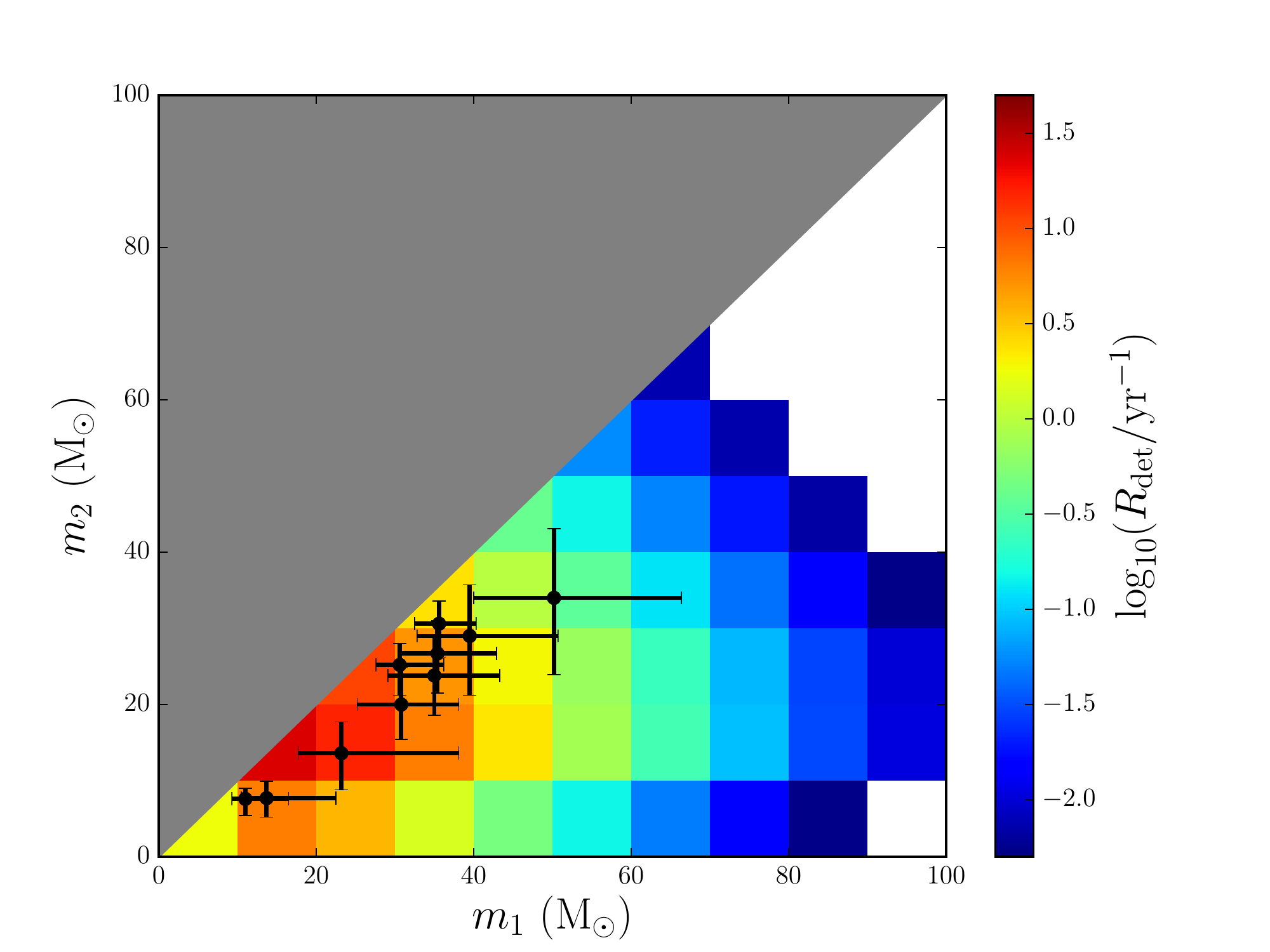}
\includegraphics[width=0.49\textwidth]{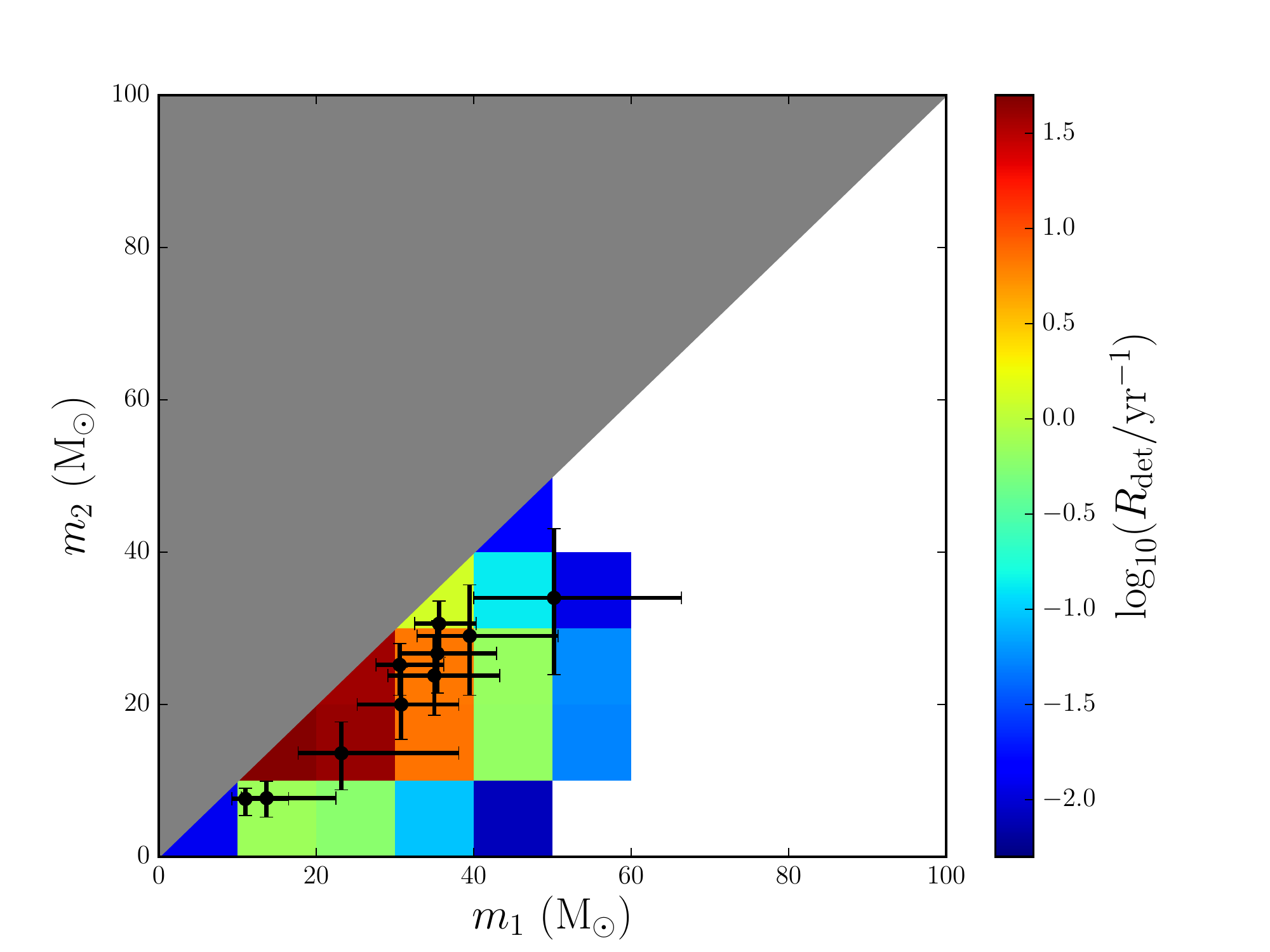}
\includegraphics[width=0.49\textwidth]{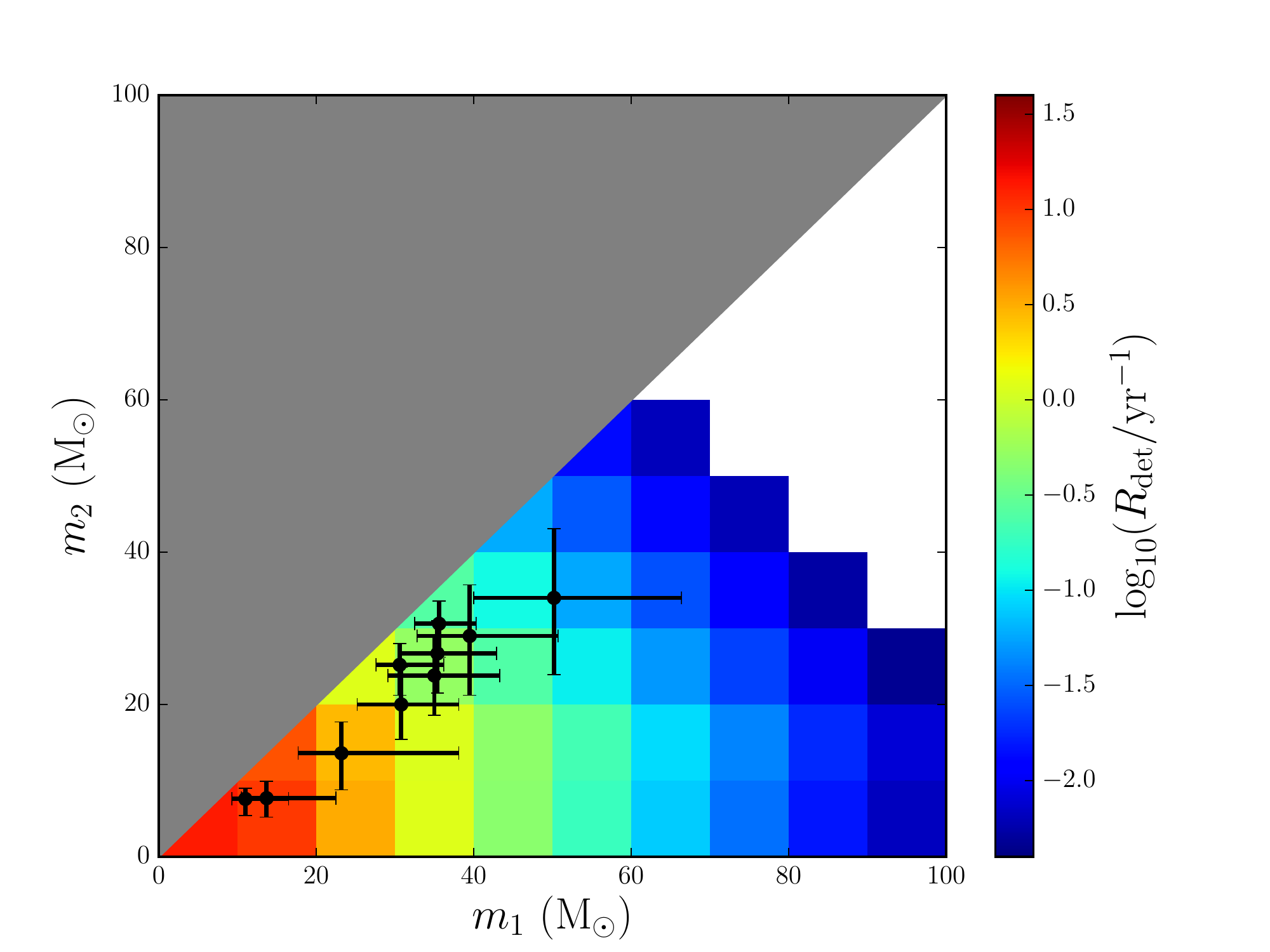}
\includegraphics[width=0.49\textwidth]{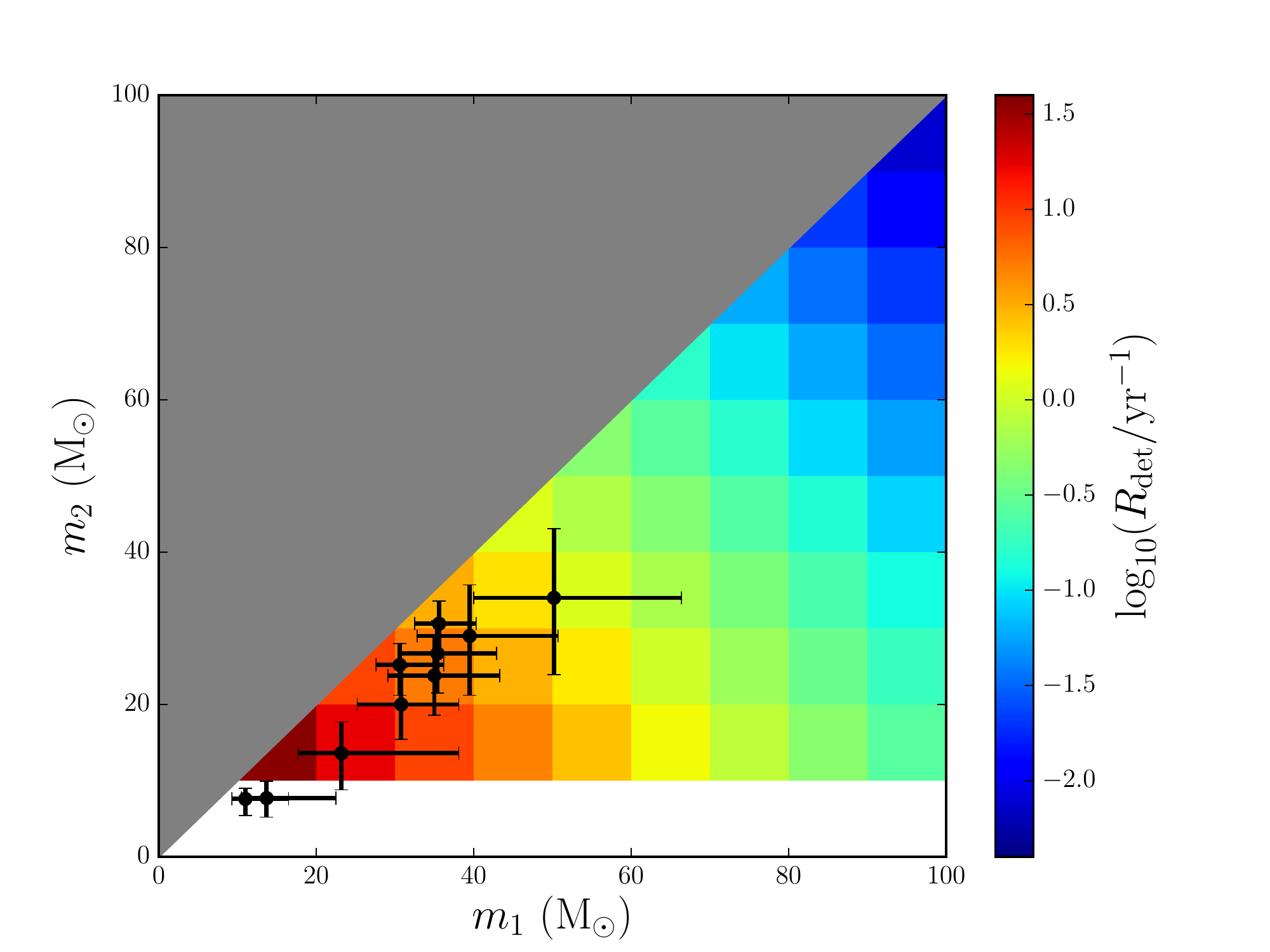}
\caption{2D Merger rate distributions in individual masses for the four distributions considered with the O1O2 sensitivity. The top row has the lognormal mass distribution with widths $\sigma=0.6$~(left) and $0.3$~(right), and the bottom row has the power-law mass distribution with minimum mass $5\ \Msun$~(left) and $10\ \Msun$~(right). All plots have $f_\PBH=10^{-2}$. The white area indicates no significant merger rate and the grey triangle indicates the case $m_2<m_1$. The LIGO values and their 90\% confidence limits are shown in black.}
\label{fig:O1O2-rvm1m2-all-dists}
\end{figure}

2D distributions were also produced for this sensitivity. The first of these, $r$~vs.~($m_1$, $m_2$) is shown in fig.~\ref{fig:O1O2-rvm1m2-all-dists} for the four mass distributions considered (the two lognormal distributions with different widths, and the two power-law distributions with different minimum masses). The black points show the LIGO values from the ten events and their 90\% confidence ranges (note that these are 1D marginalised error bars, and the full contour would not just follow the shape of the errors). All the data points lie in the lower triangle due to the LIGO analysis imposing $m_2<m_1$.

It can be seen by eye that some of these PBH models fail quite badly to match the observed locations of the LIGO events, whereas others look more acceptable. A statistical procedure is required to compare the observed and predicted distributions on the $m_1$-$m_2$ plane so that we can quantify the acceptability of the different models. This is relatively straightforward, but is made harder by the complex shape of the LIGO event posteriors in the $m_1$-$m_2$ plane that depends on the true parameters of the event (since these dictate the SNR of the detection). We are currently pursuing a Bayesian inference of the PBH scenario using the LIGO events, but we present here a simplified attempt to quantify how well the PBH model fits the data. We attempt to capture the posteriors by assuming that all events have measurement error distributions that are independently lognormal in $m_1$ and $m_2$, adopting a typical rms of 0.2 in $\ln m$ (the final $p$-values are not highly sensitive to this choice). With this assumption, we can smooth our distribution on the $m_1$-$m_2$ plane and hence convert it to a 2D function from which the observed data can be treated as random error-free samples (note that the smoothing is performed in the symmetric $m_1$-$m_2$ plane before folding to impose the convention $m_2<m_1$, and preserves the total merger rate over the $m_1$-$m_2$ plane).

In order to carry out the statistical comparison, there are two relevant questions: (1) is the observed number of events consistent with expectation? and (2) is the distribution over the \mbox{$m_1$-$m_2$} plane correct? For the former, we have 10 events, and the likelihood of this number is to be computed using a Poisson distribution based on the predicted number from integration over the $m_1$-$m_2$ plane and multiplying by the observing time. As the Poisson distribution is exponentially sensitive to the expected number $\mu$, the probability can drop off rapidly as $\mu$ moves away from the observed number of events. The expected value for each case is shown in table~\ref{tab:pValues}, with the corresponding probabilities adjacent. For the 2D distribution, we normalise the rate distribution over the $m_1$-$m_2$ plane to obtain a 2D probability density. This could then be compared with the data using the 2D Kolmogorov-Smirnov (KS) test \cite{Peacock_1983}. The only drawback with this approach is that the KS test is rather insensitive to whether a few points lie in an area of the plane with zero density, as does seem to be the case here. We therefore prefer a simpler statistic, which is just the overall likelihood of the data (product of the 2D density at the location of the 10 events). The expected distribution of this statistic can be readily obtained by drawing 10 points independently and at random from the 2D distribution multiple times. In this way, we can identify models whose likelihood is sufficiently low that they can be ruled out at an interesting level of significance. Thus we obtain two two-tailed frequentist $p$-values based on the absolute number of events and on their distribution. For the present purpose, it is probably the second of these that is of more interest, since it addresses directly the initial question of whether the PBH model can consistently generate nearly equal mass binaries.

\begin{table}[H]
\centering
\caption{Probability of the LIGO results based on the number of observed mergers $N=10$ with observing time $T=0.46$ yr, and their distribution in the $m_1$-$m_2$ plane for four PBH mass distributions: A = Lognormal ($m_c=20\ \Msun$, $\sigma=0.6$), B = Lognormal ($m_c=20\ \Msun$, $\sigma=0.3$), C = Power-law ($m_\text{min}=5\ \Msun$, $\alpha=3/2$), D = Power-law ($m_\text{min}=10\ \Msun$, $\alpha=3/2$)}
\label{tab:pValues}
\begin{tabular}{cccccc|cc}
& Test & \multicolumn{4}{c|}{$N$} & \multicolumn{2}{c}{$m_1$-$m_2$} \\
& $f_\PBH$ & \multicolumn{2}{c}{$10^{-2}$} & \multicolumn{2}{c|}{$10^{-3}$} & $10^{-2}$ & $10^{-3}$ \\ \hline
Model & & Expected number & Probability & Expected number & Probability & Probability & Probability \\ \hline
A & & 31 & $1.2\times10^{-5}$ & 1.6 & $1.3\times10^{-6}$ & 0.15 & 0.37 \\
B & & 47 & $6.4\times10^{-11}$ & 1.9 & $4.1\times10^{-6}$ & $\lesssim10^{-4}$ & $5\times10^{-4}$ \\
C & & 15 & 0.12 & 0.96 & $6.3\times10^{-9}$ & $2\times10^{-4}$ & 0.30 \\
D & & 37 & $1.5\times10^{-7}$ & 2.0 & $9.3\times10^{-6}$ & 0.15 & 0.45 \\ \hline
\end{tabular}
\end{table}

The results of this exercise are collected in table~\ref{tab:pValues} for the four extended mass distributions discussed above, labelled A-D as in the caption to table~\ref{tab:pValues}, and for two values of $f_\PBH$ each ($10^{-2}$ and $10^{-3}$). It can be seen that only model C is compatible at the 5\% level with the LIGO number of observed mergers for either of the values of $f_\PBH$. For the other models, $f_\PBH=10^{-2}$ significantly overproduces merger events, while for all the models $f_\PBH=10^{-3}$ does not produce enough, assuming that all the LIGO events are of primordial origin. For any of these models, it will be possible to choose a value of $f_\PBH$ between $10^{-2}$ and $10^{-3}$ that will match the observed number of events. The test of the $m_1$-$m_2$ plane shows better agreement, with only model B and model C at $f_\PBH=10^{-2}$ disfavoured at the 5\% level. These three probabilities are limited by shot noise due to the number of samples. While the global factor of $f_\PBH$ has been normalised out, there is still a non-trivial degeneracy between this parameter and the shape of the merger rate distribution in the $m_1$-$m_2$ plane. The $p$-values for model C vary greatly between the $10^{-2}$ and $10^{-3}$ case, due to the rapid suppression of the merger rate for higher values of $m_1$ and $m_2$ with $f_\PBH=10^{-2}$. This is likely to be an effect of the broadness problem discussed in section \ref{ssec:Power-law}, and so the small $p$-value for model C with $f_\PBH=10^{-2}$ should be taken with some hesitation. It is clear however, that a PBH scenario can explain the shape of the merger rate distribution in the $m_1$-$m_2$ plane and, with an appropriate $f_\PBH$ for normalisation, the total number of observed mergers.

\section{Conclusions}
Since the LIGO/Virgo collaborations have begun the era of direct gravitational wave detection, there has been great interest in the origin of the black holes whose mergers they have detected. We have focused on observational methods to discriminate between primordial and astrophysical black holes. We have explored a larger range of observables as a probe of the PBH scenario than have been greatly explored previously, with a particular focus on the mass ratio of the BHs that merged. Astrophysical BHs which form from a common envelope may dynamically equalise their masses and hence predict $q\approx1$, while still forming a large range of masses between different binary pairs \cite{Marchant_2016,Rodriguez_2016,O_Leary_2016,Kovetz_2017}. In contrast, PBHs form before they become part of a binary system, suggesting that $q\sim1$ is only likely to occur if the PBH mass distribution is narrow. However, a narrow mass distribution may be in tension with the broad range of chirp (or total) masses observed in the 10 binary black hole merger events detected by LIGO/Virgo to date.  

In order to analyse this problem, we have made merger rate distributions incorporating the LIGO detectability for the O1O2 LIGO/Virgo sensitivity curves, and compared to the LIGO data from the O1 and O2 runs. A rough analysis shows that the LIGO data have begun to apply constraints on the form and parameters of the PBH mass distribution, which is only possible using the detectable merger rate. Three types of PBH mass distribution were considered: lognormal, power-law, and monochromatic, although the monochromatic distribution is already ruled out by the variation in masses detected by LIGO. Table \ref{tab:pValues} shows the results of this analysis, indicating that the narrow lognormal (model B) is disfavoured at the 5\% level.

We have also calculated the expected detectable merger rate distributions at LIGO design sensitivity for the mass ratio $q$, redshift $z$, total mass $M$ and the chirp mass $\mathcal{M}$, and compared the results to distributions generated for astrophysical BHs. These distributions, which take into account the detection probability of any given merger, overcome the problem of comparing the intrinsic PBH merger rate to the intrinsic merger rate estimated by LIGO, each of which assumes a different mass distribution for the component black holes. The LIGO estimation method is described in section VII of \cite{LIGO_2019_GWTC-1}.

With the many new events expected to be detected by LIGO in the future, the PBH mass distribution can be probed in great detail, and following on from the methods developed in this paper, the best fits for any form of the PBH mass distribution can be found. A complication in these fits would be the uncertainty of the source of any given merger. In the future, a fit simultaneously incorporating the two potential BH populations (primordial and astrophysical) should be made, and it may be possible to rule out all of the BHs being primordial in origin. One potential discriminant is the spin of the BHs, which is expected to be negligibly small for PBHs formed during radiation domination \cite{Fernandez_2019}. Even if the spin cannot discriminate, it must be taken into account for astrophysical BHs, which may have significant spins. We are currently pursuing a fully Bayesian inference of the PBH merger scenario with current (and future) LIGO data.

Uncertainties in the PBH merger rate calculation remain an open issue. The method applied in this paper from \cite{Raidal_2019} builds a strong framework for the merger rate calculation, but there are further considerations, such as the torquing effects from matter and radiation perturbations \cite{Garriga_2019}, and the uncertainty of how frequently binaries are disrupted between formation and merger \cite{AliHaimoud_2017,Raidal_2019}. We have also found that the current calculation cannot be used in the case of a very broad mass distribution. Detailed simulations and further analytic developments of PBH binary formation, disruption, and merger events are essential to ensure that the fits to the current and future LIGO data are accurate.

\section*{Acknowledgements}
AG thanks Ville Vaskonen, Davide Gerosa, and Zu-Cheng Chen for discussions and clarification. AH thanks Derek Inman for helpful discussions. AG is funded by a Royal Society Studentship by means of a Royal Society Enhancement Award. CB is funded by a Royal Society University Research Fellowship. AH is supported by an STFC Consolidated Grant.

\newpage
\appendix
\section*{Appendix - Additional Plots}
\renewcommand\thefigure{A.\arabic{figure}}
\renewcommand{\theHfigure}{A.\arabic{figure}}
\setcounter{figure}{0}
\subsection*{Design sensitivity plots}
\begin{figure}[H]
\centering
\includegraphics[width=0.48\textwidth]{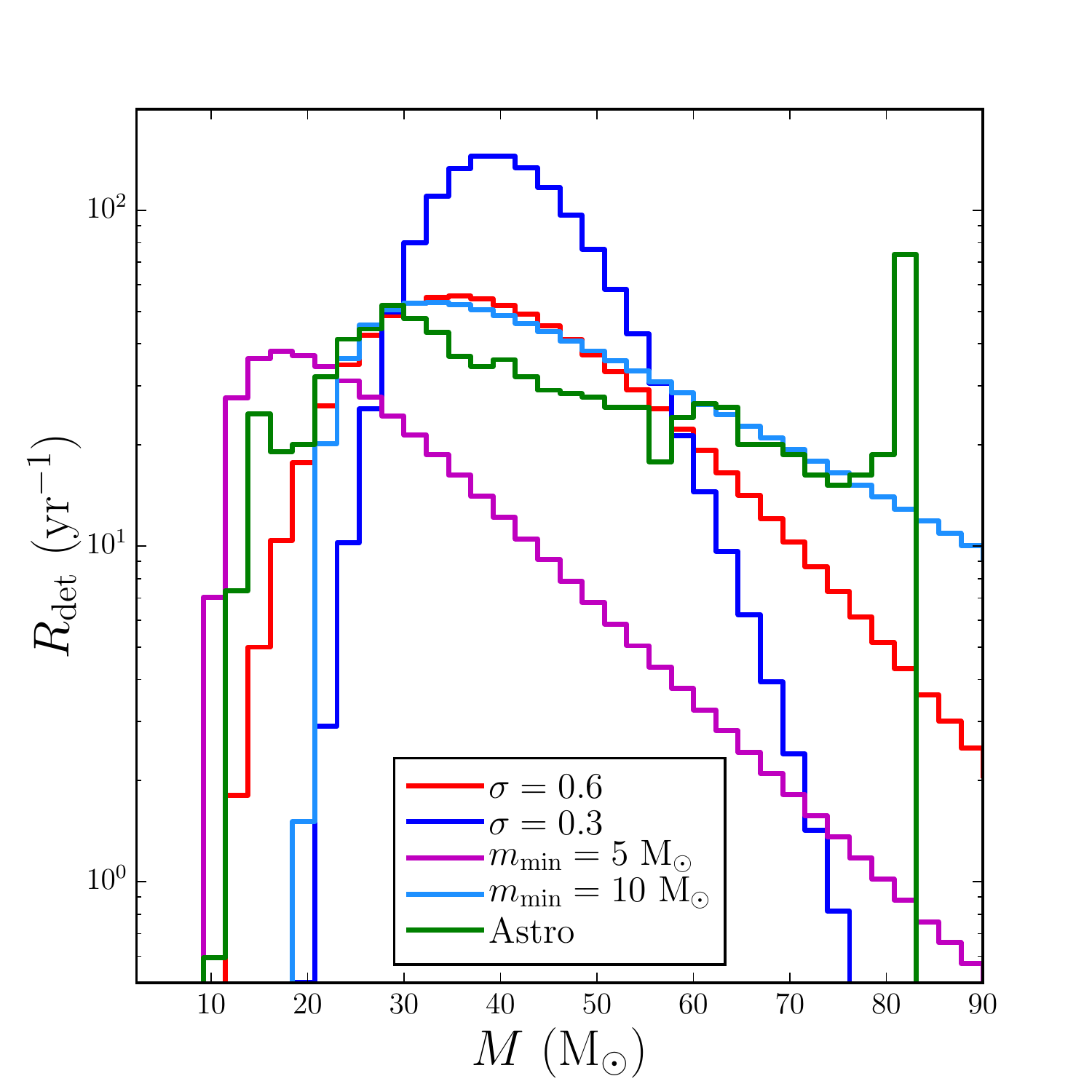}
\includegraphics[width=0.48\textwidth]{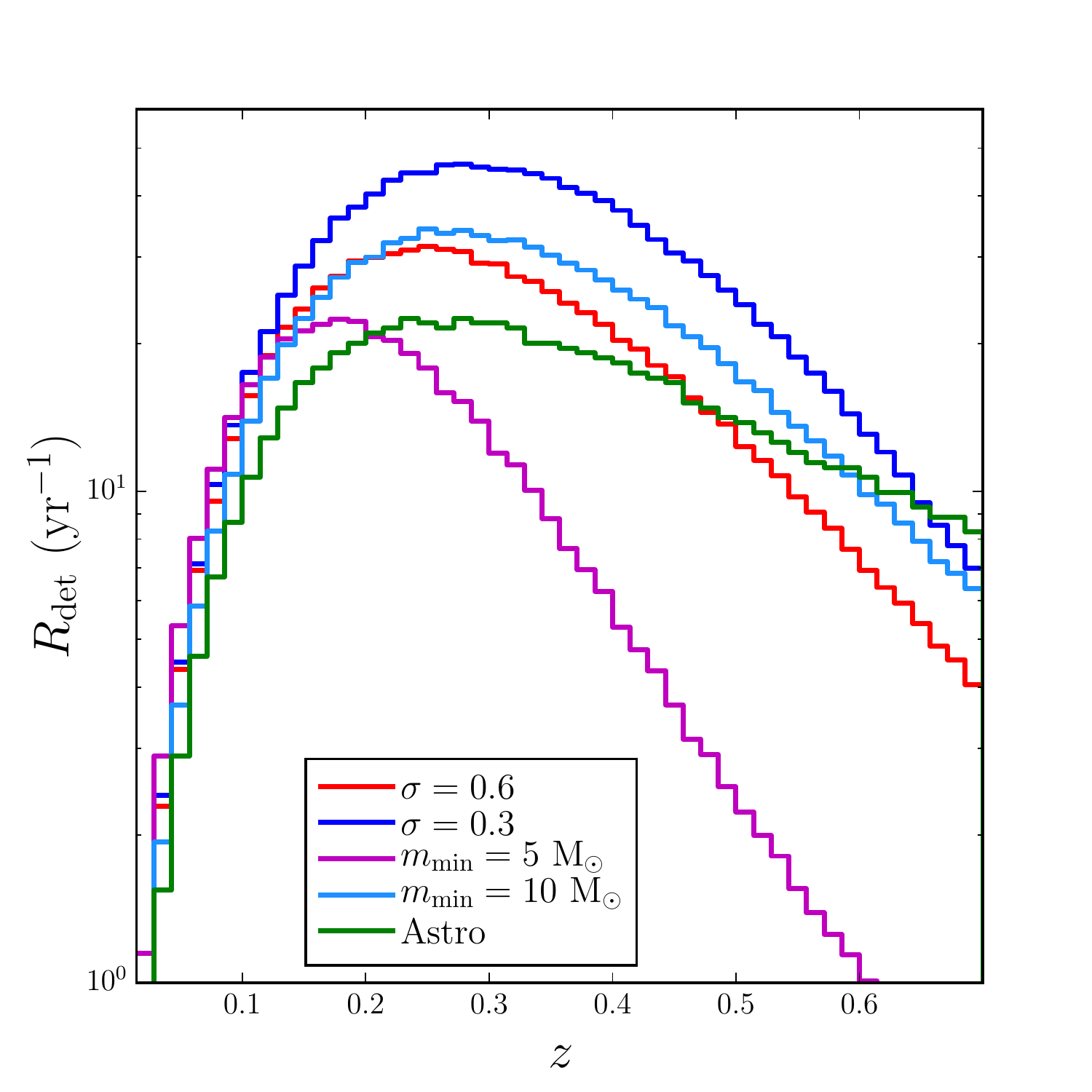}
\vspace*{-2em}
\includegraphics[width=0.48\textwidth]{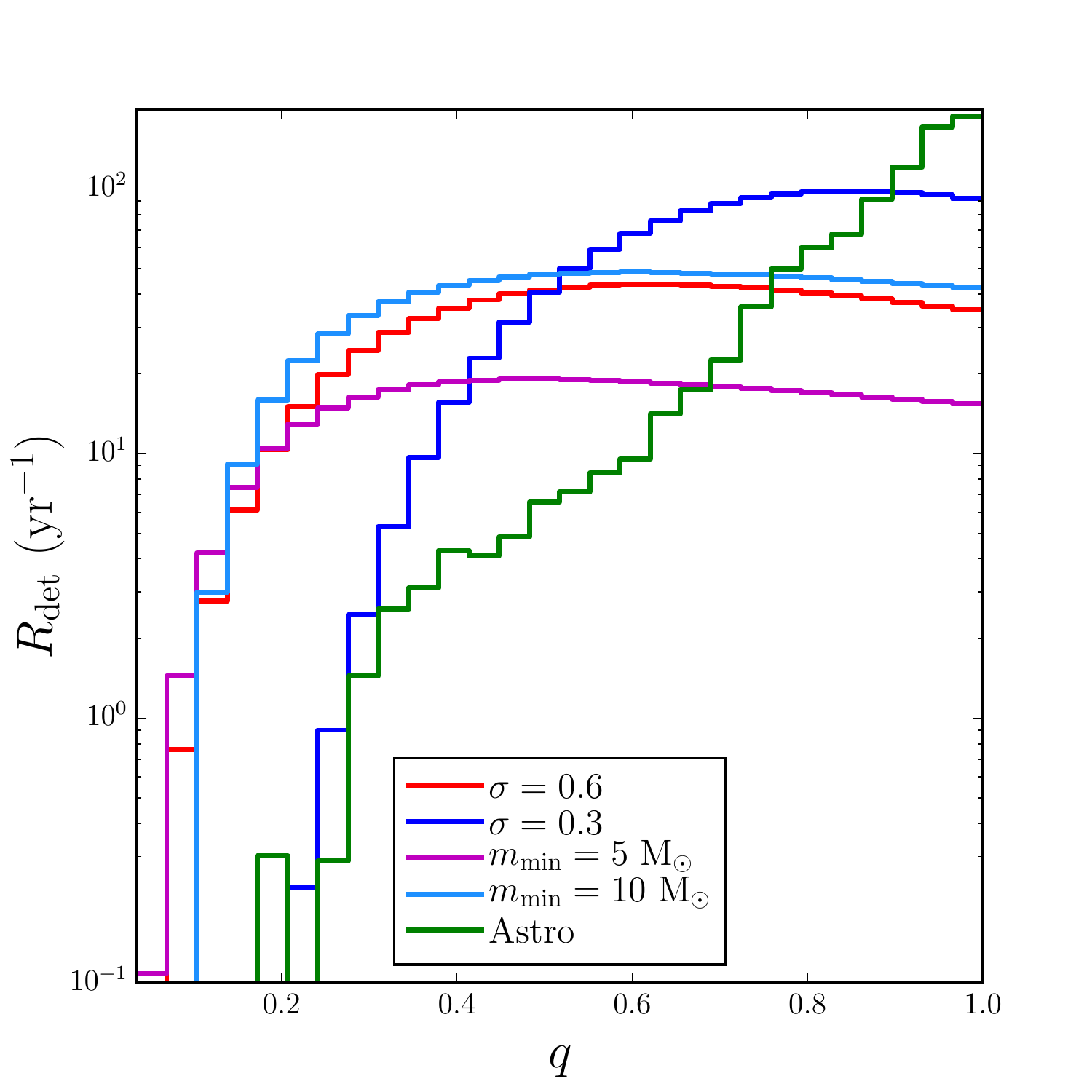}
\includegraphics[width=0.48\textwidth]{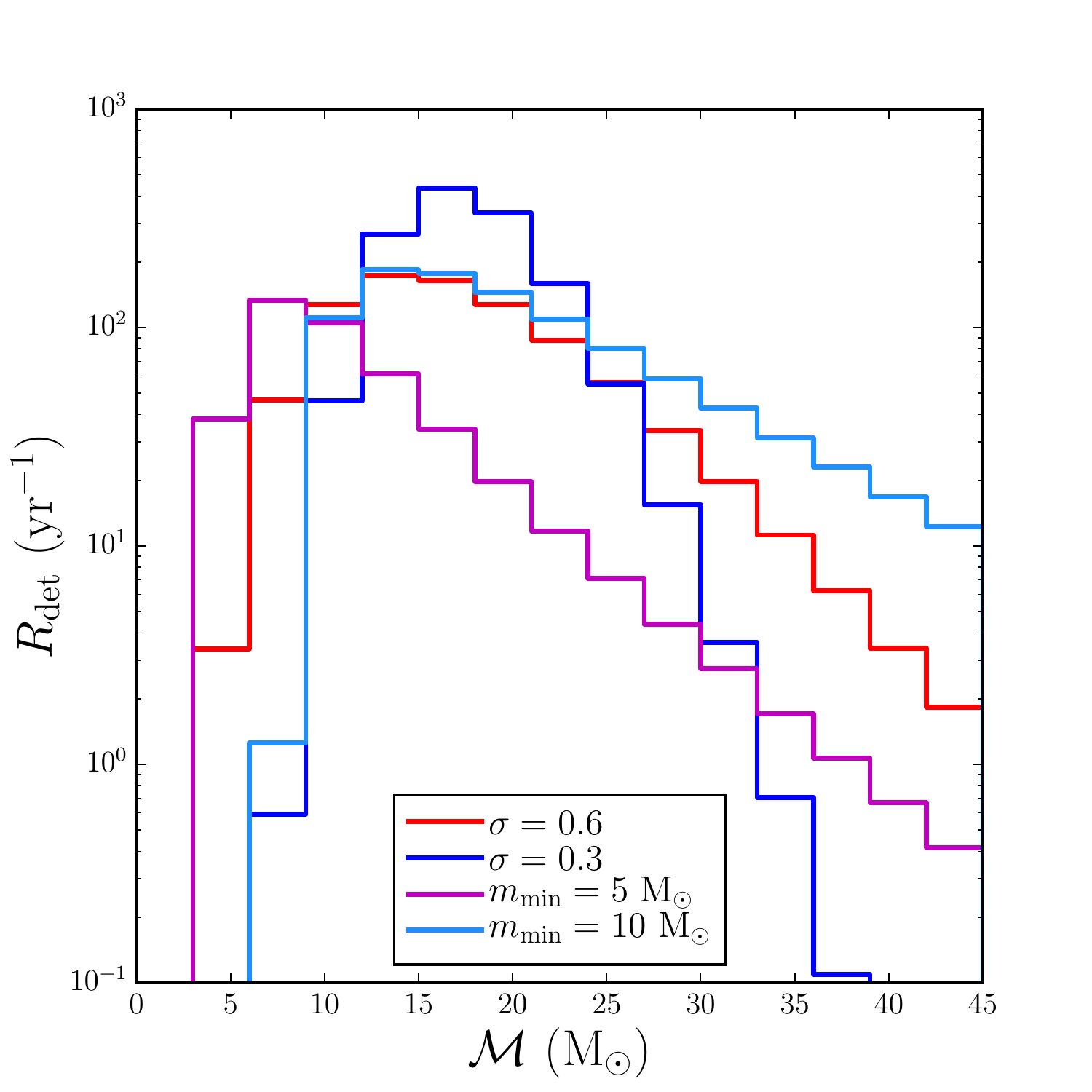}
\caption{Merger rate distributions in total mass $M$, redshift $z$, mass ratio $q$, and chirp mass $\mathcal{M}$ for a lognormal mass distribution with $\sigma=0.3$ and $\sigma=0.6$, and a power-law mass distribution with $m_\mathrm{min}=5\ \Msun$ and $m_\mathrm{min}=10\ \Msun$. The distributions for astrophysical black holes from \cite{Gerosa_2019} are shown in green for the first three plots. All plots have $f_\PBH=10^{-2}$.}
\label{fig:Design-Combined}
\end{figure}

Figure \ref{fig:Design-Combined} shows the design sensitivity merger rate distributions for the two lognormal mass distributions and the two power-law distributions. The astrophysical distributions are shown in green.

\begin{figure}[H]
\centering
\includegraphics[width=0.49\textwidth]{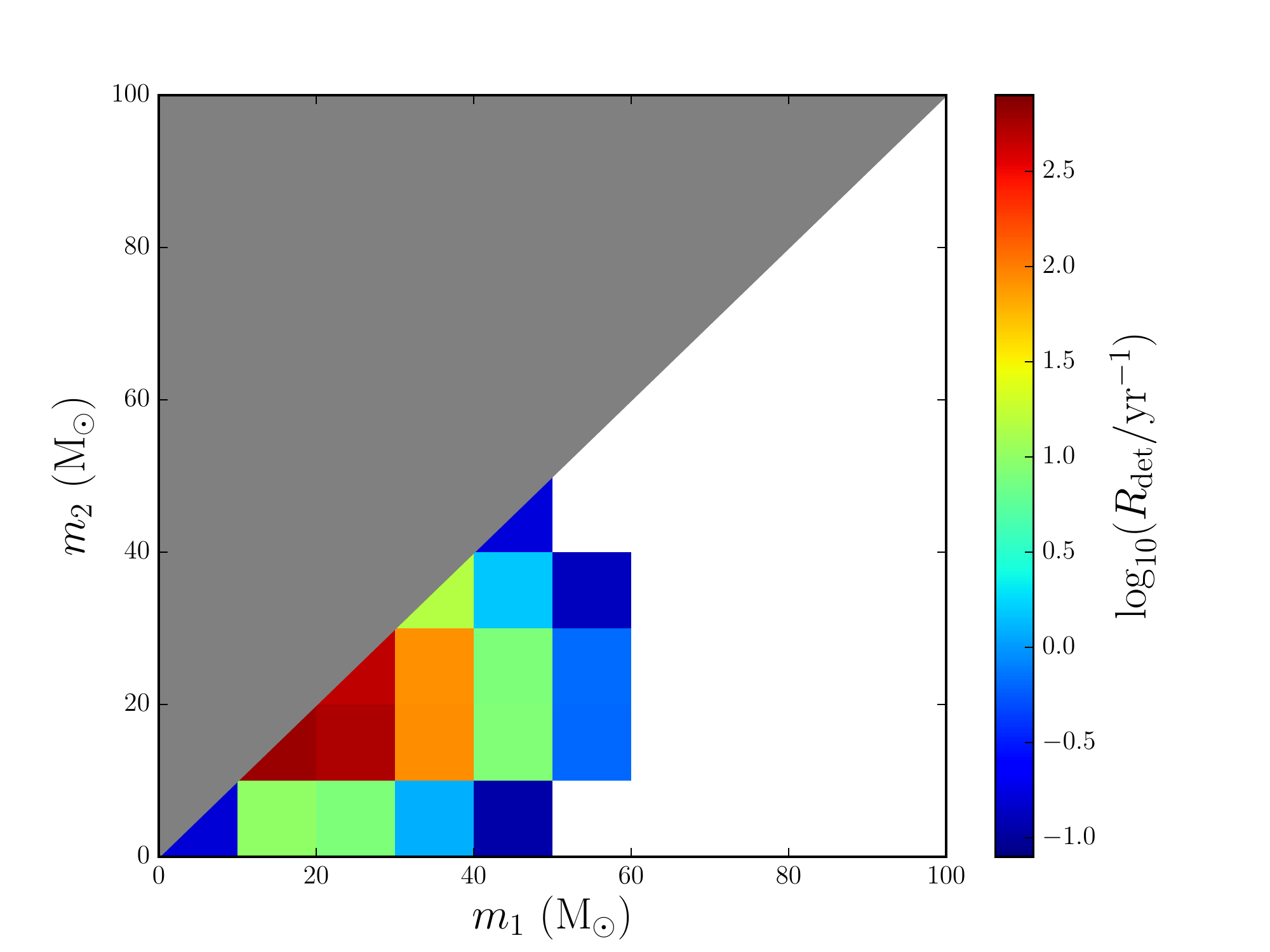}
\includegraphics[width=0.49\textwidth]{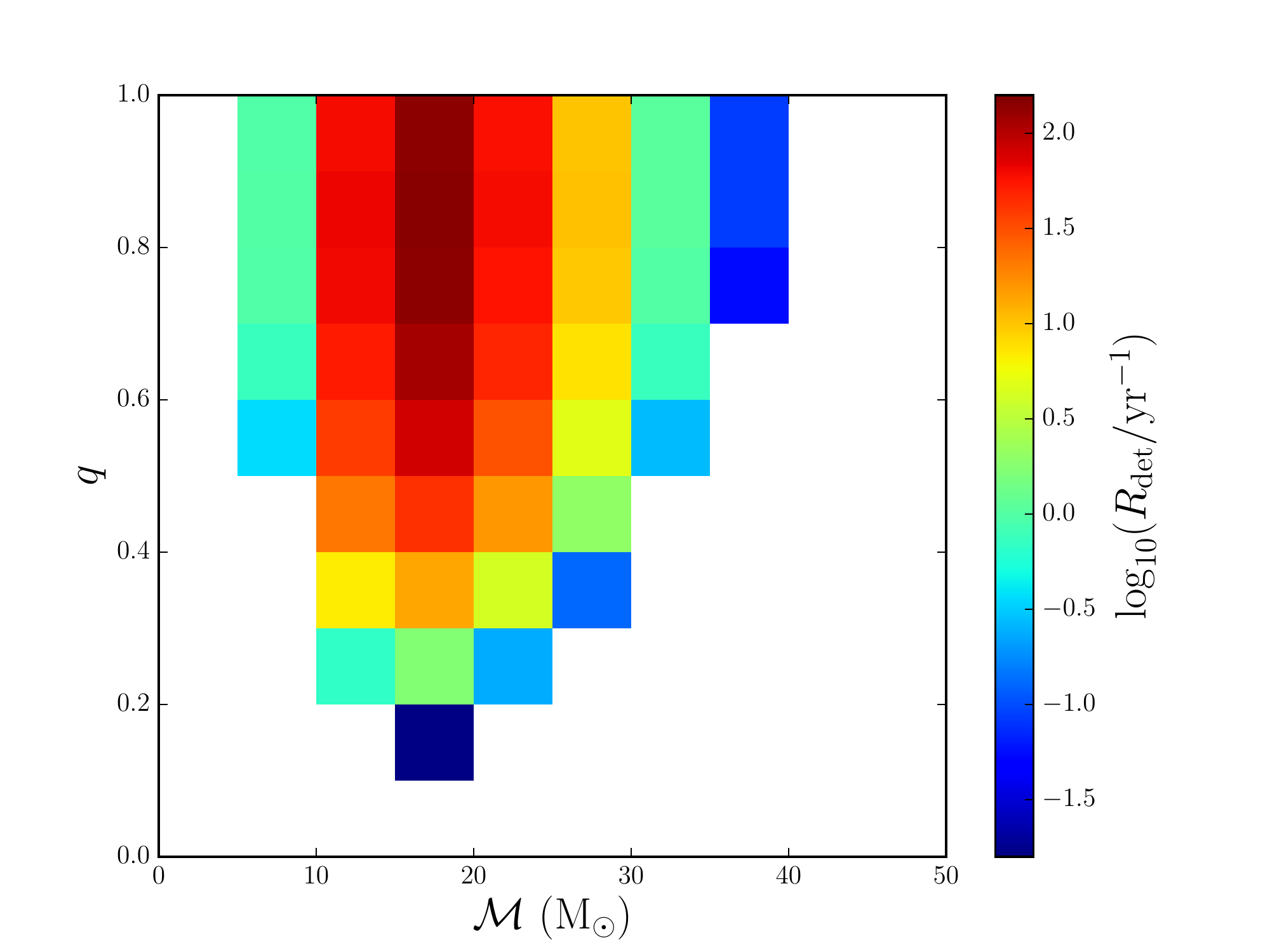}
\includegraphics[width=0.49\textwidth]{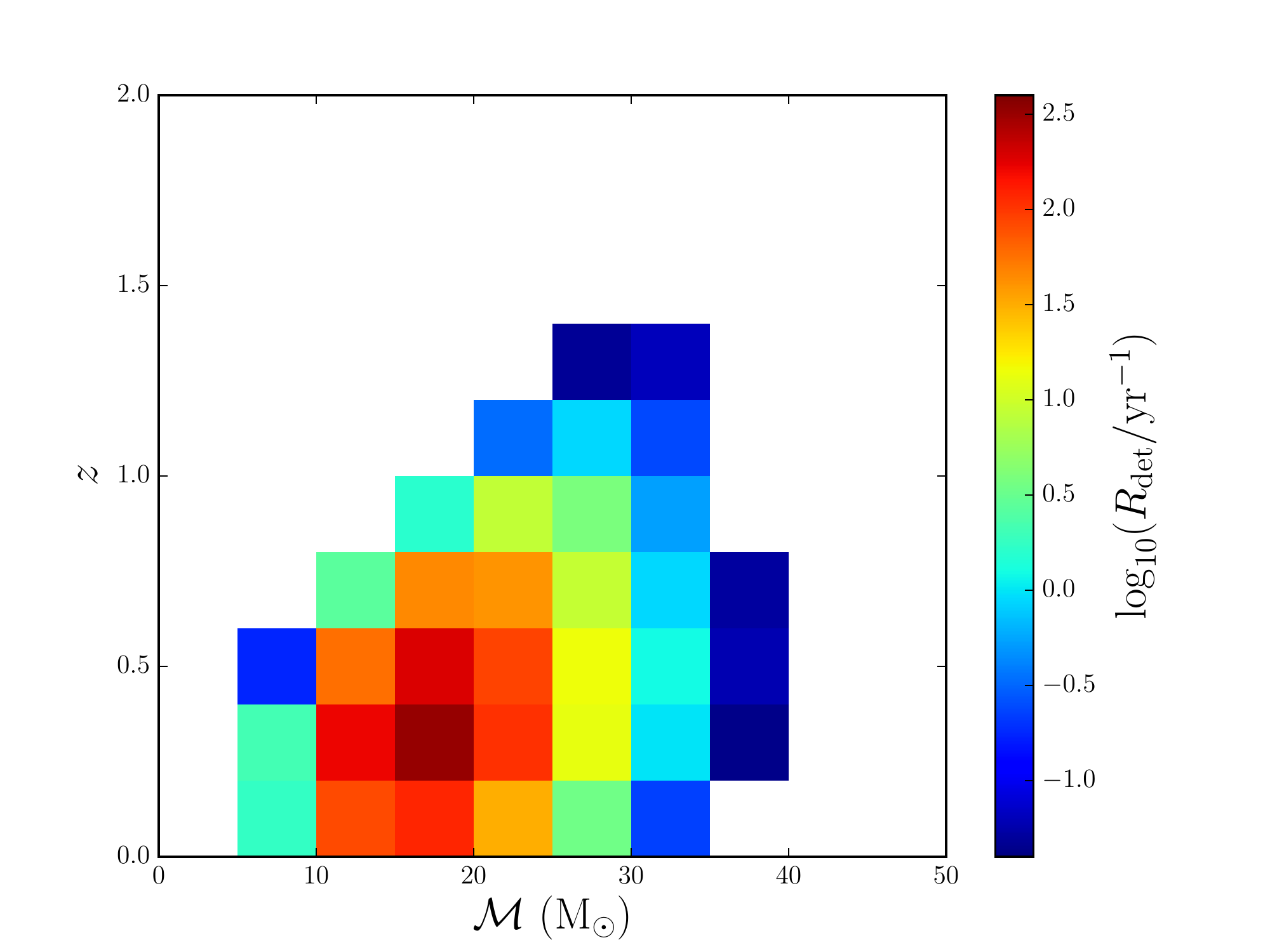}
\includegraphics[width=0.49\textwidth]{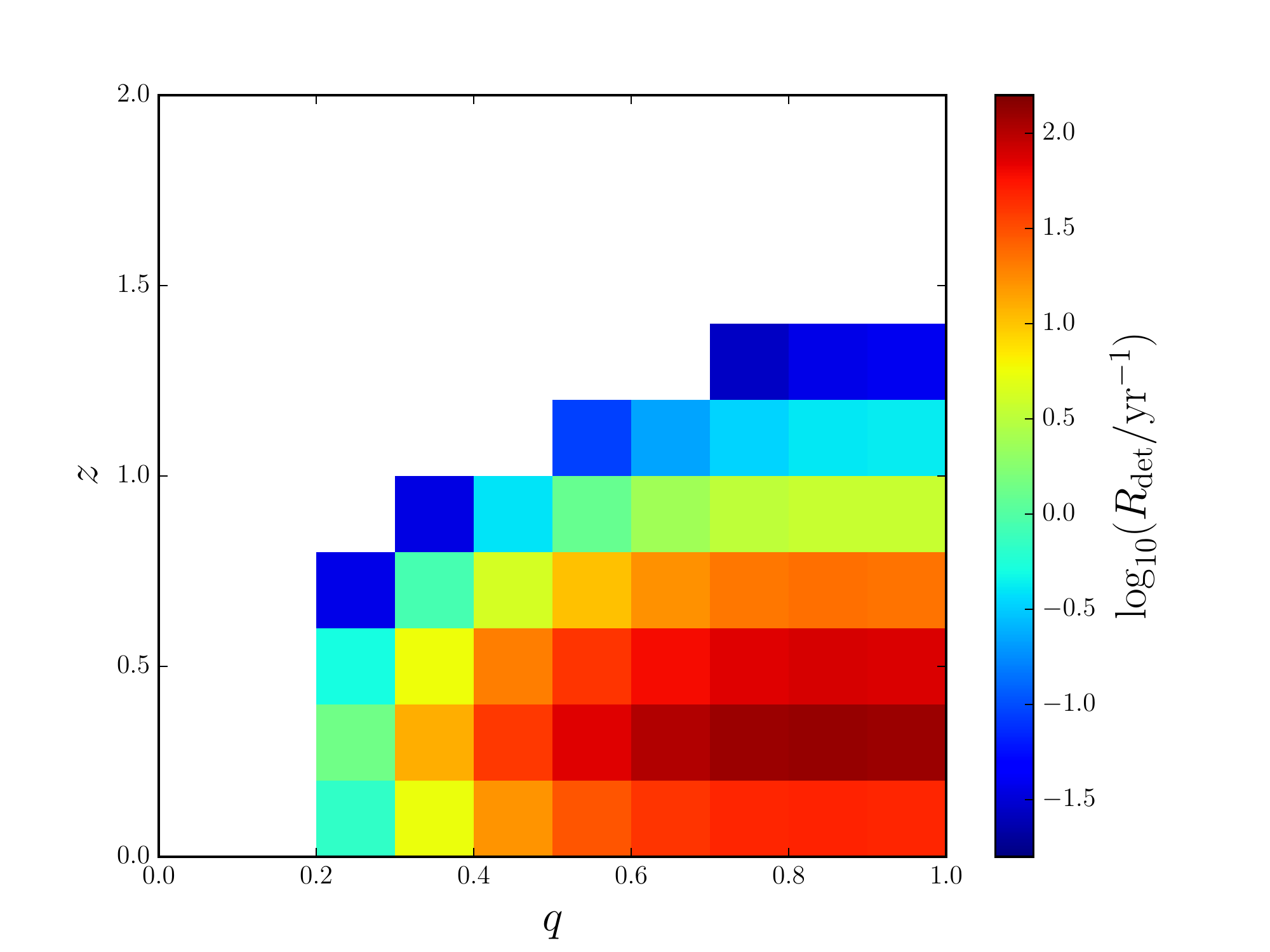}
\caption{2D Merger rate distributions in individual masses, mass ratio $q$, chirp mass $\mathcal{M}$ and redshift $z$ for a lognormal distribution with $\sigma=0.3$. All plots have $f_\PBH=10^{-2}$. The white area indicates no significant merger rate and the grey triangle indicates the case $m_2<m_1$. The colorbar limits are the same as in fig.~\ref{fig:Design-Lognormal-2D-s06} for comparison.}
\label{fig:Design-Lognormal-2D-s03}
\end{figure}

\begin{figure}[H]
\centering
\includegraphics[width=0.49\textwidth]{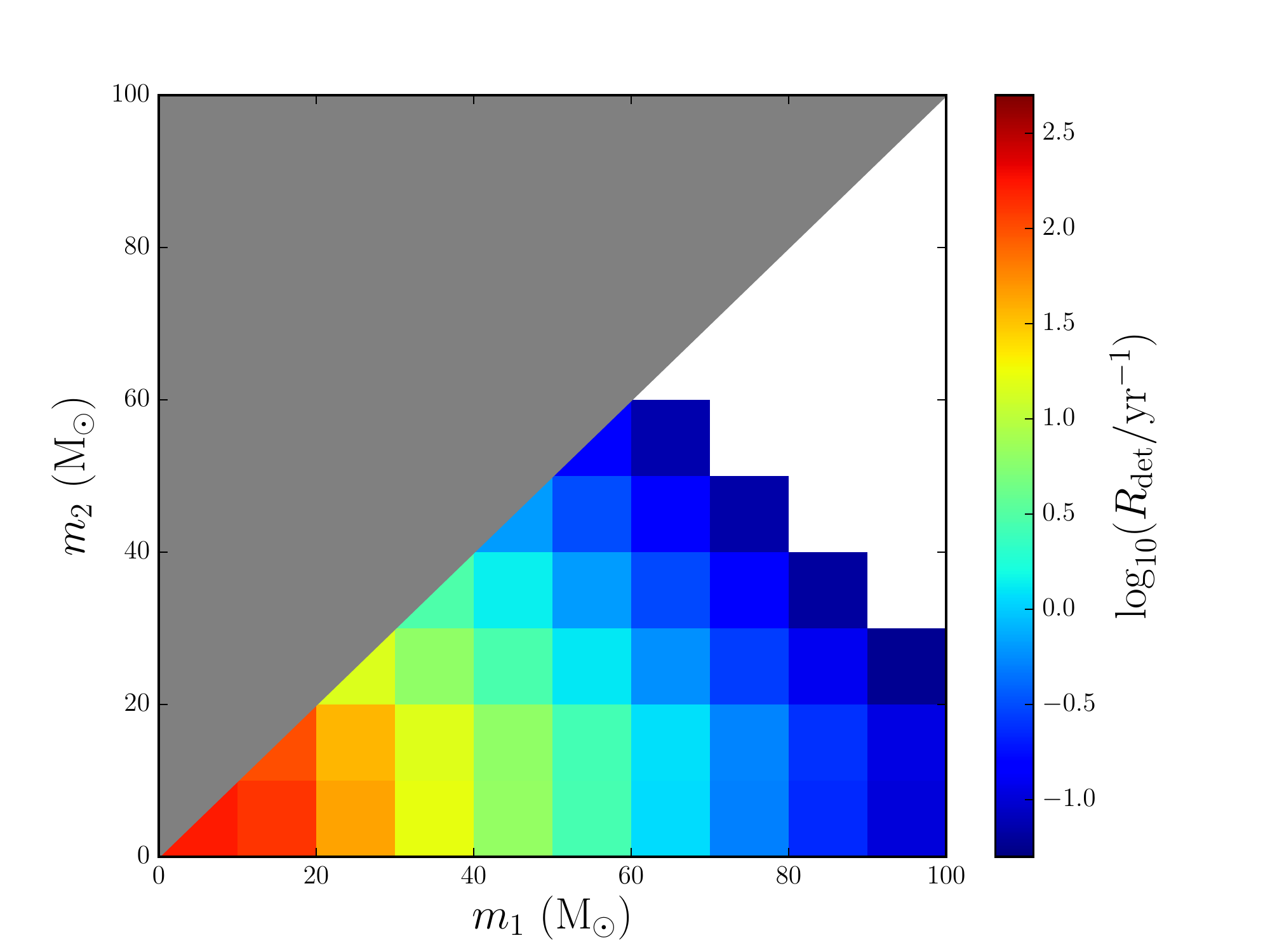}
\includegraphics[width=0.49\textwidth]{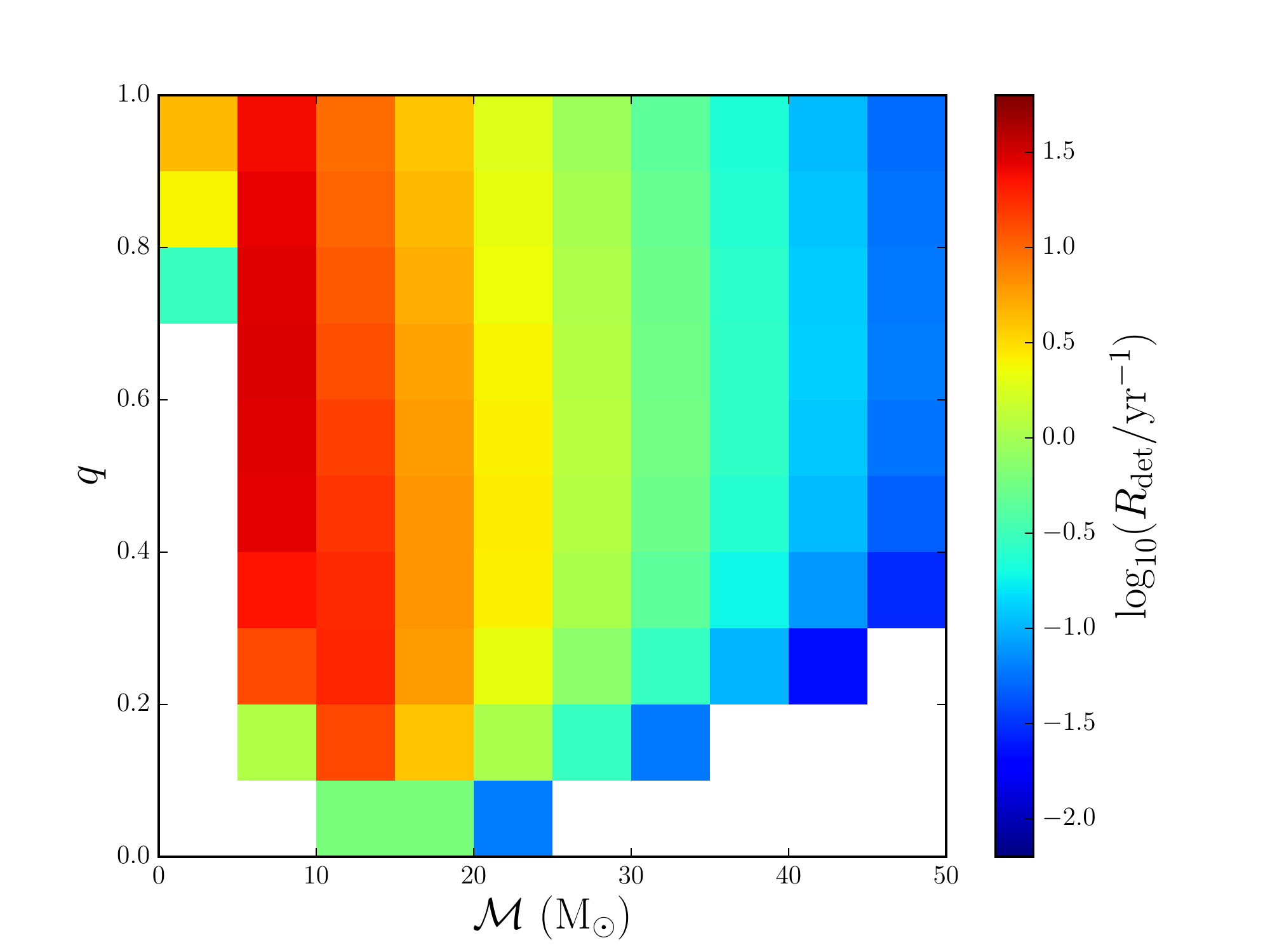}
\includegraphics[width=0.49\textwidth]{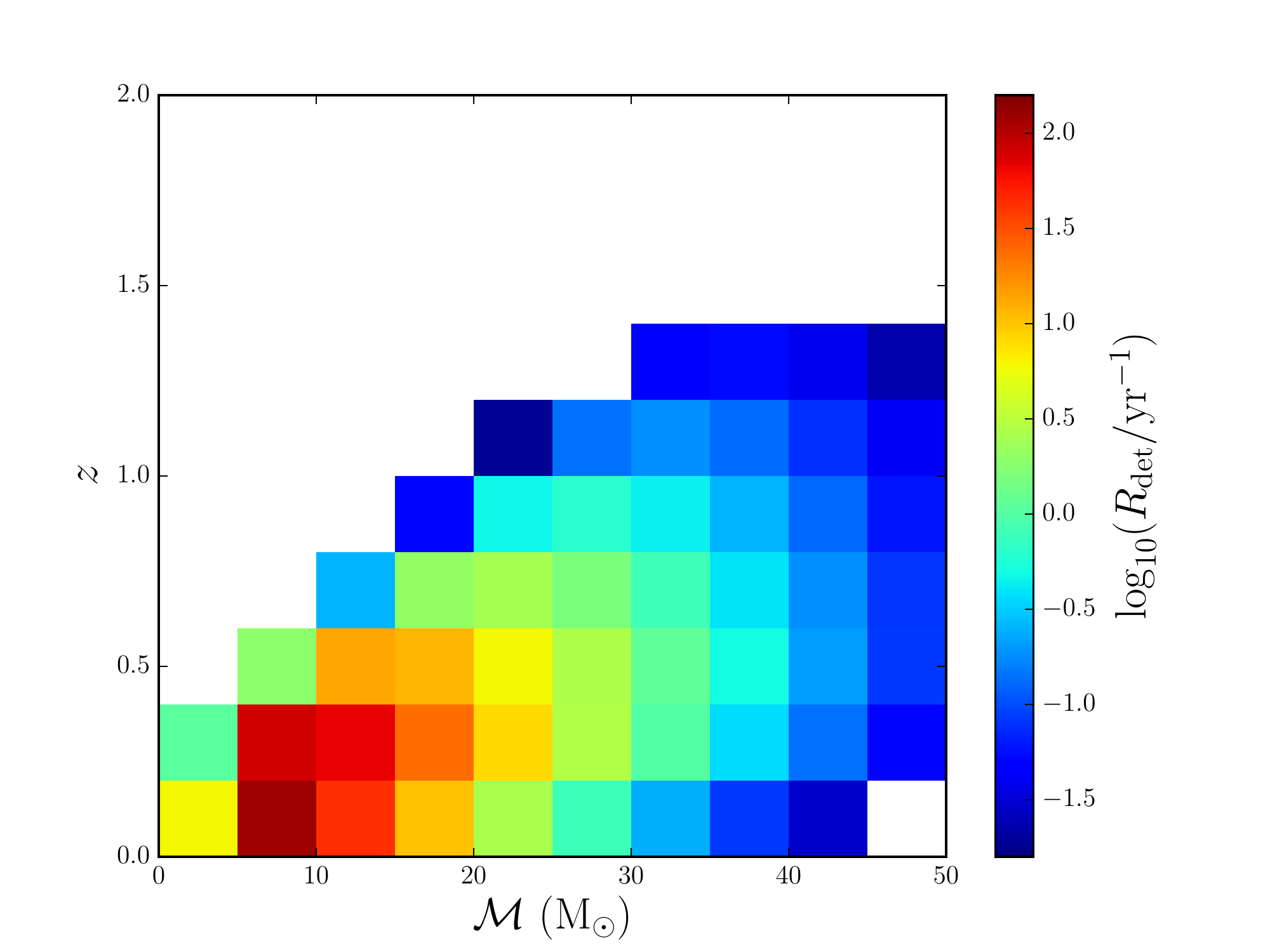}
\includegraphics[width=0.49\textwidth]{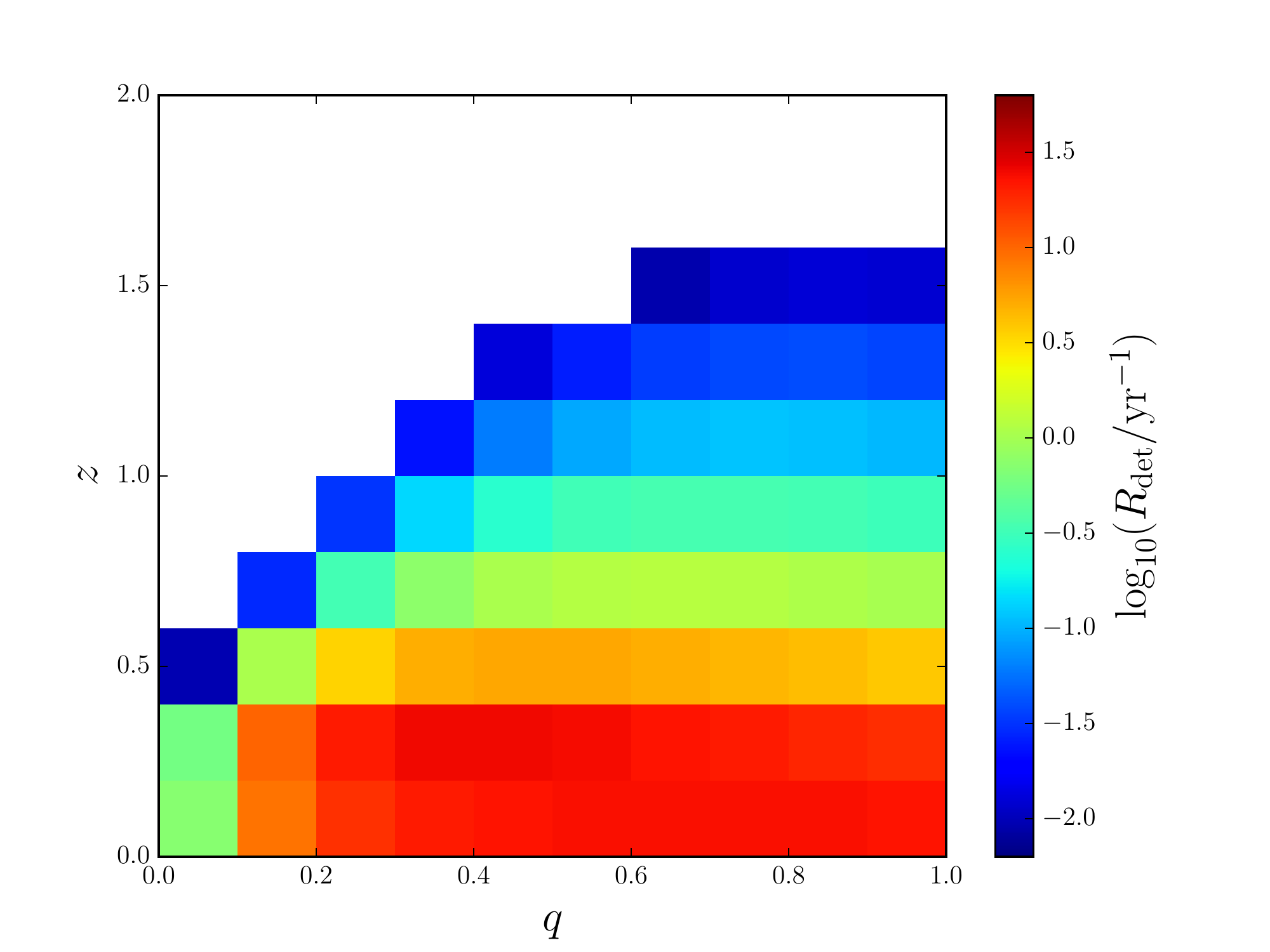}
\caption{2D Merger rate distributions in individual masses, mass ratio $q$, chirp mass $\mathcal{M}$ and redshift $z$ for a power-law distribution with $m_\mathrm{min}=5\ \Msun$. All plots have $f_\PBH=10^{-2}$. The white area indicates no significant merger rate and the grey triangle indicates the case $m_2<m_1$. The colorbar limits are the same as in fig.~\ref{fig:Design-PL-2D-mMin10} for comparison.}
\label{fig:Design-PL-2D-mMin5}
\end{figure}

\begin{figure}[H]
\centering
\includegraphics[width=0.49\textwidth]{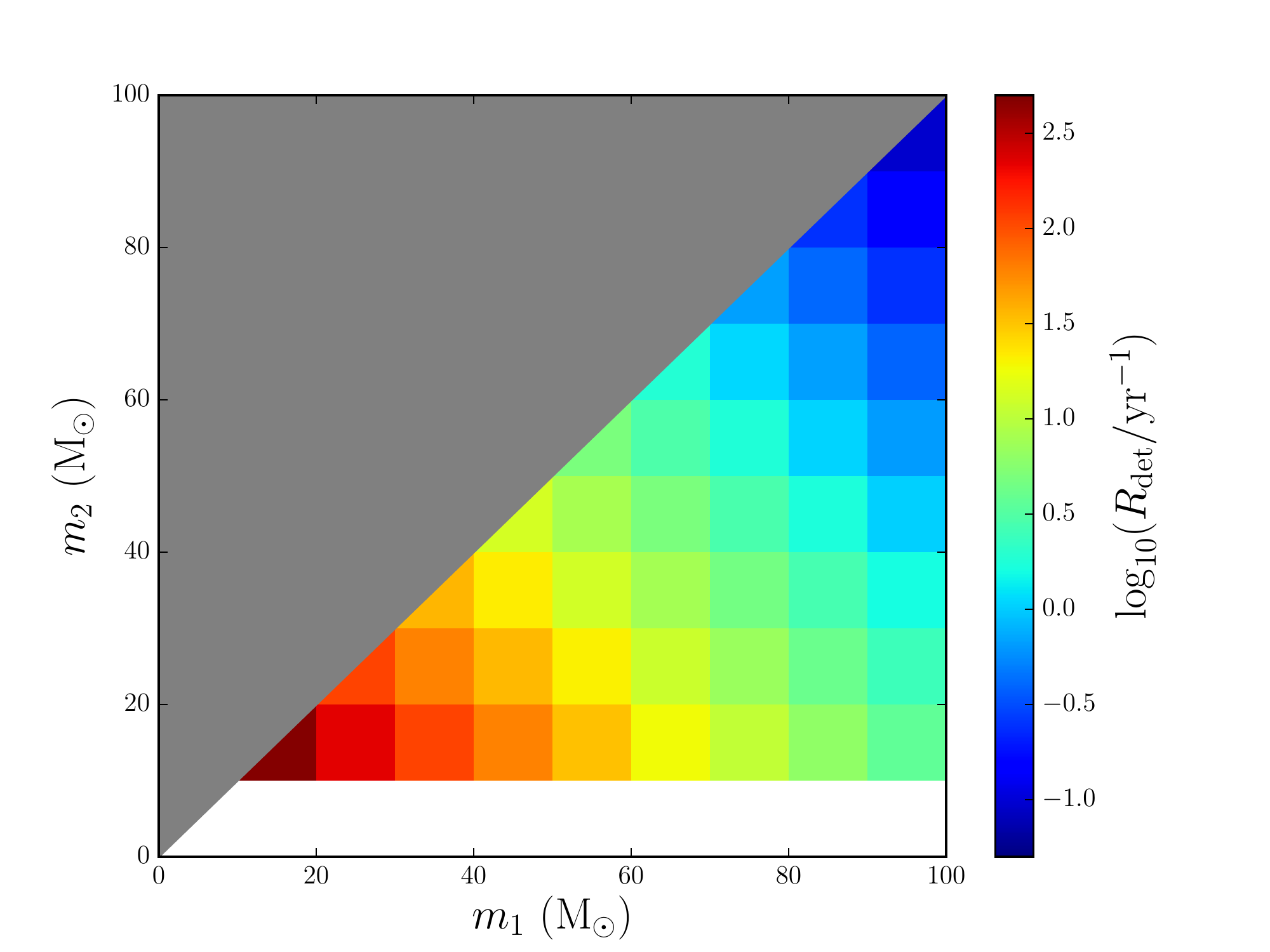}
\includegraphics[width=0.49\textwidth]{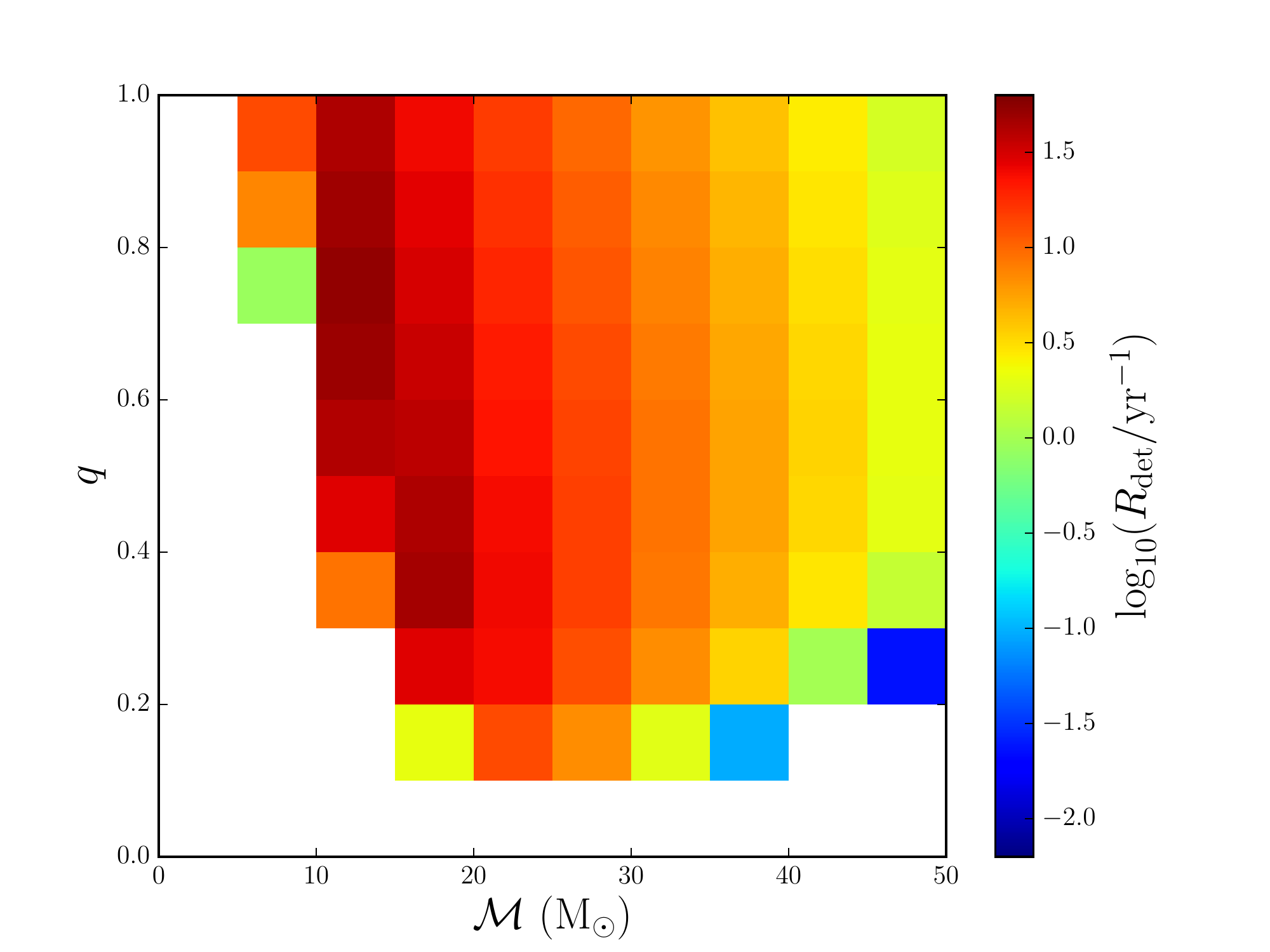}
\includegraphics[width=0.49\textwidth]{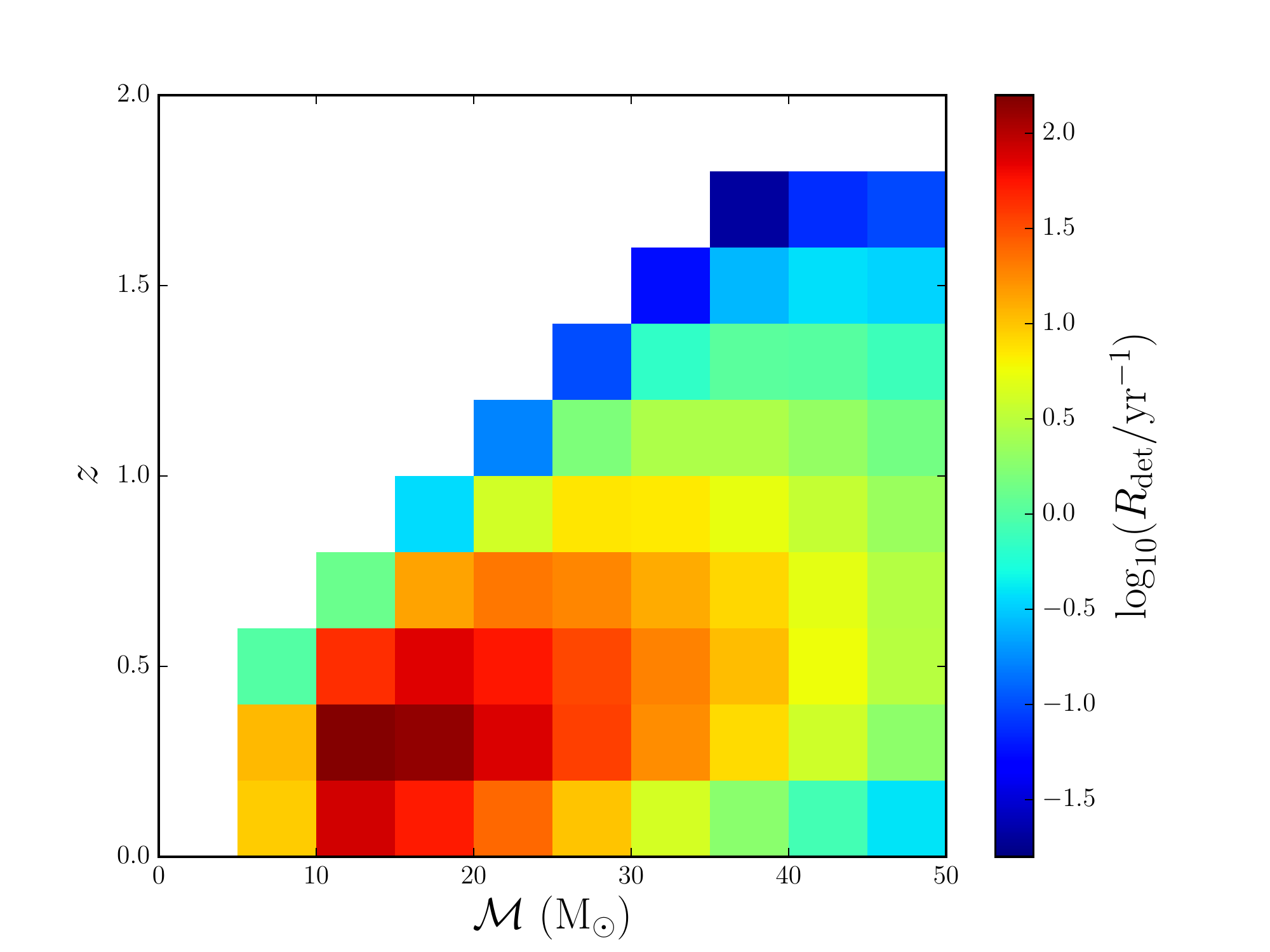}
\includegraphics[width=0.49\textwidth]{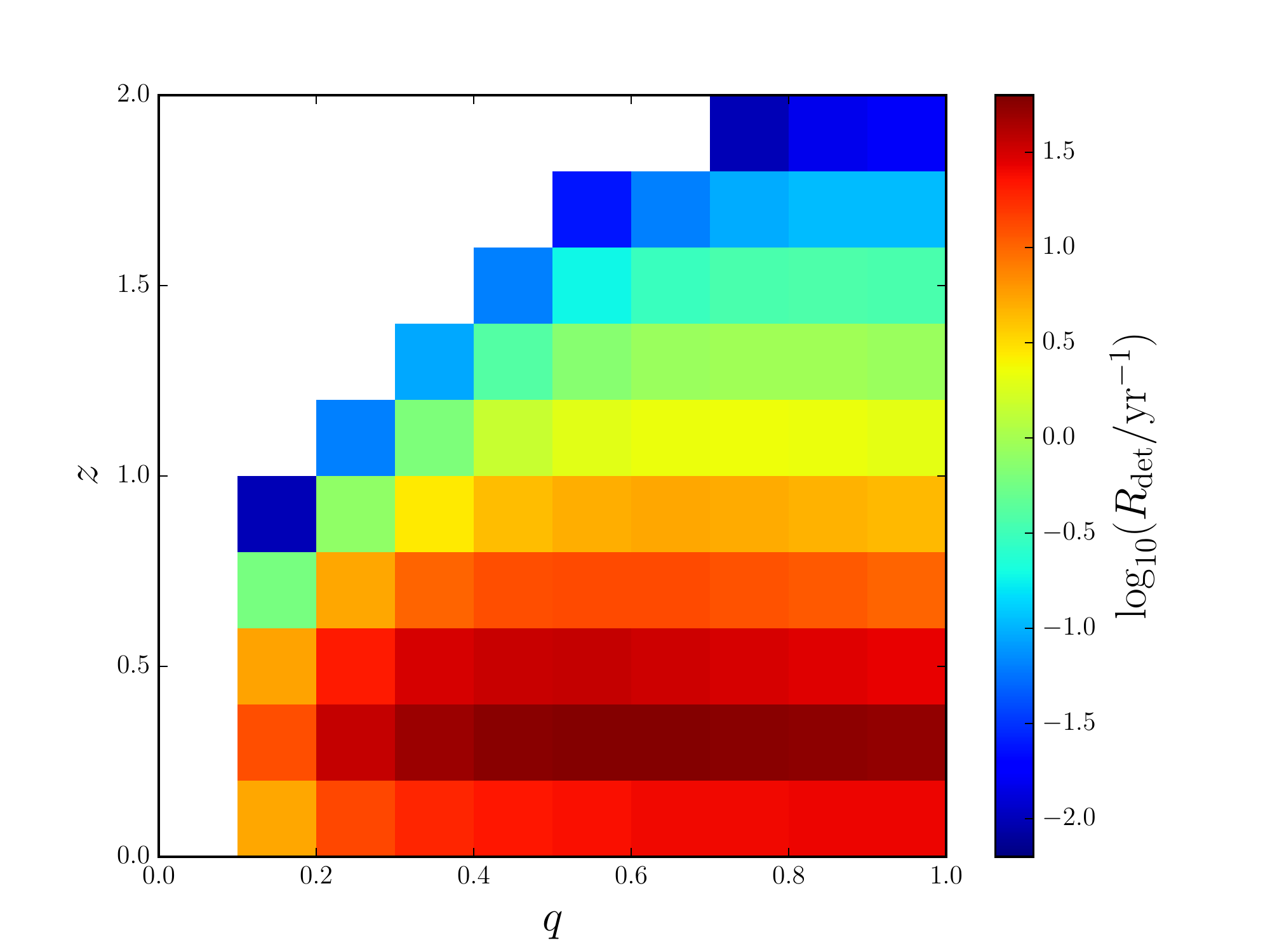}
\caption{2D Merger rate distributions in individual masses, mass ratio $q$, chirp mass $\mathcal{M}$ and redshift $z$ for a power-law distribution with $m_\mathrm{min}=10\ \Msun$. All plots have $f_\PBH=10^{-2}$. The white area indicates no significant merger rate and the grey triangle indicates the case $m_2<m_1$. The colorbar limits are the same as in fig.~\ref{fig:Design-PL-2D-mMin5} for comparison.}
\label{fig:Design-PL-2D-mMin10}
\end{figure}

\newpage
Figures \ref{fig:Design-Lognormal-2D-s03} to \ref{fig:Design-PL-2D-mMin10} show the design sensitivity 2D distributions for a lognormal mass distribution with $m_c=20\ \Msun$ and $\sigma=0.3$, and two power-law mass distributions with $\alpha=3/2$ and $m_\mathrm{min}=5\ \Msun$ and $10\ \Msun$ respectively.

\newpage
\subsection*{O1O2 sensitivity plots}
\begin{figure}[H]
\centering
\includegraphics[width=0.49\textwidth]{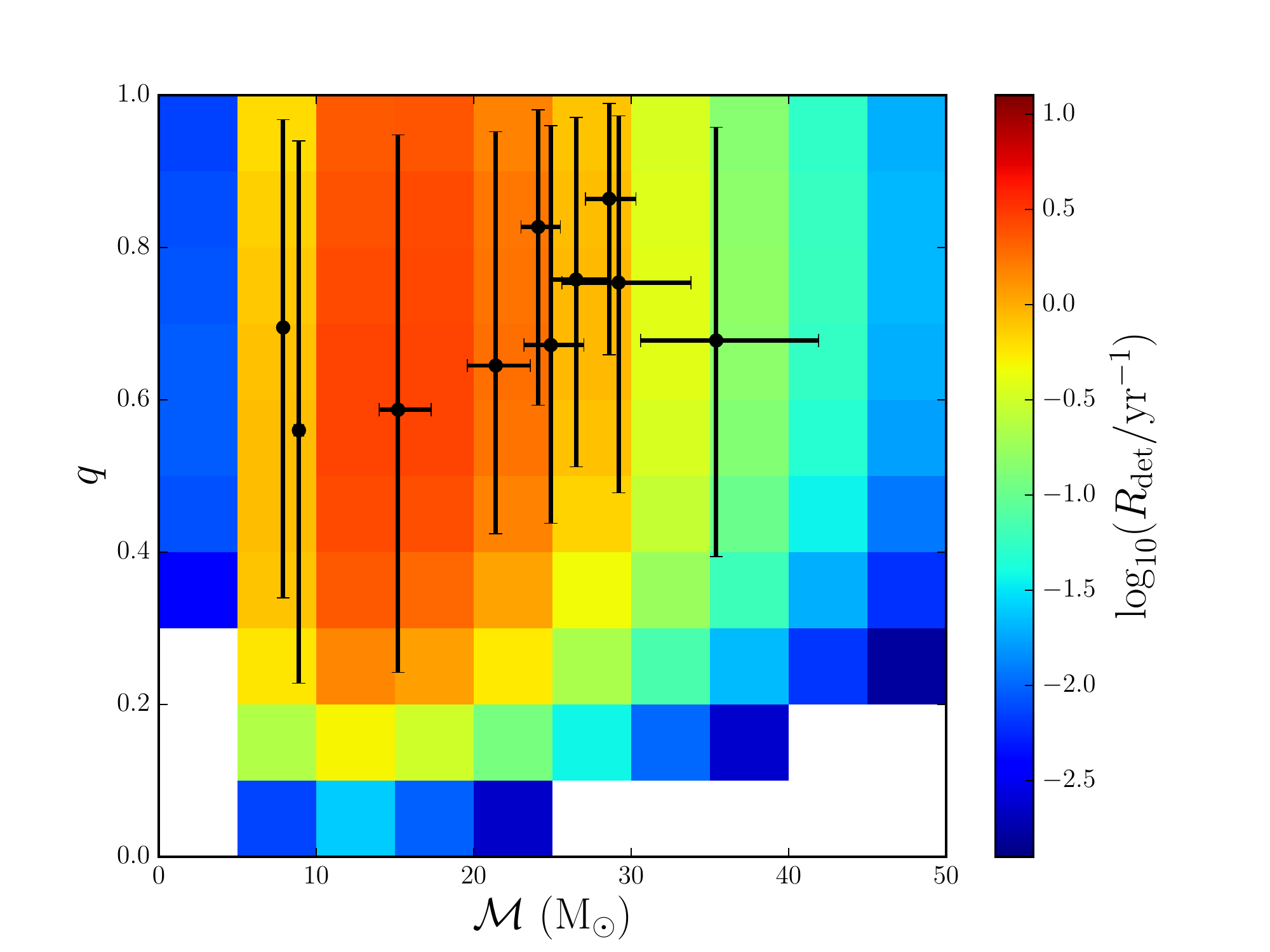}
\includegraphics[width=0.49\textwidth]{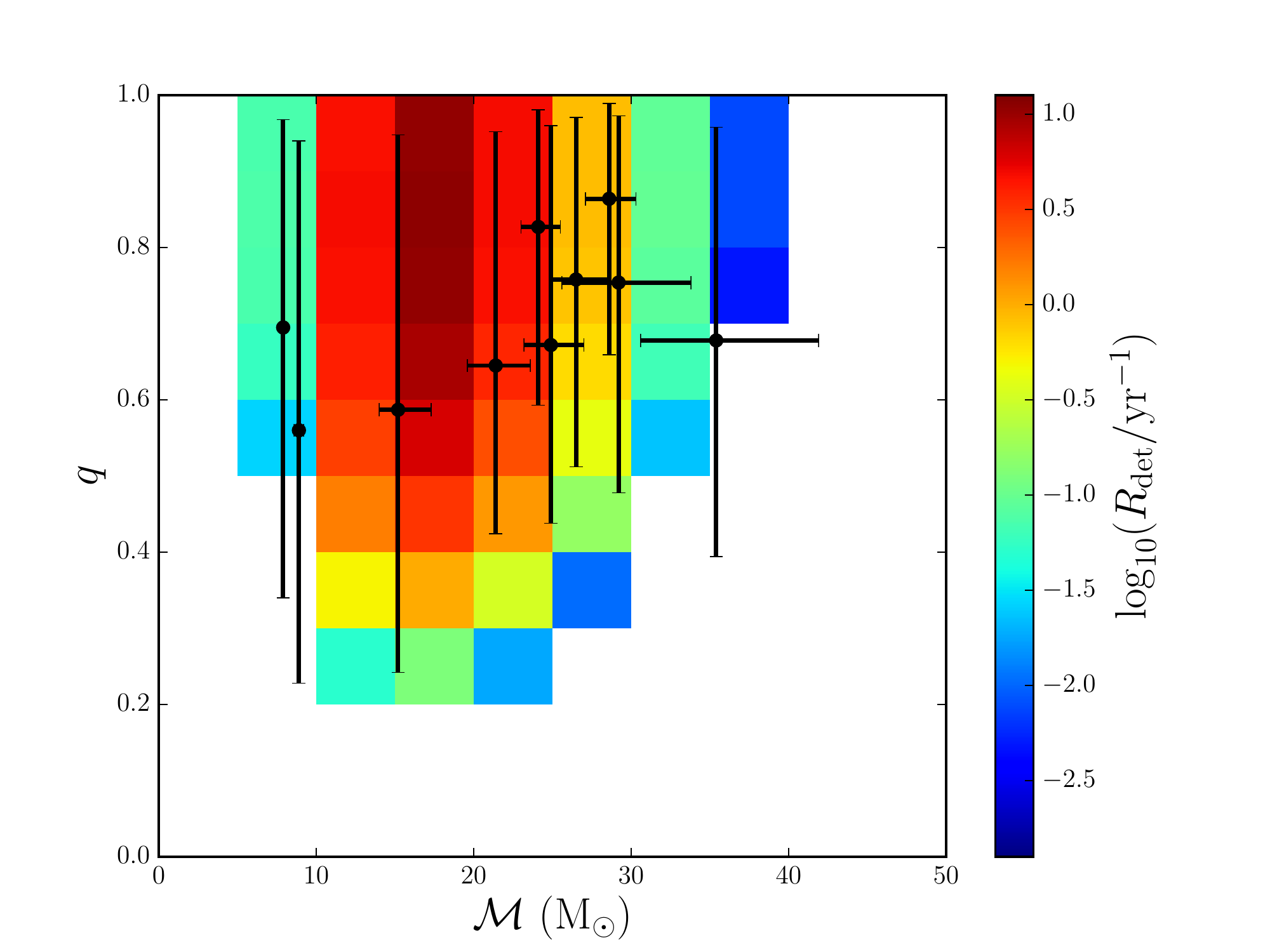}
\includegraphics[width=0.49\textwidth]{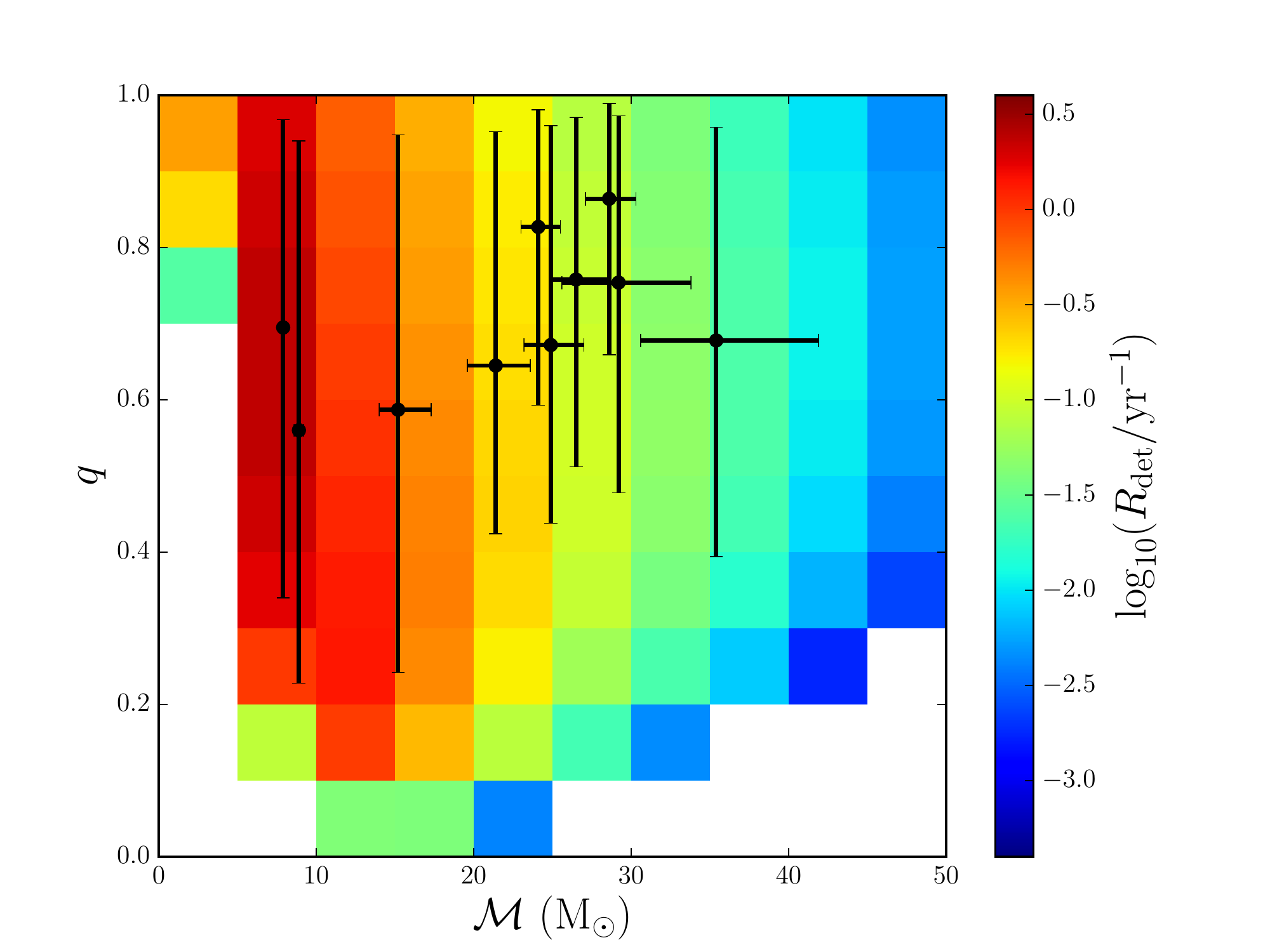}
\includegraphics[width=0.49\textwidth]{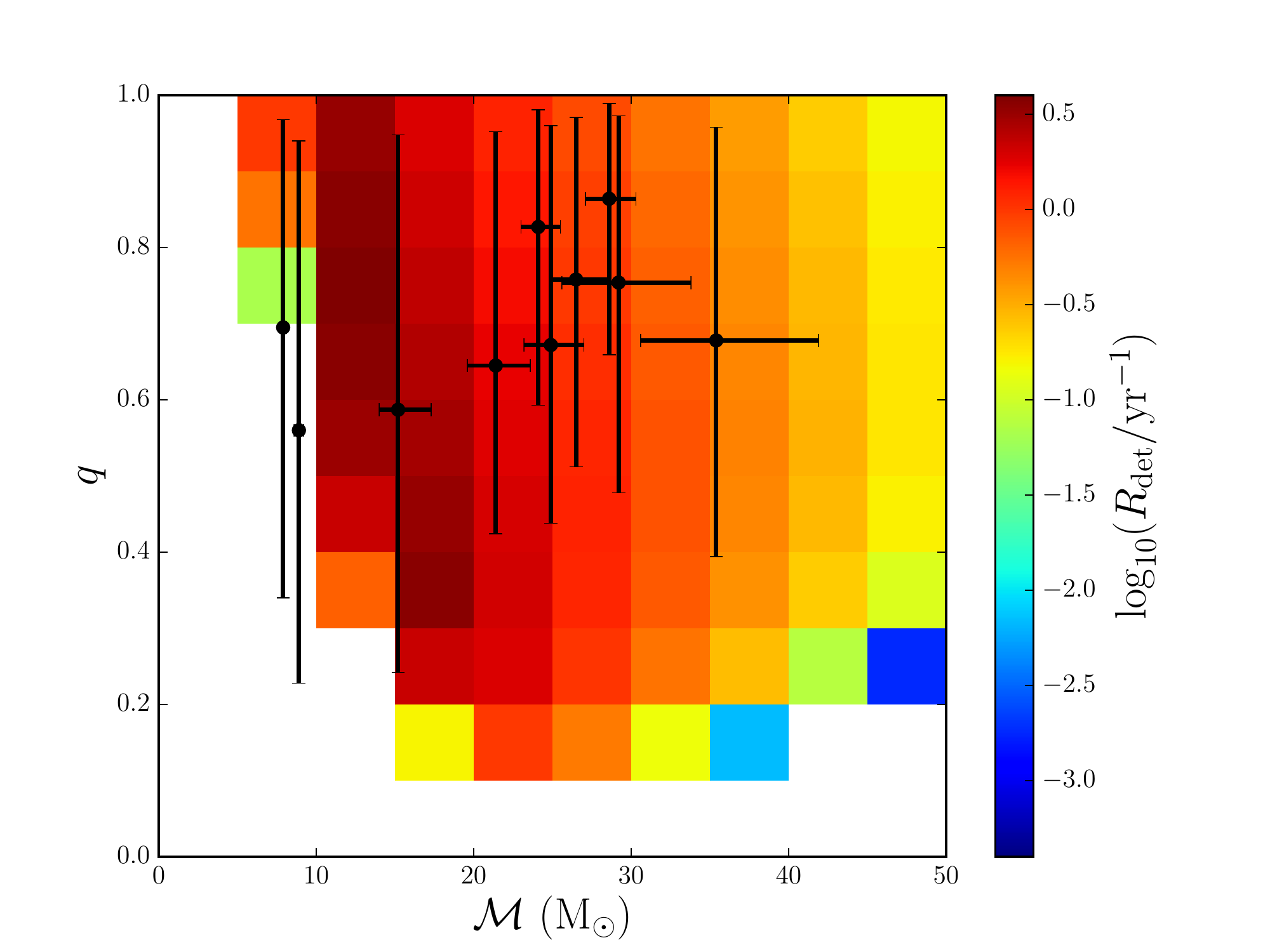}
\caption{2D Merger rate distributions in mass ratio $q$ and chirp mass $\mathcal{M}$ for the four distributions considered with the O1O2 sensitivity. The top row has the lognormal mass distribution with widths $\sigma=0.6$~(left) and $0.3$~(right), and the bottom row has the power-law mass distribution with minimum mass $5\ \Msun$~(left) and $10\ \Msun$~(right). All plots have $f_\PBH=10^{-2}$. The white area indicates no significant merger rate. The colorbar limits are the same for the top two plots (lognormal distribution), and for the bottom two plots (power-law distribution). The LIGO values and their 90\% confidence limits are shown in black.}
\label{fig:O1O2-rvqMc-all-dists}
\end{figure}

\begin{figure}[H]
\centering
\includegraphics[width=0.49\textwidth]{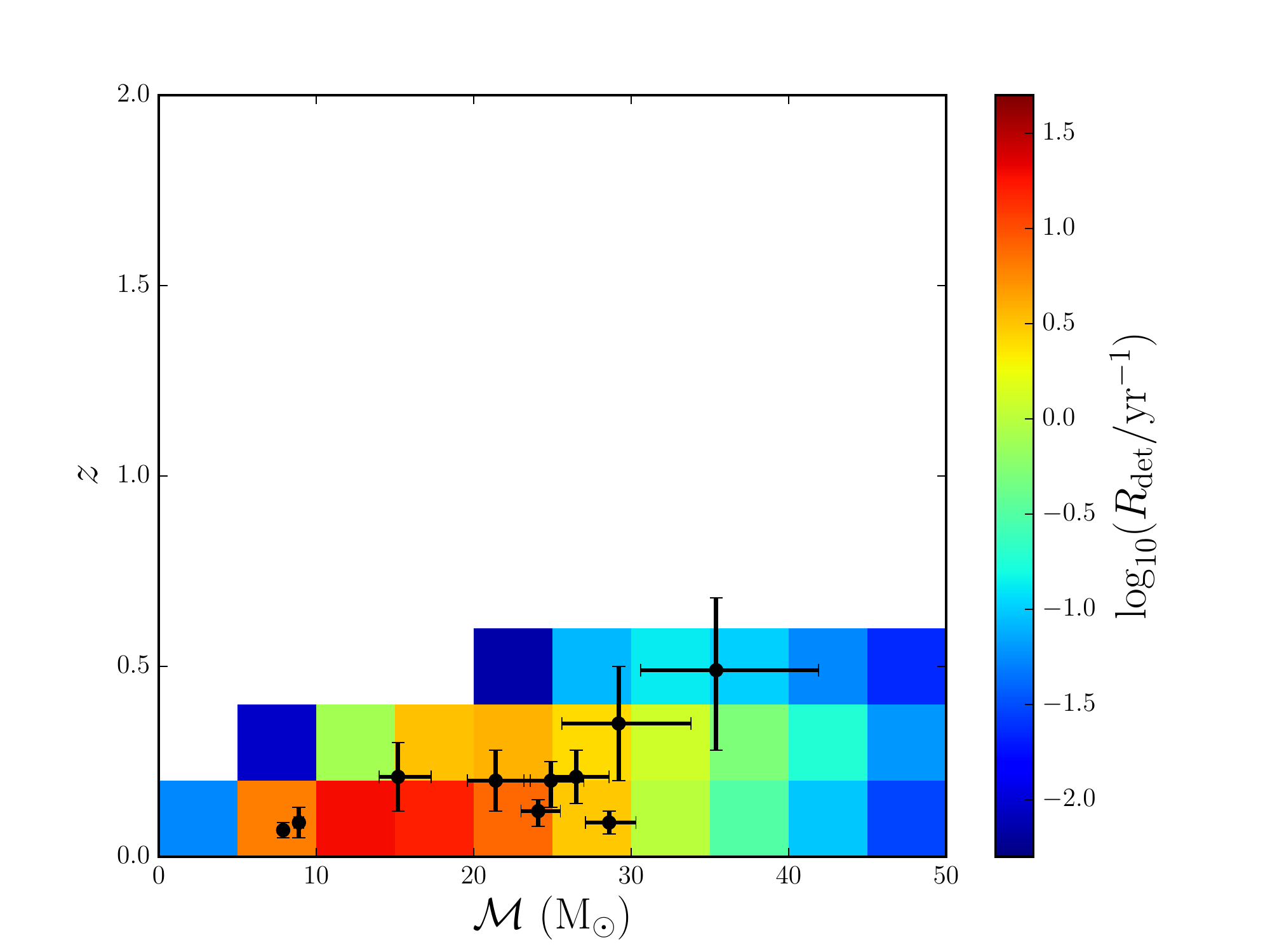}
\includegraphics[width=0.49\textwidth]{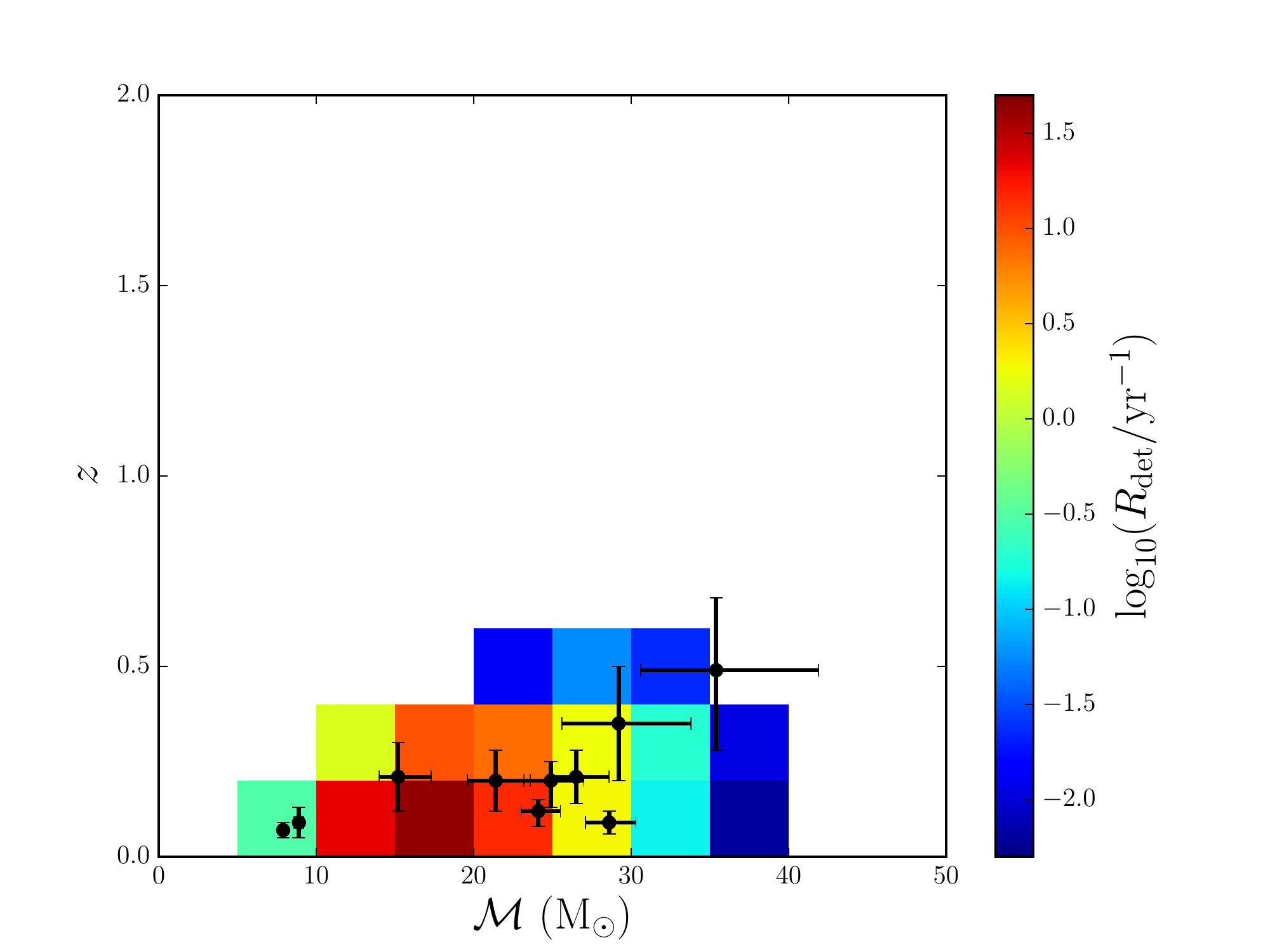}
\includegraphics[width=0.49\textwidth]{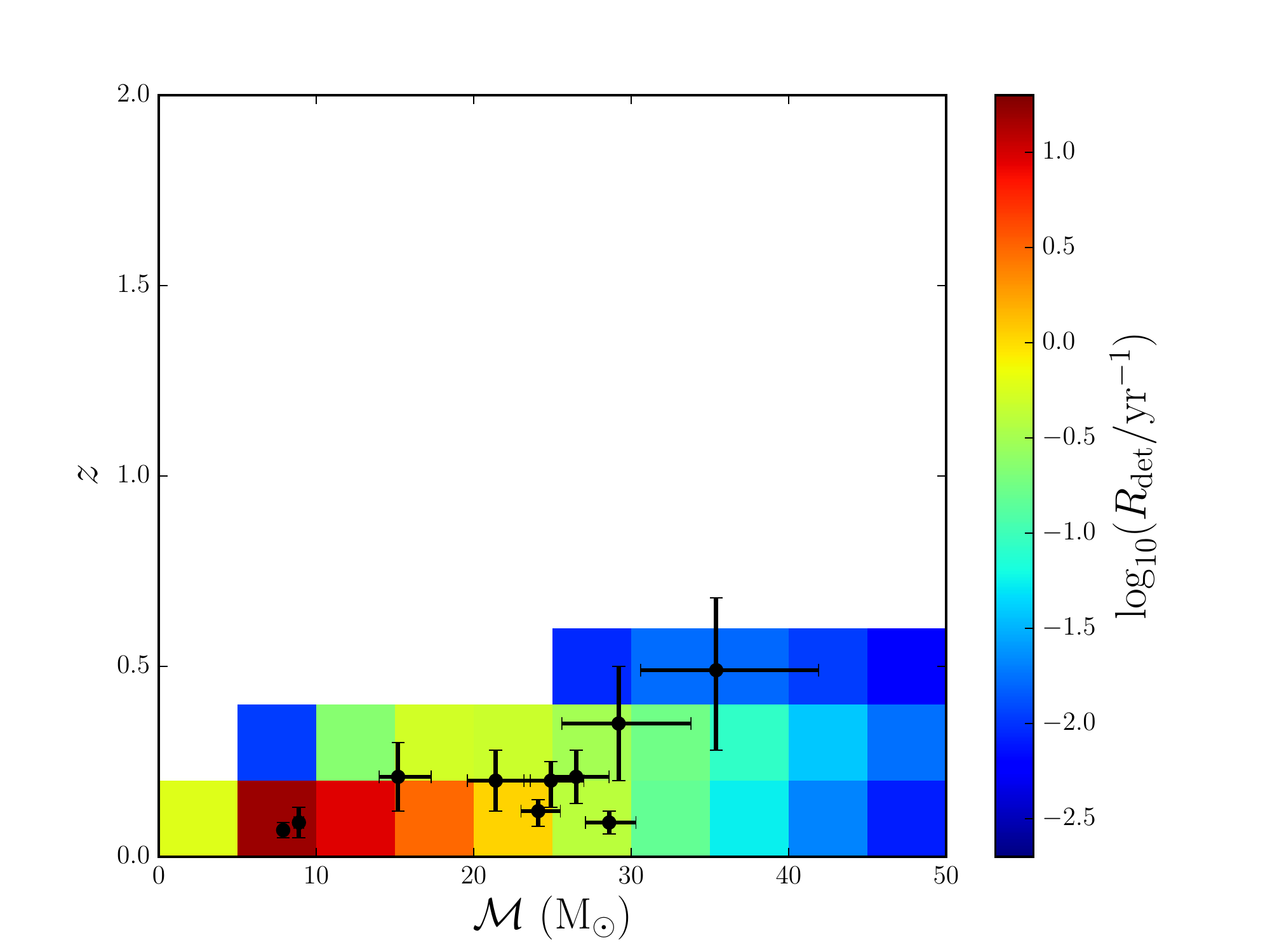}
\includegraphics[width=0.49\textwidth]{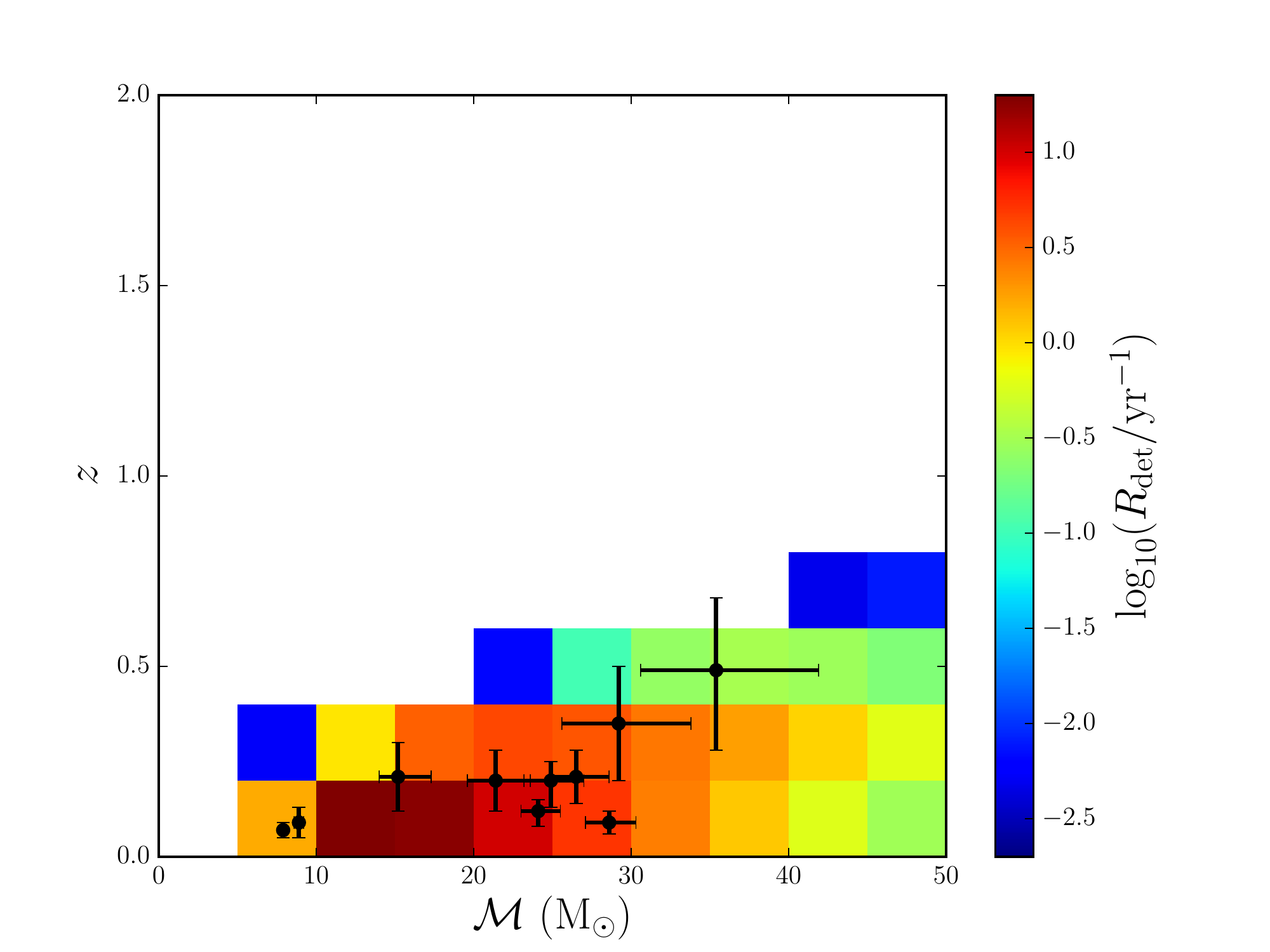}
\caption{2D Merger rate distributions in redshift $z$ and chirp mass $\mathcal{M}$ for the four distributions considered with the O1O2 sensitivity. The top row has the lognormal mass distribution with widths $\sigma=0.6$~(left) and $0.3$~(right), and the bottom row has the power-law mass distribution with minimum mass $5\ \Msun$~(left) and $10\ \Msun$~(right). All plots have $f_\PBH=10^{-2}$. The white area indicates no significant merger rate. The colorbar limits are the same for the top two plots (lognormal distribution), and for the bottom two plots (power-law distribution). The LIGO values and their 90\% confidence limits are shown in black.}
\label{fig:O1O2-rvzMc-all-dists}
\end{figure}

\begin{figure}[H]
\centering
\includegraphics[width=0.49\textwidth]{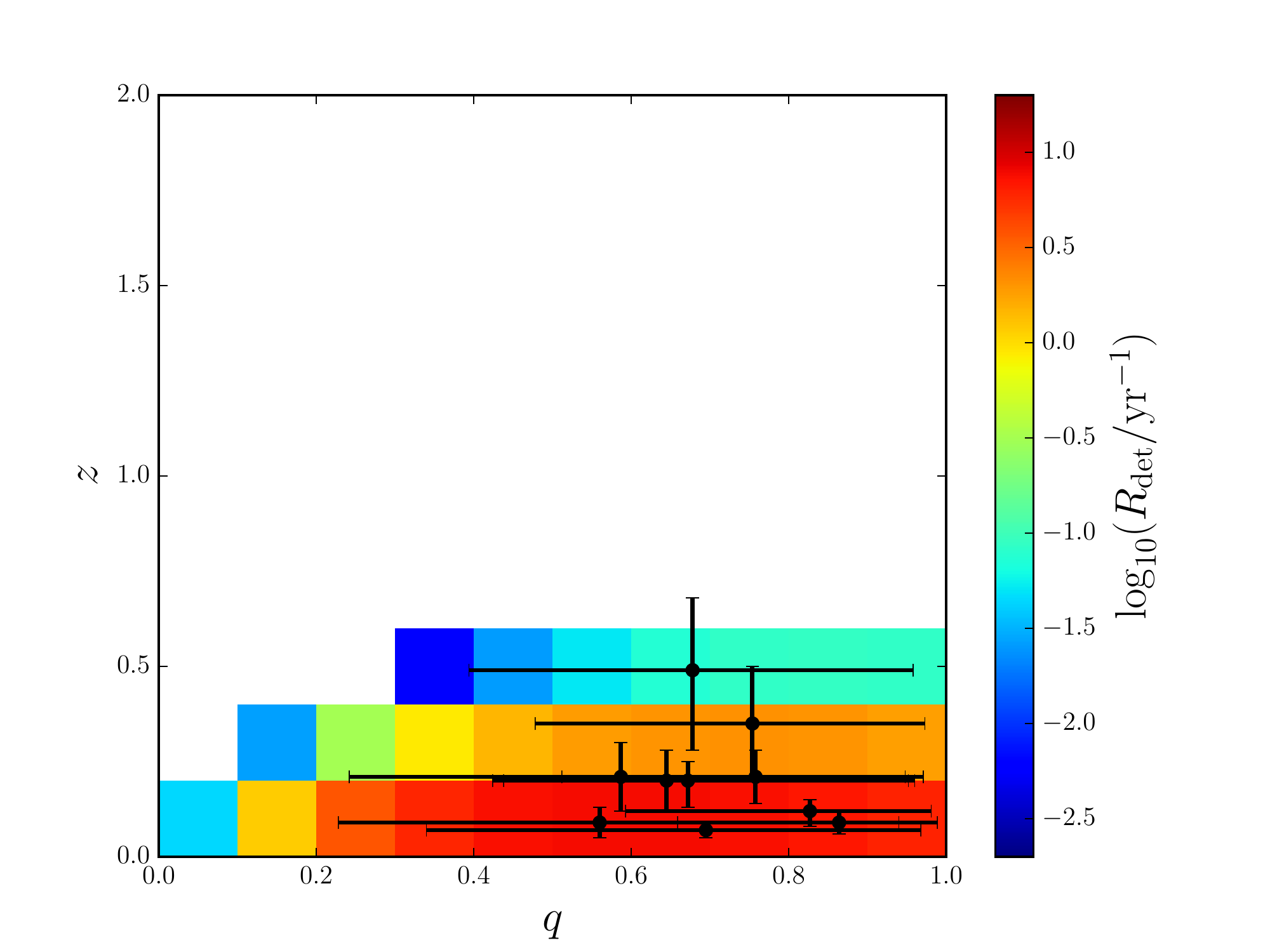}
\includegraphics[width=0.49\textwidth]{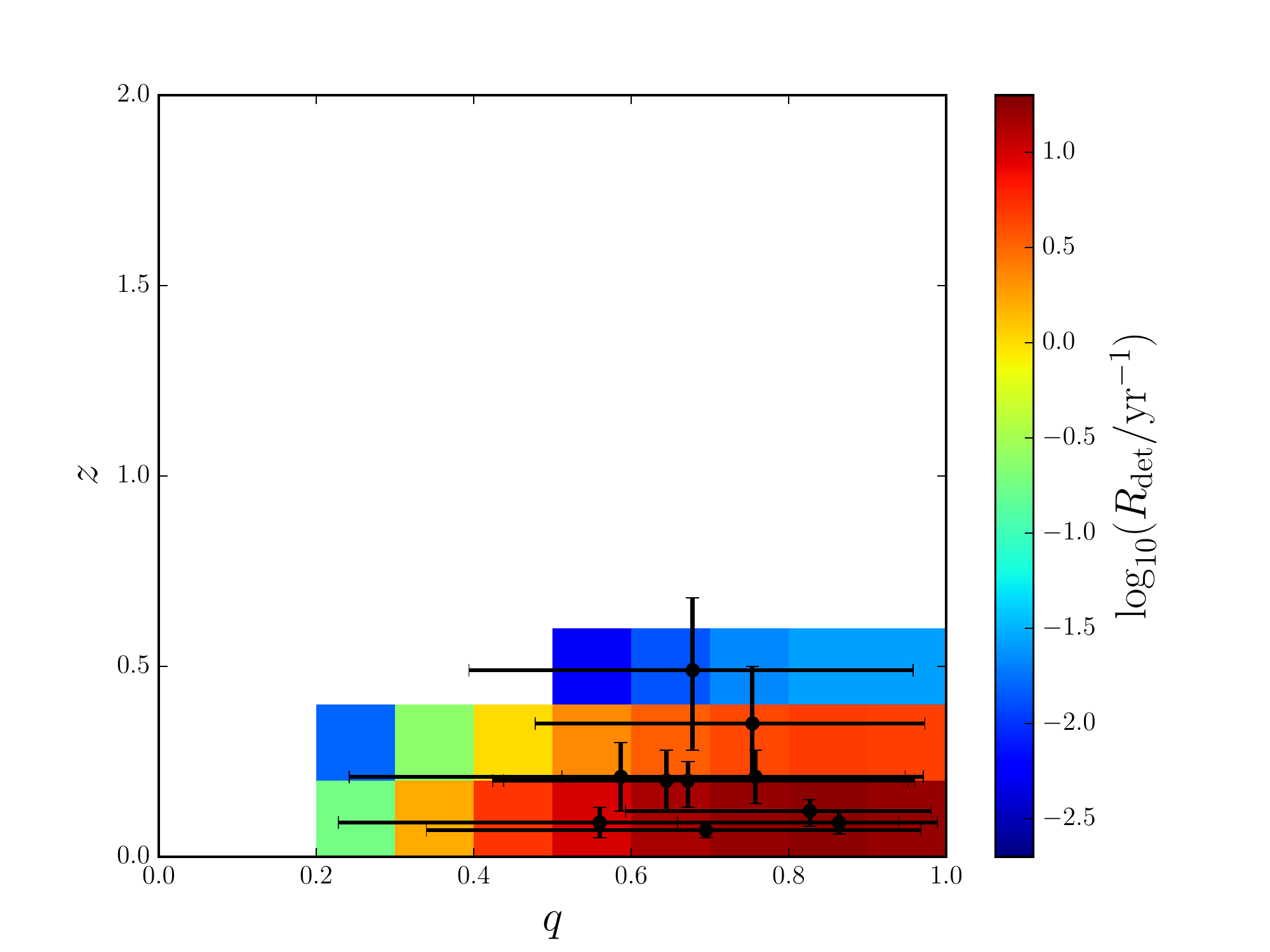}
\includegraphics[width=0.49\textwidth]{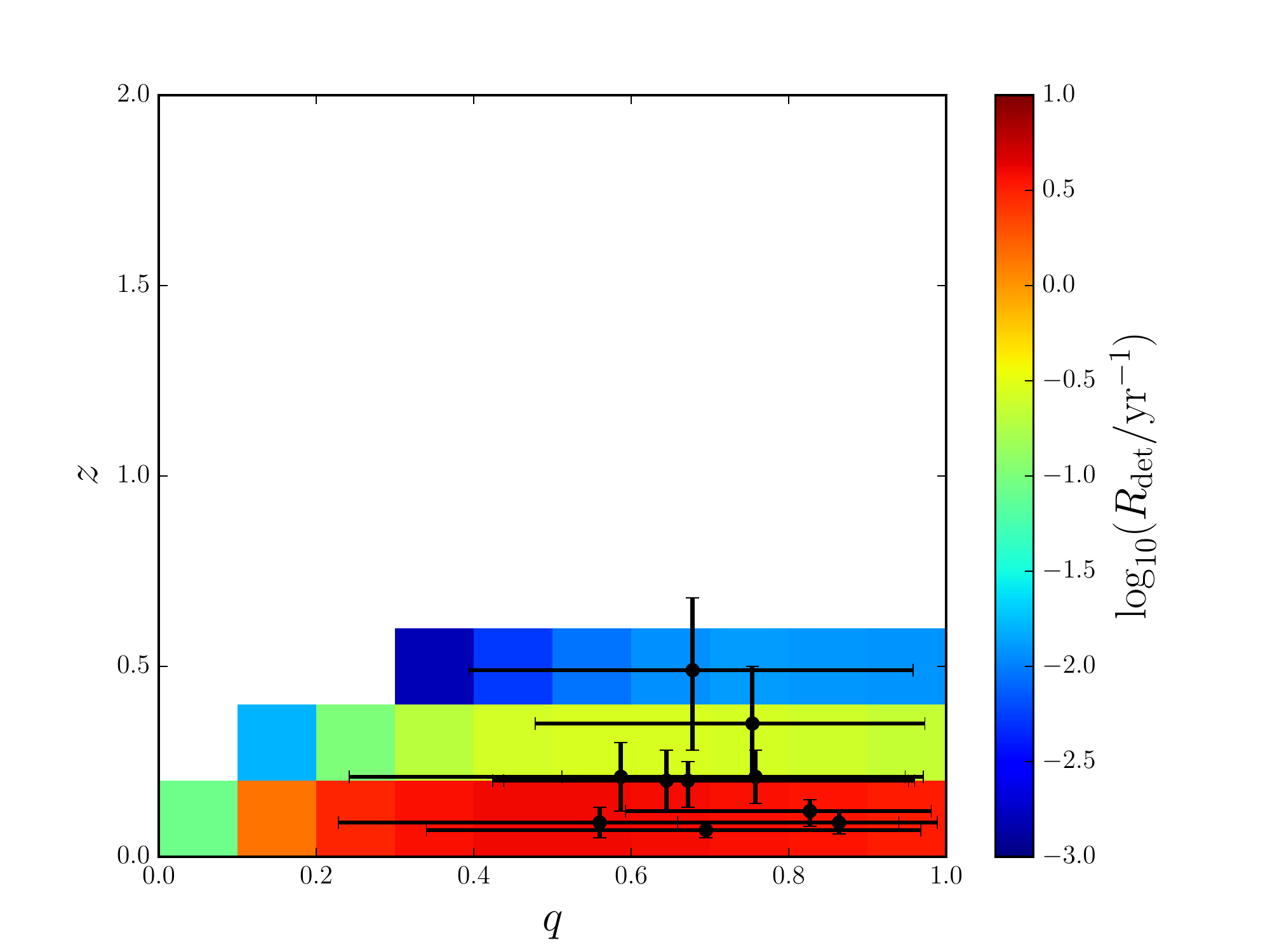}
\includegraphics[width=0.49\textwidth]{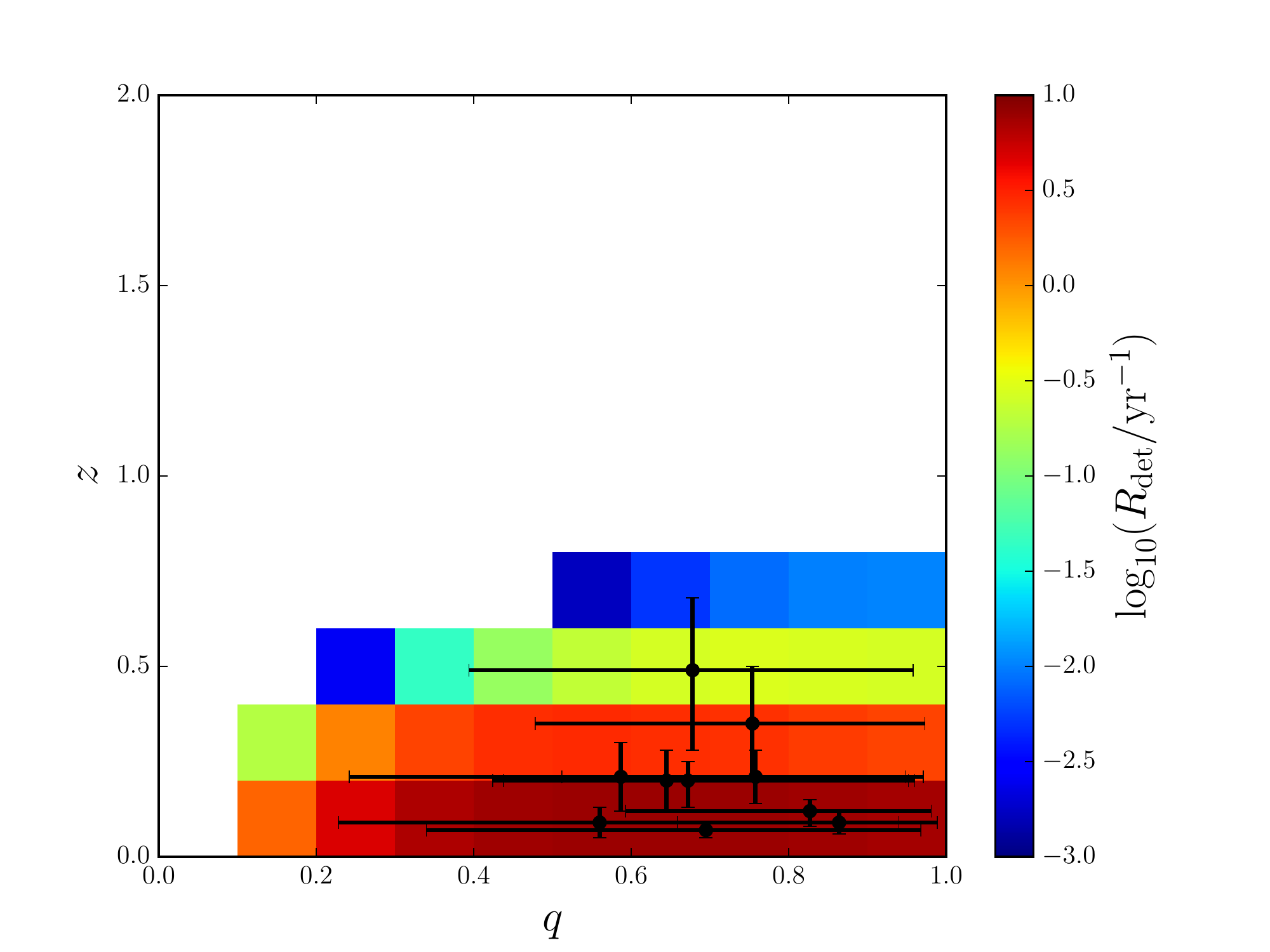}
\caption{2D Merger rate distributions in redshift $z$ and mass ratio $q$ for the four distributions considered with the O1O2 sensitivity. The top row has the lognormal mass distribution with widths $\sigma=0.6$~(left) and $0.3$~(right), and the bottom row has the power-law mass distribution with minimum mass $5\ \Msun$~(left) and $10\ \Msun$~(right). All plots have $f_\PBH=10^{-2}$. The white area indicates no significant merger rate. The colorbar limits are the same for the top two plots (lognormal distribution), and for the bottom two plots (power-law distribution). The LIGO values and their 90\% confidence limits are shown in black.}
\label{fig:O1O2-rvzq-all-dists}
\end{figure}

\vspace{-2.5em}
Figures \ref{fig:O1O2-rvqMc-all-dists} to \ref{fig:O1O2-rvzq-all-dists} show the O1O2 sensitivity 2D distributions for the four mass distributions in the observables $(\mathcal{M}$, $q)$, $(\mathcal{M}$, $z)$, and $(q$, $z)$ respectively. The LIGO values and their 1D marginalised 90\% confidence limits are shown by the black dots and their error bars.
\newpage
\bibliographystyle{JHEP-edit} 
\bibliography{MergerRatePaper}{}

\end{document}